\documentclass[twocolumn,preprintnumbers,amsmath,amssymb,nofootinbib,epsfig,bmamsfonts,yfonts,superscriptaddress]{revtex4}
\usepackage{graphicx}
\usepackage{enumerate}
\relax

\usepackage[normalem]{ulem}

\begin{document}

\title{Flow in AA and pA as an interplay of fluid-like and non-fluid like excitations}

\author{Aleksi Kurkela}
\email{a.k@cern.ch}
\affiliation{Theoretical Physics Department, CERN, CH-1211 Gen\`eve 23, Switzerland}
\affiliation{Faculty of Science and Technology, University of Stavanger, 4036 Stavanger, Norway}
\author{Urs Achim Wiedemann}
\email{urs.wiedemann@cern.ch}
\affiliation{Theoretical Physics Department, CERN, CH-1211 Gen\`eve 23, Switzerland}
\author{Bin Wu}
\email{b.wu@cern.ch}
\affiliation{Theoretical Physics Department, CERN, CH-1211 Gen\`eve 23, Switzerland}

\preprint{CERN-TH-2019-066}

\begin{abstract} 
To study the microscopic structure of quark-gluon plasma, data from hadronic collisions must be
confronted with models that go beyond fluid dynamics. Here, we study a simple kinetic theory model
that encompasses fluid dynamics but contains also particle-like excitations in a boost invariant setting with 
no symmetries in the transverse plane and with large initial momentum asymmetries. We determine the
relative weight of fluid dynamical and particle like excitations as a function of system size and energy density
by comparing kinetic transport to results from the 0th, 1st and 2nd order gradient expansion of viscous fluid dynamics. 
We then confront this kinetic theory with data on azimuthal flow coefficients over a wide centrality range in 
PbPb collisions at the LHC, in AuAu collisions at RHIC, and in pPb collisions at the LHC. 
Evidence is presented that non-hydrodynamic excitations make the dominant contribution to collective flow 
signals in pPb collisions at the LHC and contribute significantly to flow in peripheral nucleus-nucleus collisions, 
while fluid-like excitations dominate collectivity in central nucleus-nucleus collisions at collider energies. 
\end{abstract}

\maketitle

\section{Introduction}

Azimuthal asymmetries $v_n$ in soft transverse multi-particle distributions have been measured in detail in PbPb, pPb and pp 
collisions at the LHC~\cite{ALICE:2011ab,Acharya:2018lmh,Khachatryan:2015waa,Sirunyan:2017uyl,Abelev:2014mda,Aaboud:2017acw}.
At lower energies they have been studied in nucleus-nucleus and deuteron-nucleus collisions at RHIC ~\cite{Back:2004mh,Adare:2014keg,Adamczyk:2015xjc}, and at yet lower center-of-mass energies reached in fixed-target 
experiments, {\it e.g.},~\cite{Alt:2003ab}. For sufficiently central nucleus-nucleus collisions at collider energies, model analyses support a fluid-dynamic interpretation of the
measured asymmetries~\cite{Teaney:2009qa,Heinz:2013th}. The recently established persistence of sizeable azimuthal asymmetries in the smaller pPb and pp
collision systems~\cite{Khachatryan:2015waa,Sirunyan:2017uyl,Abelev:2014mda,Aaboud:2017acw}
now raises the question of whether
small transient fluid droplets are also formed in these systems, or whether other physics mechanisms are at the origin of the observed signatures of collectivity. 
The main purpose of the present work is to address a particular variant of this central question: {\it To what extent are fluid-dynamic or particle-like 
excitations at the origin of the flow phenomena observed in proton-proton, proton-nucleus and nucleus-nucleus collisions? And how does the 
interplay between these two sources of collectivity change as a function of system size and energy density?}

\begin{figure}[t]
\includegraphics[width=0.23\textwidth]{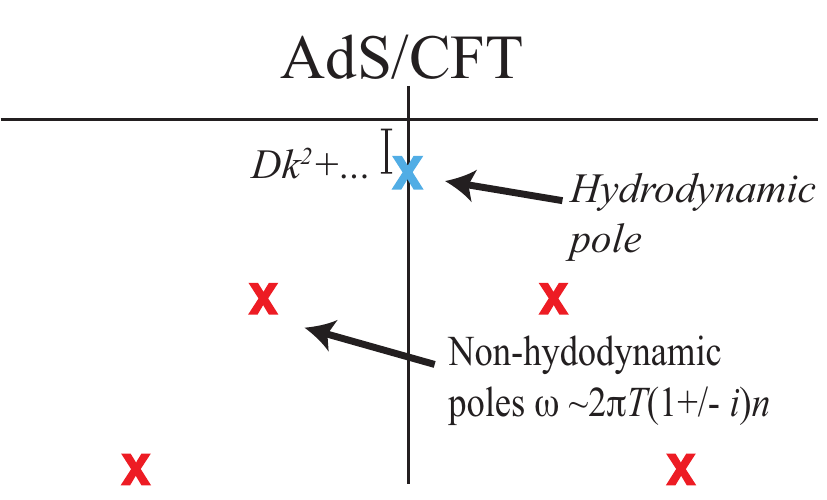}
\includegraphics[width=0.23\textwidth]{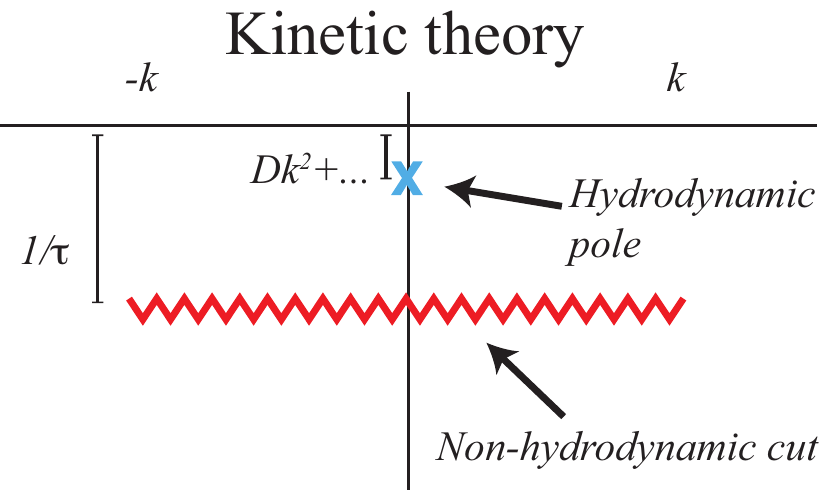}\\
\vspace{0.01\textheight}
\includegraphics[width=0.23\textwidth]{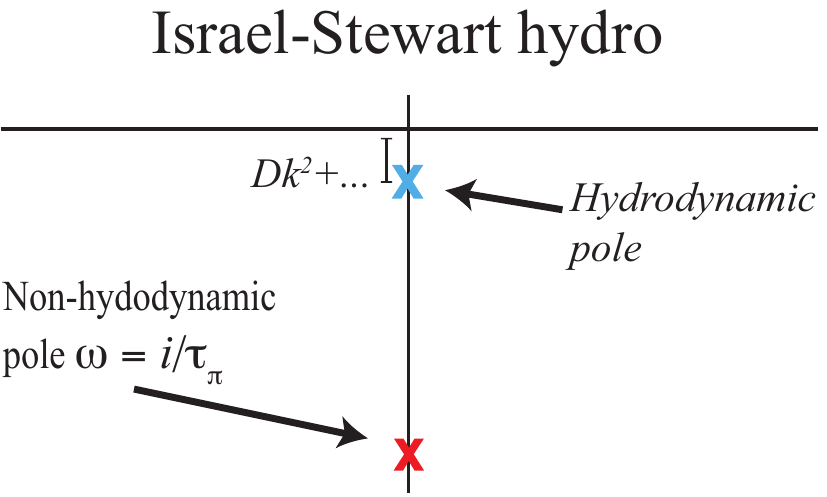}
\includegraphics[width=0.23\textwidth]{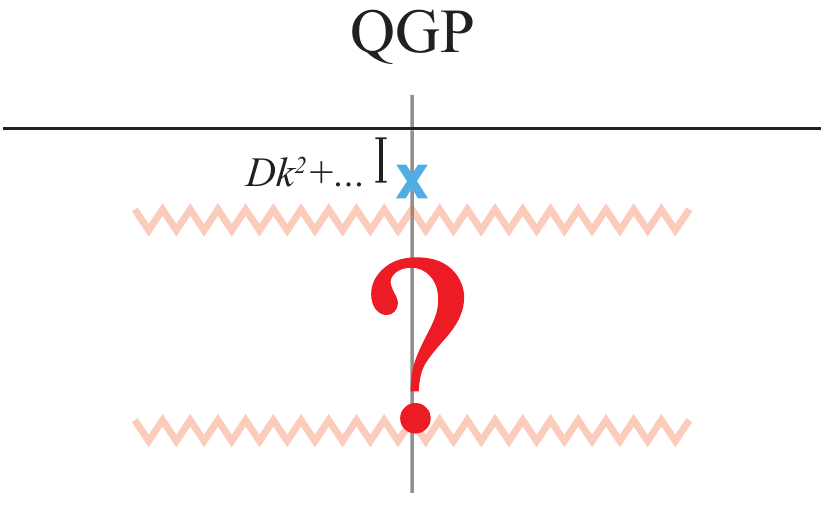}
\caption{
Different analytic structures appearing in microscopic models of fluids. The presence of hydrodynamic 
poles in the long wavelength limit is a common feature of all Lorentz symmetric theories with self-interactions. On the contrary, the 
structure of the non-hydrodynamic sector closely reflects the underlying microscopic degrees of freedom. 
}
\label{fig1}
\end{figure}

To set the stage for these questions, we recall that any self-interacting matter 
that does not spontaneously break Lorentz symmetry carries both fluid-dynamic and non-fluid dynamic
excitations. 
Analysis of the retarded two-point correlation functions $G_R^{\mu\nu,\alpha\beta}(x;t) = \langle [ T^{\mu\nu}(x,t), \, T^{\alpha\beta}(0,0)]\rangle$
is one way of characterizing the excitations that a system can carry. In particular, the Fourier transform $G_R^{\mu\nu,\alpha\beta}(k,\omega)$, understood
as a function of complex frequency $\omega$, displays non-analytic structures that can be related unambiguously to excitable physical degrees of freedom.
Fluid dynamic excitations correspond to poles at positions $\omega_{\rm hyd}(k)$ close to the real axis in the negative imaginary 
$\omega$ half plane. In all known stable and causal dynamics, they are supplemented by other non-analytic structures whose precise nature depends on the microscopic dynamics. Prominent examples are sketched in Fig. \ref{fig1}.
For instance, for the non-abelian plasma of ${\cal N}=4$ super-Yang-Mills theory in the large-$N_c$ limit at strong coupling, $G_R^{\mu\nu,\alpha\beta}(k,\omega)$ displays---in addition to
fluid dynamic poles---a tower of quasi-normal modes that are located 
deep in the negative complex plane at depths $\propto - i 2\pi T  n$, $n\in 1,2,3,...$~\cite{Son:2002sd,Starinets:2002br,Hartnoll:2005ju}; 
somewhat distorted versions of this non-fluid dynamic pole structure are found in a class of related, strongly coupled quantum field theories with gravity duals~\cite{Grozdanov:2016vgg,Grozdanov:2016fkt,Grozdanov:2018gfx}.
In contrast, kinetic theory in the
relaxation time approximation (RTA) supplements fluid-dynamic excitations with a branch-cut of quasi-particle excitations that is located at a depth $-i/\tau_R$ set 
by the relaxation time $\tau_R$~\cite{Romatschke:2015gic}; scale-dependent versions of this RTA allow for more general but related non-analytic structures~\cite{Kurkela:2017xis}.
Also Israel-Stewart dynamics~\cite{Israel:1979wp} is more than viscous fluid dynamics. The ad-hoc requirement that the shear-viscous tensor relaxes to the 
constitutive fluid-dynamic 1st-order relation on a timescale $\tau_\pi$ results in a non-fluid-dynamic mode in Israel-Stewart dynamics.

Very little is known about the possible excitations in finite-temperature QCD at energy densities attained in heavy-ion collisions. 
In the limit of vanishing interactions, $G_R^{\mu\nu,\alpha\beta}(k,\omega)$ has a tower of branch cuts corresponding to different
Matsubara frequencies~\cite{Hartnoll:2005ju}, but one may only speculate whether these structures evolve gradually with the strength of $\alpha_s$.  

The experimental determination of these non-hydrodynamic structures would be tantamount to determining 
the microscopic structure of quark-gluon plasma. While experimental reality prevents
us from directly measuring the response of a static and infinite quark-gluon plasma to a linearized perturbation and directly 
accessing the non-hydrodynamic structures in Fig.~\ref{fig1}, the non-hydrodynamic sector may leave its imprint to the
dynamical evolution of the system created in physical collisions. This is especially the case if the hydrodynamic and
non-hydrodynamic modes are not well separated from each other, or even more so if the non-hydrodynamic modes are closer to the 
real axis than the hydrodynamic ones and thus dominate the evolution of the system. 

The non-hydrodynamic sector becomes more prominent in smaller systems.
Firstly, the smallest wavenumber that excitations in a system with transverse size $R$ can have is $k \geq 1/R$. 
With decreasing $R$, the spectrum of fluid-dynamic excitations in the system thus corresponds to poles that lie deeper and deeper in the negative imaginary $\omega$-plane (at positions ${\rm Im}\left[\omega_{\rm hyd} \right] \lesssim -  \textstyle\frac{\eta}{\varepsilon + p} \textstyle\frac{1}{R^2}$ for the shear channel), and thus approach the non-hydrodynamic sector.
Secondly, in a system that lives for a lifetime of $\Delta \tau$, the information of modes with ${\rm Im}(\omega)> 1/\Delta \tau$ is lost as the decay of the modes is dictated by the imaginary 
part of frequency $\exp({\rm Im}(\omega) \tau)$. 
Therefore with decreasing $\Delta \tau \sim R$, one may access an increasing amount of information about the non-hydrodynamic structures that lie deeper in the complex plane. 
For both of these reasons, systems that are small enough are inevitably dominated by physics that is not fluid-like and the specific way of how the hydrodynamic description fails for small 
systems carries the information of what lies beyond. 

A systematic comparison of system-size dependencies between models of different microscopic dynamics and 
measured azimuthal anisotropies may offer an empirical pathway to discriminate between different possibilities. 
 We note that while disagreement with the data would exclude a given model, the opposite does not hold true. 
In fact, an example of such an agreement comes from the phenomenological success of Israel-Stewart hydrodynamics in small systems
where the dynamics is dominated by the non-hydrodynamic pole governing the relaxation of the shear-stress tensor to its Navier-Stokes value. 
Apparent agreement with data surely does not imply that the microscopic dynamics of Israel-Stewart theory are at play in Nature but 
does mean that the non-hydrodynamic sector of Israel-Stewart theory at least somewhat resembles that of the true underlying dynamics. 

To what extent we may discriminate whether the underlying dynamics contains quasiparticles, quasi-normal modes of AdS black hole, or perhaps something else depends to what extent we can identify features inconsistent with data with decreasing system size. 
The most prominent feature exhibited by the small systems is the free-streaming dynamics upon which the standard modelling of pp-collisions is based. Hence, it is a natural starting point to ask how this limit is approached in the only one of the above mentioned models that does support free streaming, namely the kinetic theory.

\section{Background}
Kinetic transport has been considered since the beginning in the study of relativistic nucleus-nucleus collisions~\cite{Baym:1984np}. On the one hand, there are parton-cascade formulations~\cite{Gyulassy:1997ib,Zhang:1998tj,Zhang:1999rs,Molnar:2000jh,Molnar:2001ux} that follow the time evolution of the individual partons. On the other hand, direct solutions of the Boltzmann transport equation trace the particle distribution functions. 

Parton cascades are nowadays embedded in widely used 
phenomenological models like the AMPT~\cite{Xu:2011jm} and BAMPS~\cite{Xu:2004mz} that aim at a complete dynamical description of all stages of nucleus-nucleus collisions, and that reproduce in particular many aspects of the measured azimuthal anisotropies 
in nucleus-nucleus collisions, and their system size dependence~\cite{Koop:2015wea,Uphoff:2014cba,He:2015hfa,Greif:2017bnr}.
Parton cascades have also been used to study the Boltzmann transport equation in isolation with the aim of gaining qualitative insights into how kinetic theory approaches a hydrodynamic regime~\cite{Zhang:1998tj,Molnar:2001ux,Huovinen:2008te,El:2009vj,Gallmeister:2018mcn}, how it builds up collective flow~\cite{Gombeaud:2007ub, Drescher:2007cd,Ferini:2008he,Borghini:2010hy}, and how this affects the flow of heavy quarks~\cite{Bhaduri:2018iwr}. 
In addition to the above mentioned partonic cascades, we mention in passing that hadronic transport codes, such as \cite{Sorge:1989vt,Bleicher:1999xi,Petersen:2018jag}, play an important role in phenomenology.  

Aspects of real-time dynamics of QCD can be followed in an effective kinetic theory (EKT)~\cite{Arnold:2002zm} which provides a systematic expansion in the strong coupling constant $\alpha_s$. Transport coefficients of weak-coupling QCD have been evaluated by directly solving the corresponding Boltzmann transport equation near thermal equilibrium~\cite{Arnold:2003zc,Arnold:2006fz,Huot:2006ys,York:2008rr,Ghiglieri:2018dib,Ghiglieri:2018dgf}. Solutions to the EKT Boltzmann equation have been studied also in more complex geometries. In particular, solutions to the EKT Boltzmann equation in 1+1 dimensional longitudinally expanding systems have shown that the out-of-equilibrium systems of saturation framework (see \emph{e.g.} \cite{Gelis:2010nm})---described by classical gluodynamics at early times~\cite{Lappi:2011ju}---can be smoothly evolved to late times where the system allows for a fluid-dynamic description \cite{Baier:2000sb, Kurkela:2015qoa,Kurkela:2016vts,Kurkela:2018xxd}.  In support of its potential phenomenological relevance, this perturbative approach---if supplemented with realistic values of the coupling constant---leads to short hydrodynamization timescales~\cite{Kurkela:2015qoa} comparable to those obtained from non-perturbative strong coupling techniques~\cite{Beuf:2009cx,Chesler:2010bi,Wu:2011yd,Heller:2013oxa,Keegan:2015avk} and favored in phenomenological models. 
This approach has been extended to 2+1 dimensional solutions \cite{Keegan:2016cpi} which form the foundation of tools that can be used to match the out-of-equilibrium initial stage models to a hydrodynamic stage in nucleus-nucleus collisions~\cite{Kurkela:2018wud,Kurkela:2018vqr}. In these solutions the transverse translational symmetry is, however, broken only
by small linear perturbations limiting their use in the study of small systems which potentially do not reach the fluid-dynamic regime before disintegrating due to transverse expansion. 

In the past years there has been significant interest also in solutions to Boltzmann transport equations with simplified collision kernels, such as relaxation time approximation or diffusion-type kernels~\cite{Heller:2016rtz,Florkowski:2017jnz,Strickland:2017kux,Blaizot:2017lht}. These formulations have the advantage that due to their simpler structure they allow for more detailed analysis.  That has allowed the study of, \emph{e.g.}, the role of out-of-equilibrium attractors and their relation to the non-hydrodynamic modes \cite{Heller:2016rtz,Heller:2018qvh}. While this class of models is not directly based on QCD (cf.~\cite{Hong:2010at, Arnold:2006fz}), they do share characteristics with EKT and may be used as toy model of EKT. In addition, they may reveal qualitative features of systems with quasiparticle descriptions even in a regime where weak coupling approximations of EKT are not appropriate. 
Besides works that solve the Boltzmann transport equation perturbatively in the collision kernel around the free-streaming solution \cite{Kurkela:2018ygx,Borghini:2010hy,Romatschke:2018wgi}, these studies are still constrained to simple geometry of 1+1 dimensions with the notable exception of \cite{Behtash:2017wqg} that breaks the translational symmetry but retains a de Sitter symmetry and remains restricted to azimuthal rotational symmetry. 

The first non-perturbative solution to Boltzmann transport equation under longitudinal transverse expansion and broken transverse translational symmetry, including broken azimuthal symmetry was given in~\cite{Kurkela:2018qeb}. 
The approach in~\cite{Kurkela:2018qeb} that we document in detail in section ~\ref{sec2} starts from an essentially analytic formulation of kinetic transport in isotropization time approximation (ITA) and extends it to the calculation of azimuthal flow. The fluid-dynamic properties of the model are analytically known which allows us in section~\ref{sec3} to
disentangle unambiguously and quantitatively fluid-dynamic from non-fluid dynamic features in the dynamical evolution of kinetic theory. We thus investigate a system in which the modern
conceptual debate about a quantitatively controlled understanding of hydrodynamization that has remained largely limited so far to arguably academic, lower dimensional analytic set-ups, can be 
extended to a problem of sufficient complexity to provide insight into the current phenomenological practice. We hasten to remark that the kinetic transport theory considered here remains a
simple one-parameter model designed mainly to address conceptual questions like those posed at the beginning of this introduction. However, in an effort of pushing the relevance of this starting
point as far as possible into the realm of full phenomenological studies explored so far with multi-parameter fluid dynamic and parton cascade codes of significant complexity, we shall compare
in section~\ref{sec4} results of this one-parameter kinetic theory to results on the system size dependence of measures of collectivity in hadronic collisions at RHIC and at the LHC. 

%%%%%%%%%%%%%%%%%%%%%%%%%%%%%%%%%%%%%%%%%%%%%%%%%%%%%%%%%%%%%%%%%%%%%%%%%%%%%%%%%%%%%%%%%%%%%
\section{The model}
We solve the kinetic theory in \emph{isotropization-time approximation} (ITA) with boost invariant geometry but with 
strongly broken transverse translational symmetry. Our initial conditions have an approximate azimuthal symmetry that is broken by linear perturbations.  In this Section, we describe the model that we solve using free-streaming coordinate-system methods introduced in~\cite{Kurkela:2018qeb} that we fully document in Appendix \ref{appendA}.

\subsubsection{Isotropization-time approximation}
\label{sec2}
 Our starting point is the longitudinally boost-invariant massless kinetic transport equation \cite{Baym:1984np}
\begin{align}
\partial_\tau f + \vec{v}_\perp \cdot \partial_{\vec{x}_\perp} f -\frac{p_z}{\tau}\partial_{p_z}f = -C[f] \, .
\label{eq:1}
\end{align}
Here the distribution function $f(\tau,\vec{x}_\perp;\vec{p}_\perp,p_z)$ denotes the phase space density of massless excitations of momentum $p^{\mu}=(p, \vec{p}_\perp,p_z)$,
$p=\sqrt{\vec{p}_\perp^2 + p_z^2}$, at proper time $\tau=\sqrt{t^2 -z^2}$. $\vec{v}_\perp = \vec{p}_\perp/p$, $v_z = p_z/p$ are transverse
and longitudinal velocities, respectively.  

In the following, we will restrict our discussion to observables that can be obtained from the first $p$-integrated moment of the distribution function 
\begin{align}
F(\tau,\vec x_\perp;\Omega) \equiv \int \frac{4\pi dp\, p^3}{(2\pi)^3}f(\tau, x_\perp;p_\perp , p_z),
\label{eq2n}
\end{align}
which does not depend anymore on the magnitude of the momentum $p$, but does depend on
the direction of the propagation of particles $\vec v_\perp$ and $v_z$ collectively denoted by $\Omega$,  subject  to constraint that $v_z^2+v_\perp^2=1$.
The evolution equation of the integrated distribution function  is obtained by taking the first integral moment of eq.~(\ref{eq:1})
 \begin{equation}
 	\partial_t F + \vec{v}_\perp \cdot \partial_{\vec{x}_\perp} F - \frac{v_z(1-v_z^2)}{\tau} \partial_{v_z} F  +  \frac{4v_z^2}{\tau}F = -C[F]\, .
 	\label{eq18}
 \end{equation}
This equation closes if the collision kernel $C[F]$ can be expressed in terms of $F$. While this does not generically happen, 
it is the case in the \emph{isotropization-time approximation} (ITA). The ITA is a simple implementation of the physical picture that interactions drive the system towards isotropy even if the system is far from equilibrium and may be too short-lived to thermalize. 
The ITA therefore assumes that the 
distribution $F$ evolves  in the local rest frame $u^{\mu}(\tau,\vec x_\perp)$ to an isotropic distribution $F_{iso}$  in timescale $\tau_{iso}$,
and that the collision kernel is taken to be of the form 
 \begin{align}
 - C[F] = -\frac{(-v_\mu u^{\mu})}{\tau_{iso}}( F- F_{iso})\, ,
 \end{align}
with $v^\mu = p^\mu/p= (1,\vec v)$, and where the local rest frame is given by the Landau matching condition $u^\mu T_\mu ^\nu = - \varepsilon u^\nu$. 
  We impose conformal symmetry that guarantees that the isotropization time is related to the local energy density $\varepsilon$
 \begin{equation}
 \tau_{iso} = \frac{1}{\gamma \varepsilon^{1/4}}.
 \end{equation}
 The constant of proportionality $\gamma$ is the only model parameter in the ITA.  In particular, 
the form of the isotropic distribution
\begin{equation}
F_{iso}(\tau, x_\perp;\Omega)= \frac{\varepsilon(x_\perp, \tau)}{[-v_\mu u^\mu(x_\perp,\tau)]^4}
\end{equation}
is completely determined by Lorentz and conformal symmetry.  

As discussed already in previous work~\cite{Kurkela:2018ygx,Kurkela:2018qeb}, the
isotropization time approximation is found to reproduce the evolution of the QCD weak coupling effective kinetic theory within $~\sim 15\%$. It
shares important commonalities with the well-known relaxation time approximation, but it is free of the assumption that the
distribution function $f$ evolves directly to a thermal equilibrium. As such it is better suited for the description of non-equilibrium systems such as 
anisotropic plasmas to which a local equilibrium distribution is difficult to associate to. 

\subsubsection{Initial conditions}
\label{sec2e}
We specify initial conditions for the first momentum moments $F(\tau_0, \vec x_\perp; \phi, v_z)$ of the distribution function at initial proper time $\tau_0 \rightarrow 0$, with the azimuthal momentum angle $\phi$ defined by  $\vec v_\perp = \left(\cos\phi,\sin\phi \right)$.
As particle production is local in space, the initial momentum distribution is azimuthally symmetric at initial time $\tau_0$ for statistical ensembles of collisions. 
Therefore, $F$ cannot depend on $\phi$ initially. Moreover, since ultra-relativistic pp, pA and AA collisions are expected to
display very large momentum anisotropies between the transverse and longitudinal direction, the initial condition will be taken to be of the form
\begin{equation}
F(\tau_0, \vec x_\perp; \phi, v_z) \propto \delta(v_z)\, .
\end{equation}
The initial transverse spatial $\vec x_\perp$-distribution is given by an azimuthally isotropic
profile function $P_{\rm iso}\left( r/R \right)$, with a characteristic root mean square
transverse system size $R = \textstyle{\left( \int d^2 r \, r^2 P_{\rm iso} / \int d^2 r \, P_{\rm iso} \right)^{1/2}}$.
The isotropic profile is multiplied by an azimuthal dependence with small $\delta_{m,n}$
\begin{equation}
	P_{\rm aniso}\left(\frac{r}{R},\theta \right) = 1 + \sum_{m,n=1}^\infty \delta_{m,n}\, \frac{r^n}{R^n}\, \cos m\theta\, ,
		\label{eq40text}
\end{equation}
where $\theta$ is the coordinate space azimuthal angle $\vec x_\perp = r (\cos \theta, \sin \theta)$.
The azimuthal eccentricity of the initial spatial distribution is characterized by the harmonic coefficients $\delta_{m,n}$ that are related to the initial state eccentricities $\epsilon_m$ (\emph{cf.} eq.~(\ref{eq:epsilon})). The initial conditions are then of the form
\begin{equation}
	F(\tau_0, \vec x_\perp; \phi, v_z) = \delta(v_z)\, P_{\rm iso}\left(\frac{r}{R}\right)\, P_{\rm aniso}\left(\frac{r}{R},\theta \right) \, .
	\label{eq41text}
\end{equation}   
We consider two transverse spatial profile functions $P_{\rm iso}$:
\begin{itemize}
	\item Initial Gaussian profile \\
	\begin{equation}
	P_{\rm iso}\left(\frac{r}{R}\right) = 2 \varepsilon_0\, \exp\left[ -\frac{r^2}{R^2} \right]\, .
	\label{eq42text}
	\end{equation}
	The normalization of this isotropic distribution is chosen such that the initial energy density $\varepsilon(\tau_0, r=0)$ calculated from $F(\tau_0, \vec x_\perp; \phi, v_z)$ in
	eq. (\ref{eq41text}) satisfies $\varepsilon(\tau_0, r=0) = \varepsilon_0$. 
	\item Initial Woods-Saxon (WS) profile\\
	For a Pb nucleus, the standard Woods-Saxon parametrization of the density profile is $\rho_{\rm WS} (r,z)= 1/\left(\exp\left[\textstyle\frac{\sqrt{r^2 + z^2}-R_{\rm WS}}{\delta_{\rm WS}} \right] + 1 \right)$
	where $R_{\rm WS} = 1.12 A^{1/3} - 0.86 A^{-1/3} \vert_{A=208} = 6.49$fm,  and $\delta_{\rm WS} = 0.54 $fm. The corresponding nuclear profile function $T_{\rm WS}(r) \equiv 
	\int_{-\infty}^\infty \rho_{\rm WS}(r,z)\, dz$ is the projection of this density onto the transverse plane. Simple models of the local energy density in the transverse plane take $\varepsilon(\tau_0, r)$
	proportional to the nuclear overlap of two Pb nuclei, \emph{i.e.}, for vanishing impact parameter, proportional to  $T^2_{\rm WS}(r)$~\cite{Kolb:2001qz}. This motivates us to make the ansatz
	\begin{equation}
	P_{\rm iso}\left(\frac{r}{R}\right) = \frac{2\varepsilon_0\, T_{\rm WS}^2\left(\frac{r\, c_{\rm rms}}{R} \right)}{   T_{\rm WS}^2\left(0\right)}\, ,
	\label{eq43}
	\end{equation}
	where we choose the number $c_{\rm rms}\approx 4.42$ such that the radius mean square of $P_{\rm iso}\left(\frac{r}{R}\right) $ is properly normalized to unity.
\end{itemize}

In principle,  results can depend on the initialization time $\tau_0$.  
However, at sufficiently early times, the system must evolve close to free streaming, and varying the initial time will therefore not matter as long as $\varepsilon_0 \tau_0$ 
is kept constant. The $\tau_0 \to 0$ singularity of the initial conditions (\ref{eq49})
is in this sense benign since it can be scaled out of the solutions of kinetic theory for constant $\varepsilon_0 \tau_0$ (For more detailed discussion see Appendix \ref{sec2f}). This allows us to present results in terms of a single 
physically meaningful model parameter, \emph{the opacity} $\hat\gamma$
\begin{equation}
	\hat \gamma =R^{3/4}\gamma (\, \varepsilon_0 \tau_0)^{1/4}\, .
	\label{eq48}
\end{equation}
As explained in detail in Section \ref{sec4f}, this parameter is related to the transverse size of the system in units of the mean free path, $\frac{R}{\tau_{\rm iso}}$. It is linearly related for $\hat \gamma \ll 1$ but in opaque systems the relation is $\frac{R}{\tau_{\rm iso}} \propto \hat \gamma^{8/9}$.

\begin{figure*}
\includegraphics[width=0.24\textwidth]{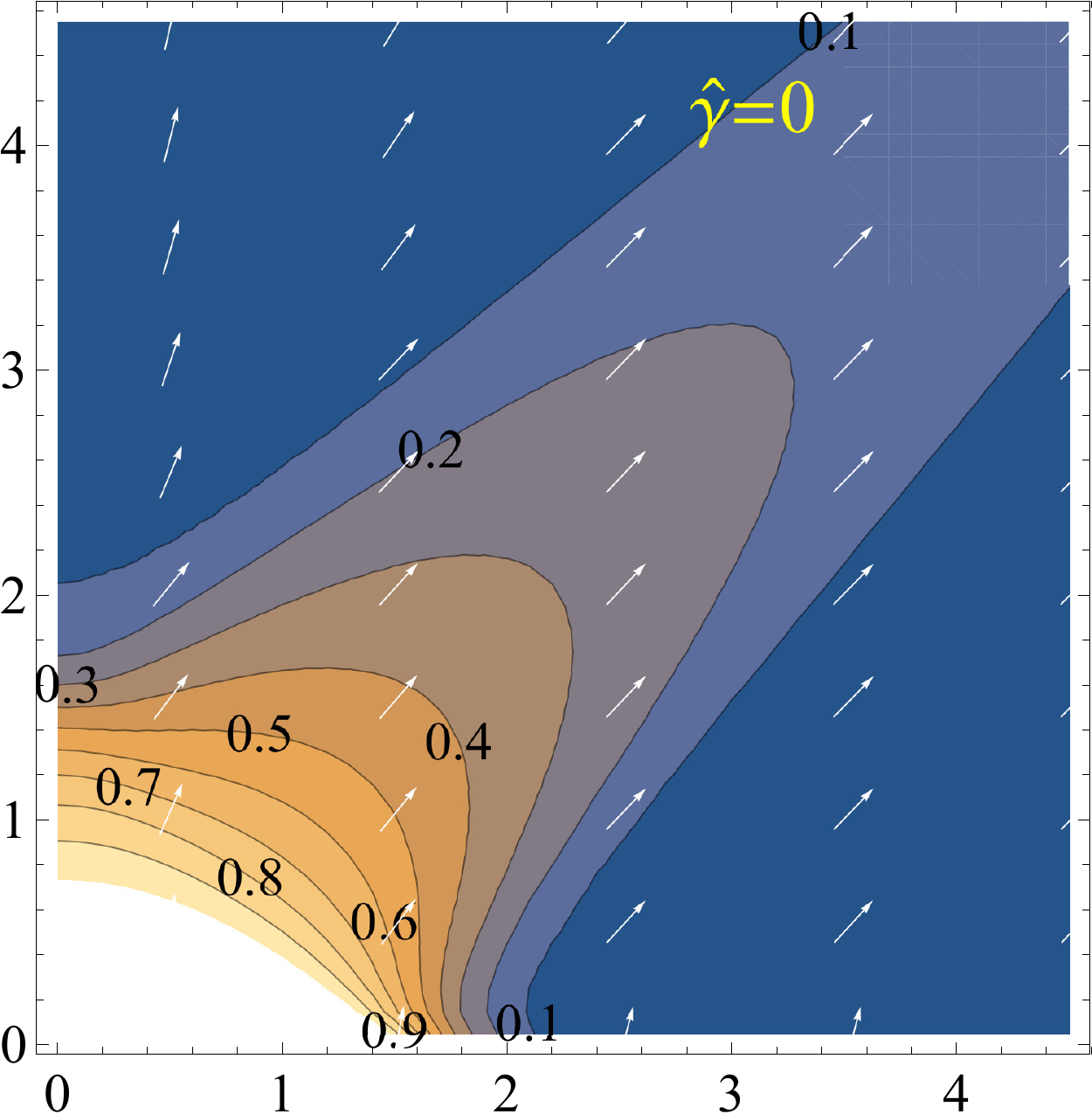}
\includegraphics[width=0.24\textwidth]{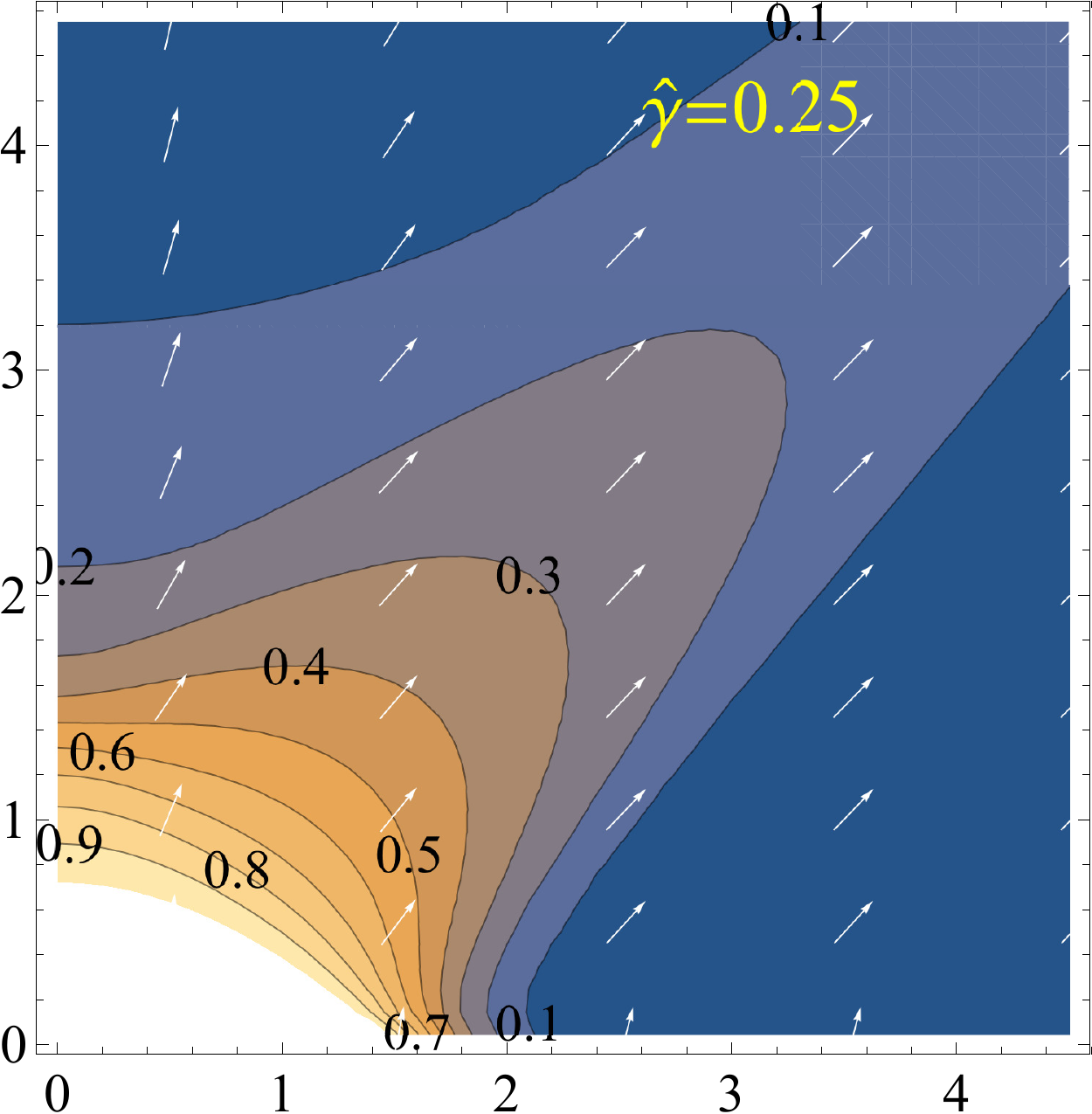}
\includegraphics[width=0.24\textwidth]{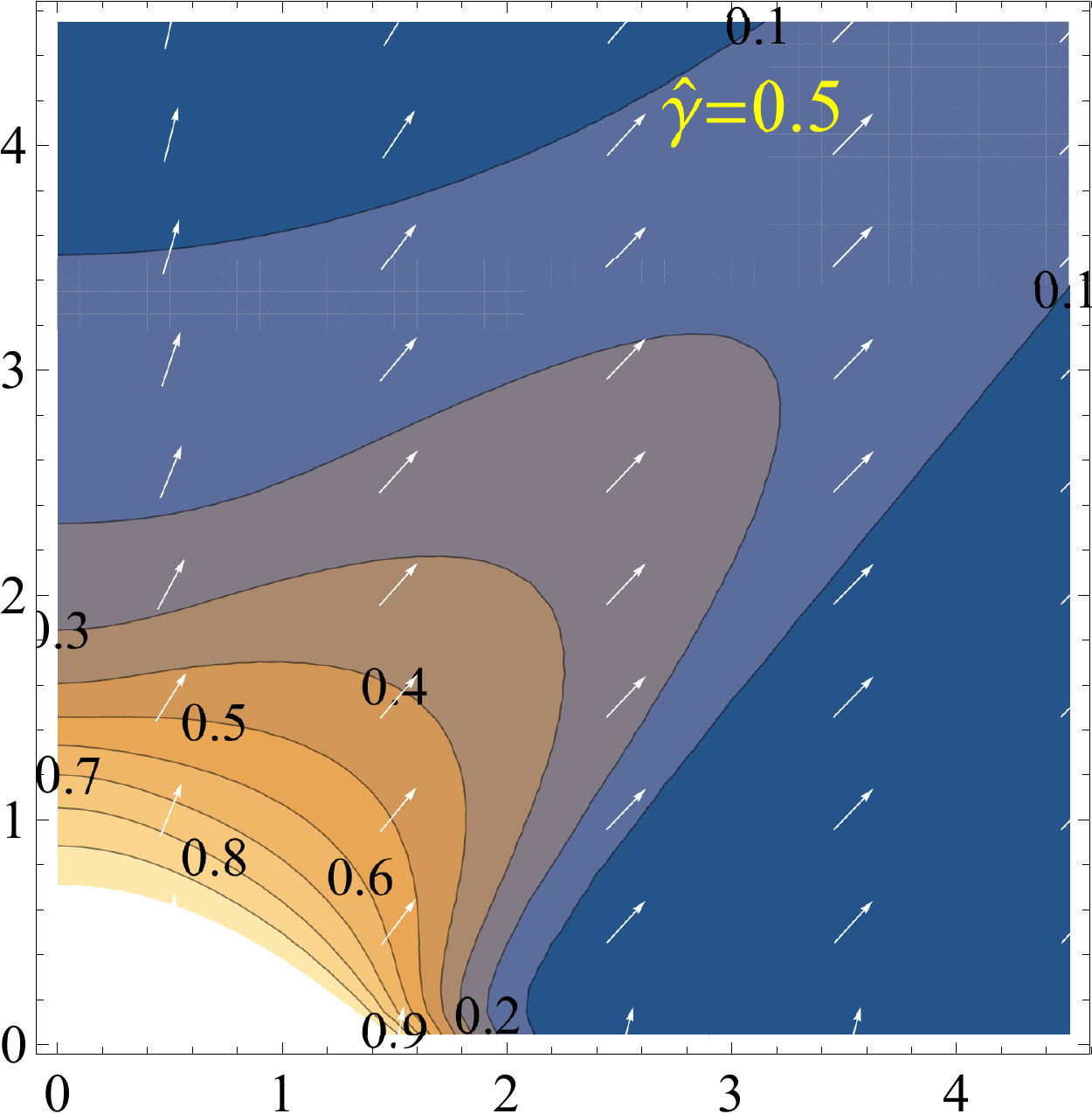}
\includegraphics[width=0.24\textwidth]{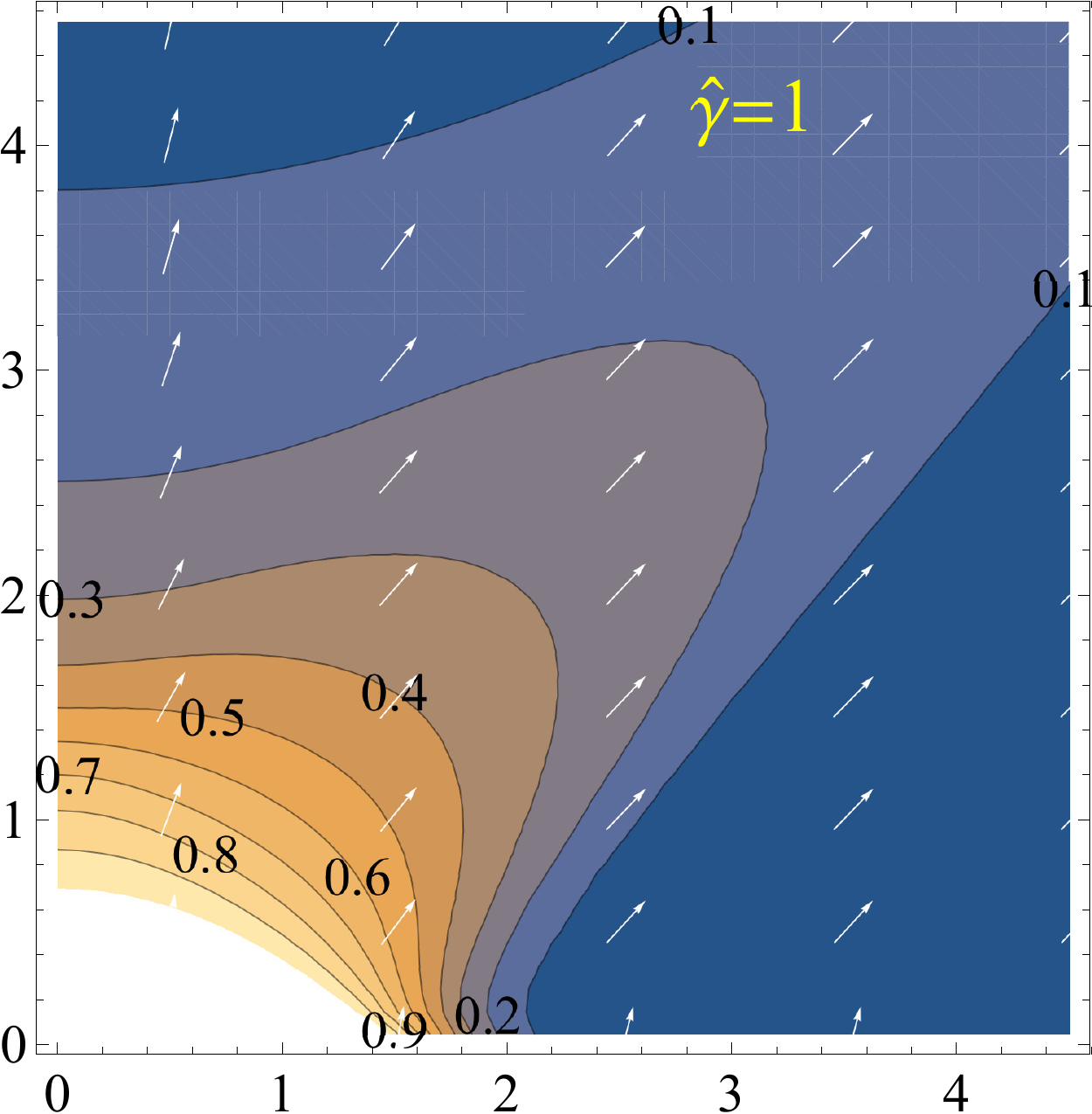}
\\
\vspace{0.5cm}
\includegraphics[width=0.24\textwidth]{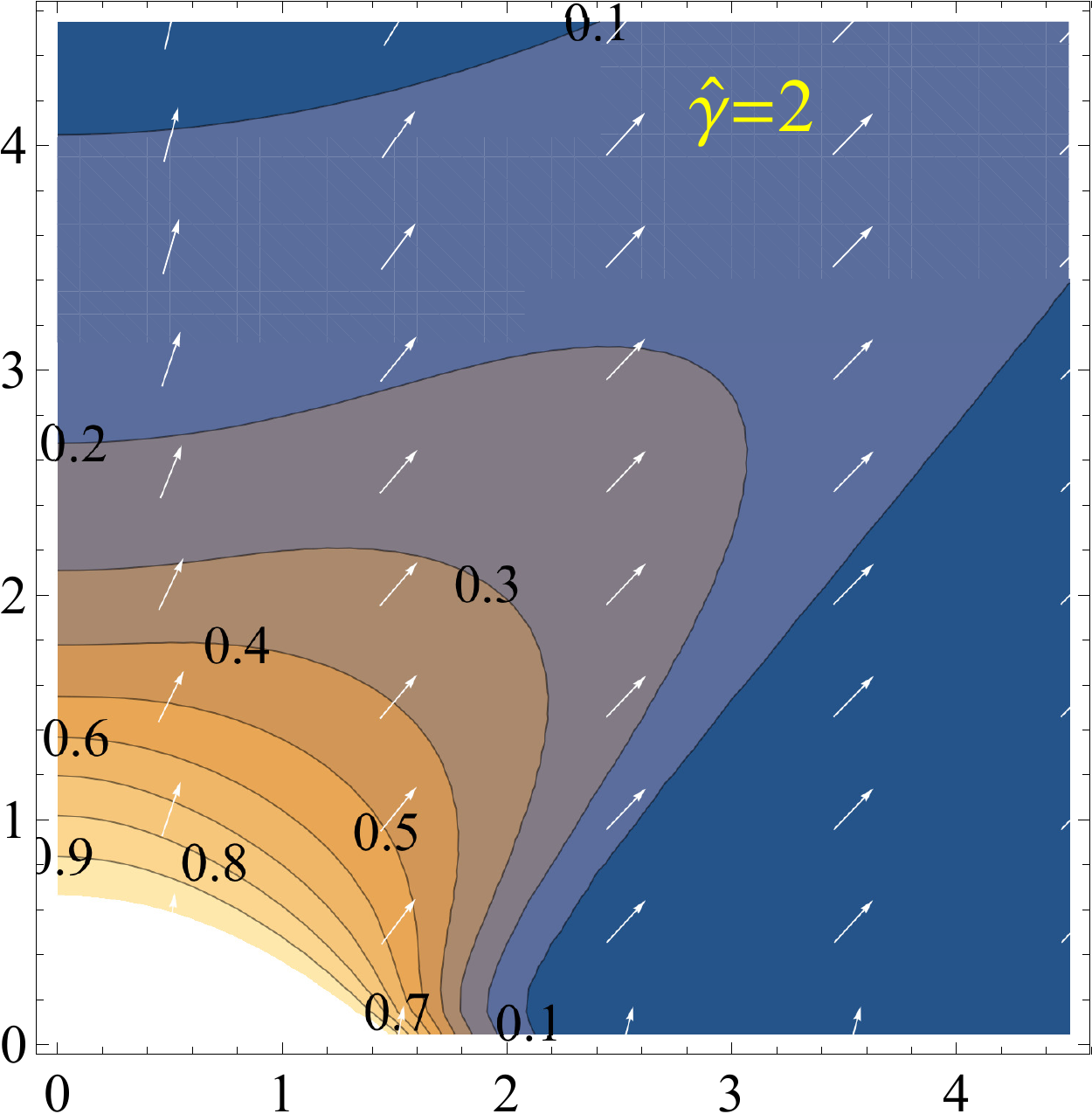}
\includegraphics[width=0.24\textwidth]{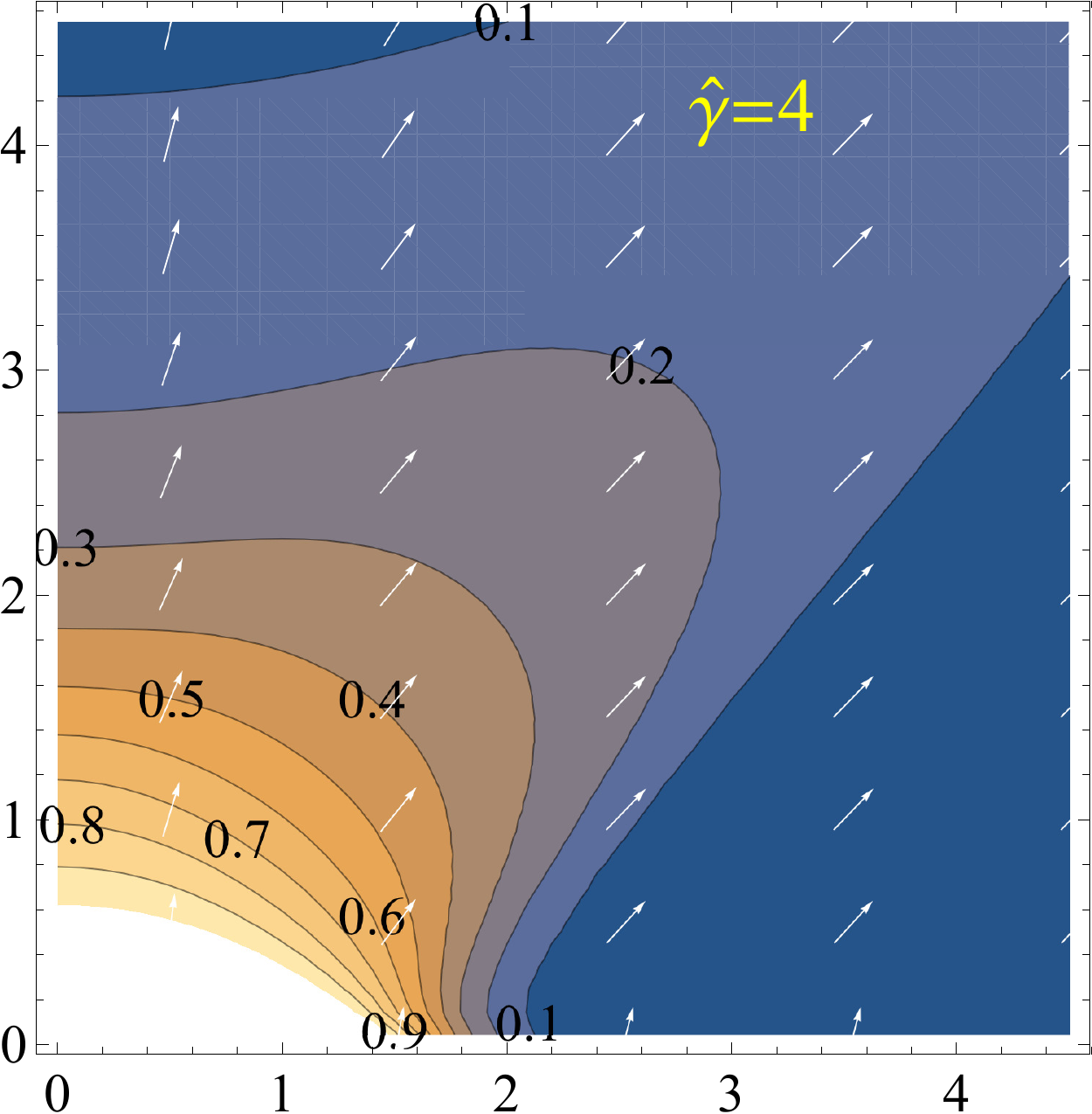}
\includegraphics[width=0.24\textwidth]{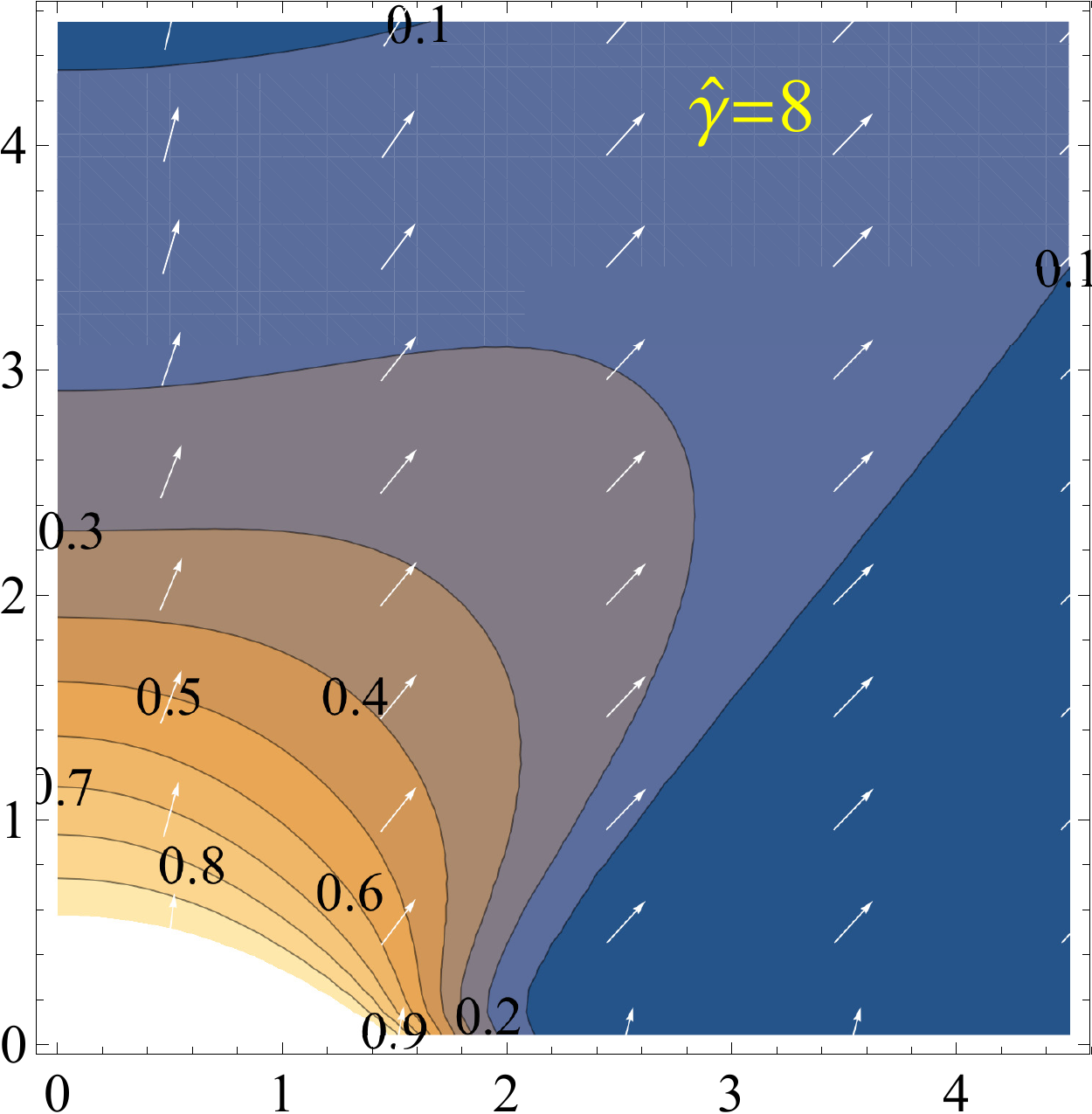}
\includegraphics[width=0.24\textwidth]{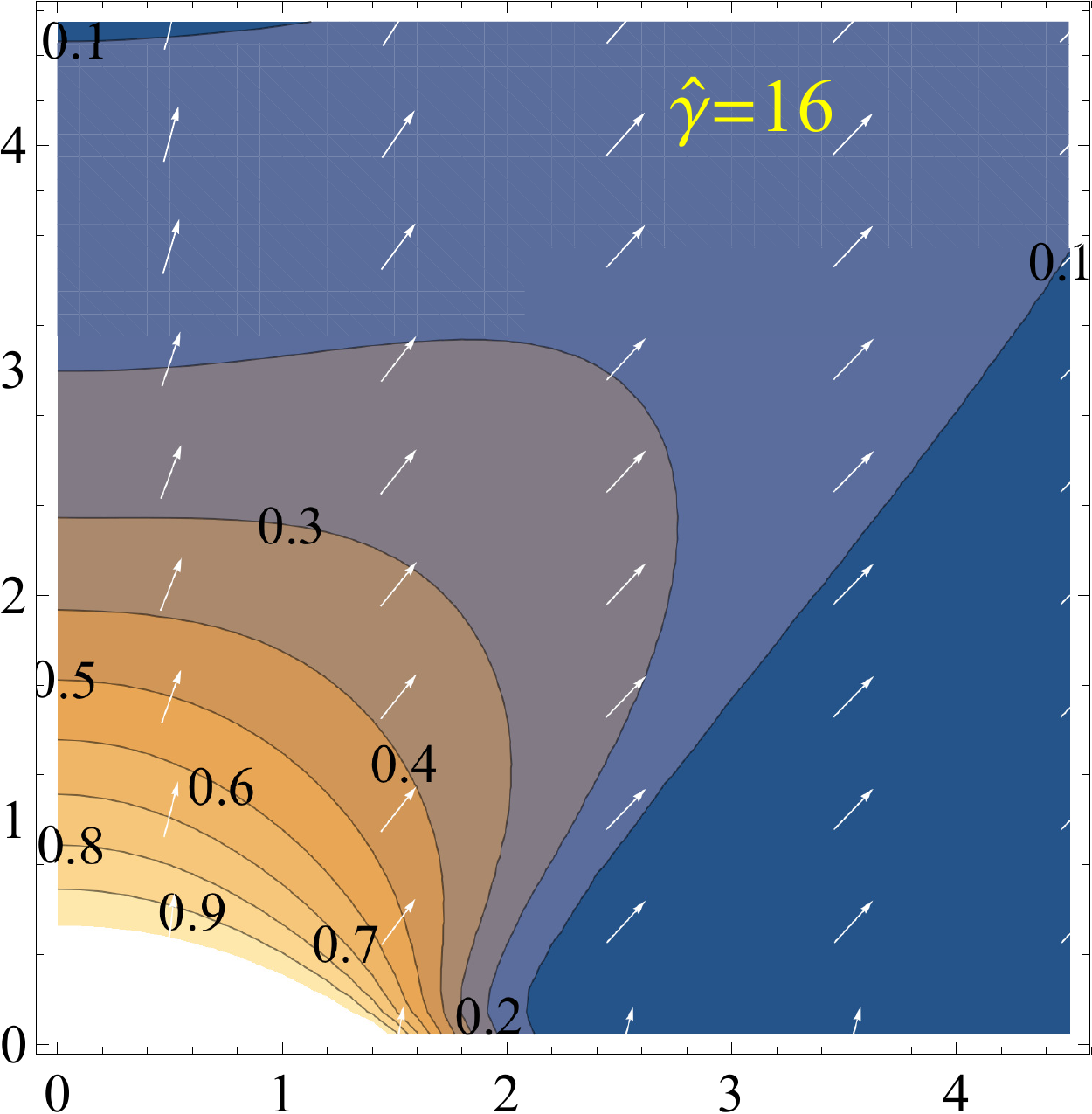}
\caption{The local energy density $\varepsilon(t,r)$ in the $(r,t)$-plane, calculated from the transport equation (\ref{eq47}) with longitudinally boost-invariant and
azimuthally symmetric initial conditions of Woods-Saxon type (\ref{eq43}). The panels scan the only model parameter, the opacity $\hat\gamma$,   from the free-streaming limit ($\hat\gamma = 0$)
up to values of $\hat\gamma$ sufficiently large to realize almost perfect fluid dynamic behavior. The contours depict isotherms, with $\varepsilon^{1/4}$ decreasing 
from one isothermal to the next by 10\% of its value along the innermost contour. White arrows denote the direction $u_\mu = (u_\tau,u_r,0,0)$ of the radial flow field. }
\label{fig2}
\end{figure*} 

\section{Numerical Results}
\label{sec3}
We analyze now in detail how the time evolution of the energy-momentum tensor varies with transverse system size between the
extreme limiting cases of free streaming and ideal fluid dynamics by varying the only available dimensionless parameter, the opacity $\hat\gamma$. In the following, at $z=0$ time $t$ and radial $r$ coordinates are plotted in units of the system size $R$.

\subsection{Time evolution of $T^{\mu\nu}_{\rm kin}$}
We consider first the time evolution of the energy-momentum tensor $T^{\mu\nu}_{\rm kin}$ for azimuthally symmetric initial conditions. This corresponds to solving the Boltzmann
transport equation (\ref{eq24}) for the zeroth harmonic $F_0$ with initial conditions (\ref{eq43}) brought into the form similar to (\ref{eq49}). The solution defines the 
energy momentum tensor (\ref{eq17}), which we shall denote with the subscript ``{\rm kin}'' to make clear that this is the result of kinetic theory. 

We start by characterizing in Fig.~\ref{fig2} the time evolution of the local energy density $\varepsilon(t,r) = u_\mu(t,r)\, T^{\mu\nu}_{\rm kin}(t,r)\, u_\nu(t,r)$ of a system initialized with the radial Woods-Saxon distribution (\ref{eq43}). To explain the main features of this figure, we note the following:
For a free-streaming gas of massless particles, all components propagate unperturbed along 
the light cone. As initially $v_z =0$, the particles can reach positions close to $r= 0$ at late times only if they originate at large radius and propagate inwards. The energy density $\varepsilon(t,r=0)$ at the center can therefore take non-negligible
values only at times $t<r_{\rm tail} \simeq R$ limited by the extent of the tail of the initial radial distribution. These particles will propagate in the $(r,t)$ diagram on a light-like trajectory
below $(r, r+r_{\rm tail})$. On the other hand, particles that start initially at the same radial position $r_{\rm tail}$ but propagate exactly outwards will show up in Fig.~\ref{fig2} 
along $(r,t)=(r+r_{\rm tail},r)$ and all particles produced initially at smaller $r$ or emitted not exactly inwards will stay above that line. A free-streaming density distribution is
therefore a density distribution that takes non-vanishing values only in the strip  limited by $(r,t)=(r, r+r_{\rm tail})$ and $(r,t)=(r+r_{\rm tail},r)$ in the $(r,t)$-diagram. This is clearly
seen for the free-streaming limit $\hat\gamma = 0$ in Fig.~\ref{fig2}. 

As one endows the system with a small interaction rate, large-angle scatterings can deflect freely propagating particles towards the neighborhood of $r=0$. As a consequence, 
a non-vanishing energy-density around $r=0$ persists for a longer time. This effect is sizeable even if scattering is infrequent: the case $\hat\gamma = 0.25$ in Fig.~\ref{fig2}
corresponds to a system size that extends only over one quarter of the in-medium path length, so most particles will escape unscattered from the collision region. Yet, compared
to the free-streaming case, the system maintains at its origin a non-vanishing energy density for a significantly longer time. 

As one increases $\hat\gamma$ and thus the collision propability per particle per system size, one finds that the general features described above evolve gradually, and
an increasing fraction of the total energy density remains for an increasing time duration in the neighborhood of the origin of the system, see Fig.~\ref{fig2}.  In this sense, the
evolution of the energy density moves with increasing $\hat\gamma$ further away from the light cone, although particles located in the tails of the system and propagating
outwards will always escape along the light cone.

\subsection{Quantifying deviations of $T^{\mu\nu}_{\rm kin}$ from fluid dynamics}

We want to understand as a function of opacity $\hat\gamma$,
to what extent fluid dynamics provides a good description of kinetic theory. To this end,  we quantify differences between kinetic 
theory and viscous fluid dynamics by comparing the energy-momentum tensor $T_{\rm kin}^{\mu\nu}$ calculated in kinetic theory to the energy momentum tensor obtained
in fluid dynamics,
\begin{equation}
T_{\rm hyd}^{\mu\nu} = \left( \varepsilon + p\right) u^\mu\, u^\nu + p\, g^{\mu\nu} + \Pi^{\mu\nu}_{\rm hyd}\, .
\label{eq52}
\end{equation}

As we use the Landau matching condition $u_\mu T^{\mu\nu}_{\rm kin} = - \varepsilon\, u^\nu$ to define energy density $\varepsilon$ and flow $u^\mu$ in the kinetic theory, 
the ideal part $T_{\rm id}^{\mu\nu} = \left( \varepsilon + p\right) u^\mu\, u^\nu + p\, g^{\mu\nu} $ is by construction the same, and
differences between kinetic theory and fluid dynamics arise only  from the shear-viscous tensor $\Pi^{\mu\nu}$. 
In a fluid dynamic gradient expansion,  $\Pi^{\mu\nu}_{\rm hyd}$ is expressed in terms of $\varepsilon$ and $u^\mu$ via 
the constitutive hydrodynamic relation~\cite{Baier:2007ix}
\begin{eqnarray}
	\Pi^{\mu\nu}_{\rm hyd} &=& \Pi^{\mu\nu}_{\rm hyd\, 1st}+ \Pi^{\mu\nu}_{\rm hyd\, 2nd} \, , \label{eq53} \\
	\Pi^{\mu\nu}_{\rm hyd\, 1st} &=& - 2 \eta \sigma^{\mu\nu}\, , \label{eq54} \\
	\Pi^{\mu\nu}_{\rm hyd\, 2nd} &=&	 2 \tau_{\Pi}\, \eta \left[ ^<D\sigma^{\mu\nu>} 
	+ \frac{1}{3} \sigma^{\mu\nu} \nabla_\alpha u^\alpha \right] \nonumber \\
	&& +	\lambda_1 \sigma^{<\mu}_\alpha\, \sigma^{\nu> \lambda} \, .
	\label{eq55}
\end{eqnarray}
Here, the brackets around indices denote
traceless, symmetrized second rank tensors,  $^{<}A^{\mu\nu>} \equiv \textstyle\frac{1}{2}\Delta^{\mu\alpha}\Delta^{\nu\beta} \left(A_{\alpha\beta} + A_{\beta\alpha} \right) $
$ - \textstyle\frac{1}{3} \Delta^{\mu\nu}\Delta^{\alpha\beta} A_{\alpha\beta} $. The terms of first and second order in fluid dynamic gradients are denoted by 
$\Pi^{\mu\nu}_{\rm hyd\, 1st}$ and $\Pi^{\mu\nu}_{\rm hyd\, 2nd}$, respectively. The projector on the subspace orthogonal to the flow vector $u^\mu$ is 
$\Delta^{\mu\nu} \equiv u^\mu u^\nu + g^{\mu\nu}$, and to first order in gradients, 
$\Pi^{\mu\nu}_{\rm hyd} $ is proportional to the tensor
\begin{equation}
	\sigma^{\mu\nu} =  \Big\{\frac{1}{2}\left[ \Delta^{\mu\alpha} \nabla_\alpha u^\nu {+} \Delta^{\nu\alpha} \nabla_\alpha u^\mu \right] 
				{-} \frac{1}{3}\Delta^{\mu\nu} \nabla_\alpha u^\alpha \Big\}\, .
				\label{eq56}
\end{equation}
The first and second order hydrodynamic coefficients in (\ref{eq55})
have been computed elsewhere (see \emph{e.g.} eq.(11) in Ref.~\cite{Heller:2016rtz}) and they read
\begin{eqnarray}
	\eta &=& \frac{1}{5} \frac{ \varepsilon + p }{\gamma \varepsilon^{1/4}} \, , \label{eq57}\\
	\tau_{\Pi} &=& \frac{5\, \eta}{\left(\varepsilon+p \right)}  \, , 
	\label{eq58}\\
	 \lambda_1 &=& \frac{27}{5} \frac{\eta^2}{\left(\varepsilon+p \right)}  \, .
	 \label{eq59}
\end{eqnarray}
Locally in space and time, we characterize the quality $Q_2(t,r)$ of the agreement between fluid dynamic constitutive relation and kinetic theory by the quantity
\begin{equation}
	Q_2(t,r) =  \sqrt{  \frac{ \left( T_{\rm kin}- T_{\rm hyd} \right)^{\mu\nu}   \left( T_{\rm kin} - T_{\rm hyd} \right)_{\mu\nu} }{ \left(T_{\rm id} \right)^{\mu\nu}    \left(T_{\rm id} \right)_{\mu\nu}    }  }\, ,
	\label{eq60}
\end{equation}
where all fields on the right hand side are understood as functions of $t$ and $r$. Since $\left(T_{\rm kin}- T_{\rm hyd} \right)^{\mu\nu} = \Pi^{\mu\nu}_{\rm kin} - \Pi^{\mu\nu}_{\rm hyd}$, 
and since the shear viscous tensor is transverse, $u_\mu \Pi^{\mu\nu} = 0$, only the spatial components of $\left( T_{\rm kin}- T_{\rm hyd} \right)^{\mu\nu}$  can be non-zero in the 
local rest frame. Therefore,  the contractions under the square root in (\ref{eq60}) yields a sum over squares with positive coefficients. Thus, $Q_2(t,r)$ is a positive definite measure 
of the difference between $T_{\rm kin}^{\mu\nu}$ and $T_{\rm hyd}^{\mu\nu}$. 
At the space-time point $(t,r)$, it measures the differences between the shear viscous tensors of kinetic theory and hydrodynamics in
units of the local energy density, $T_{\rm id}^{\mu\nu} {T_{\rm id}}_{\mu\nu} = \textstyle\frac{4}{3} \varepsilon(t,r)$.

\begin{figure*}
\begin{tabular}{cc}
\begin{tabular}{cccc}
 & Ideal & 1st oder & 2nd order \\
 \\
$\hat\gamma=1 $\hspace{0.5cm} 
& 
\noindent\parbox[c]{0.25\hsize}{\includegraphics[width=0.25\textwidth]{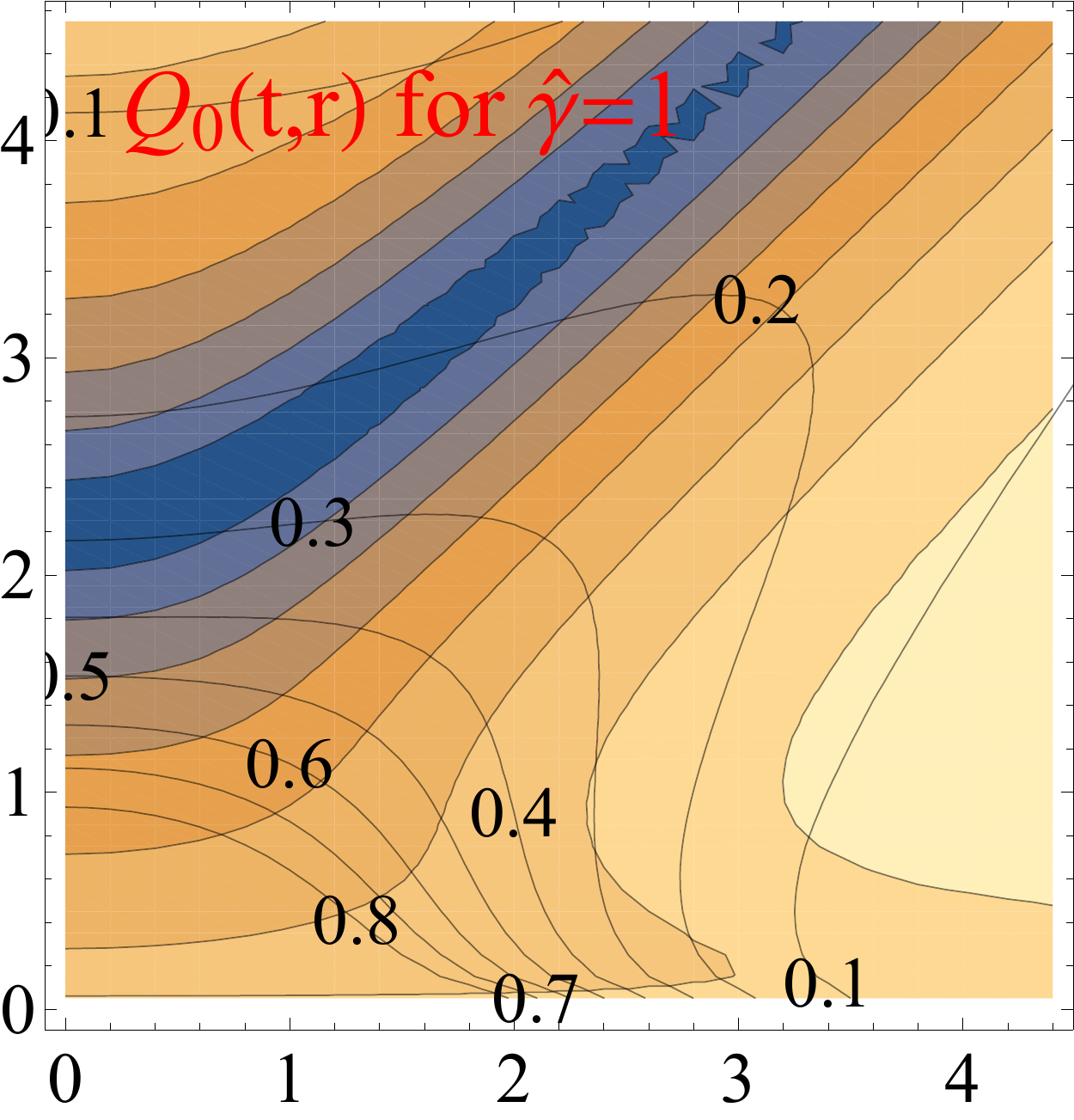}}
&
\noindent\parbox[c]{0.25\hsize}{\includegraphics[width=0.25\textwidth]{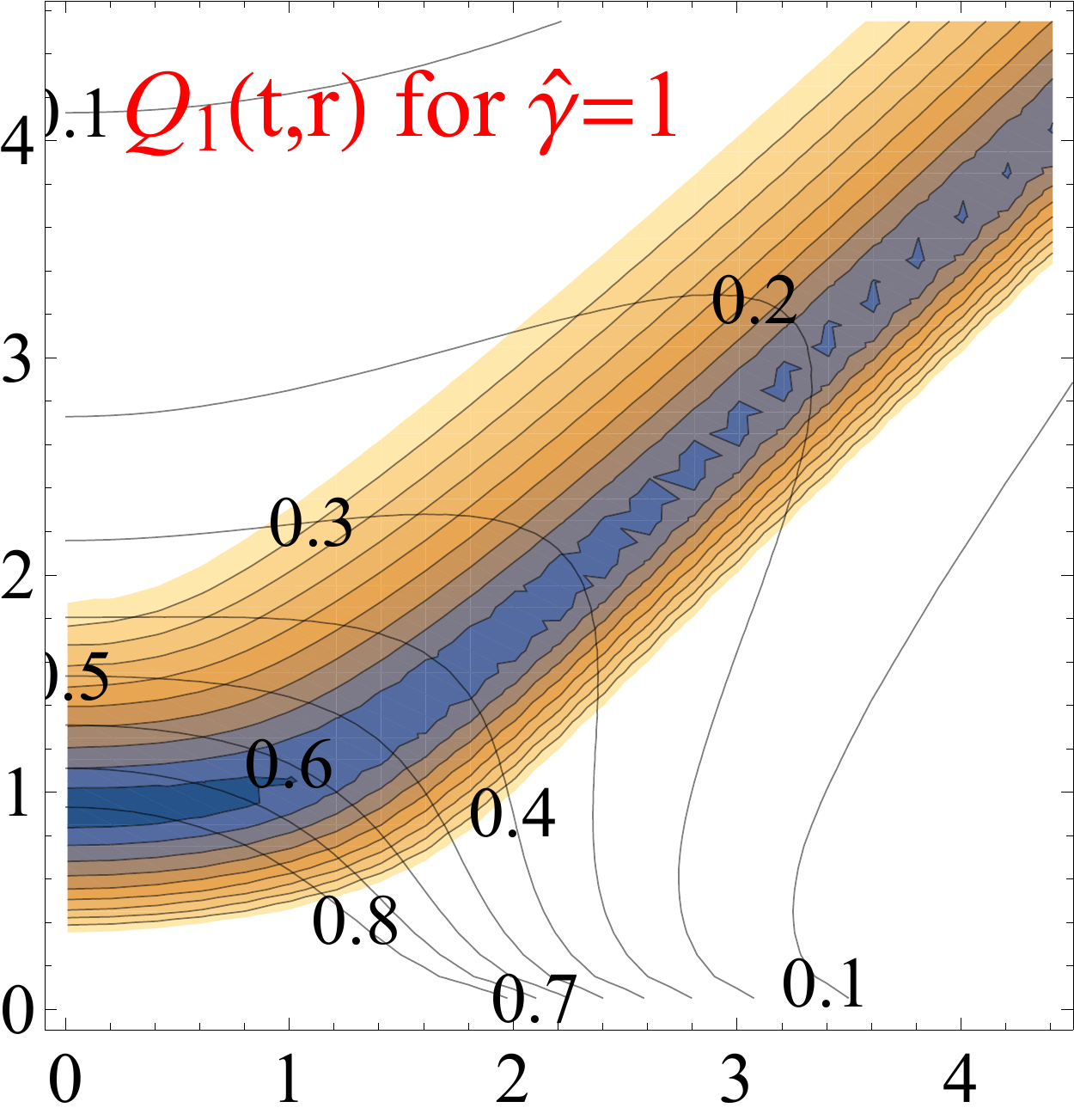}}
&
\noindent\parbox[c]{0.25\hsize}{\includegraphics[width=0.25\textwidth]{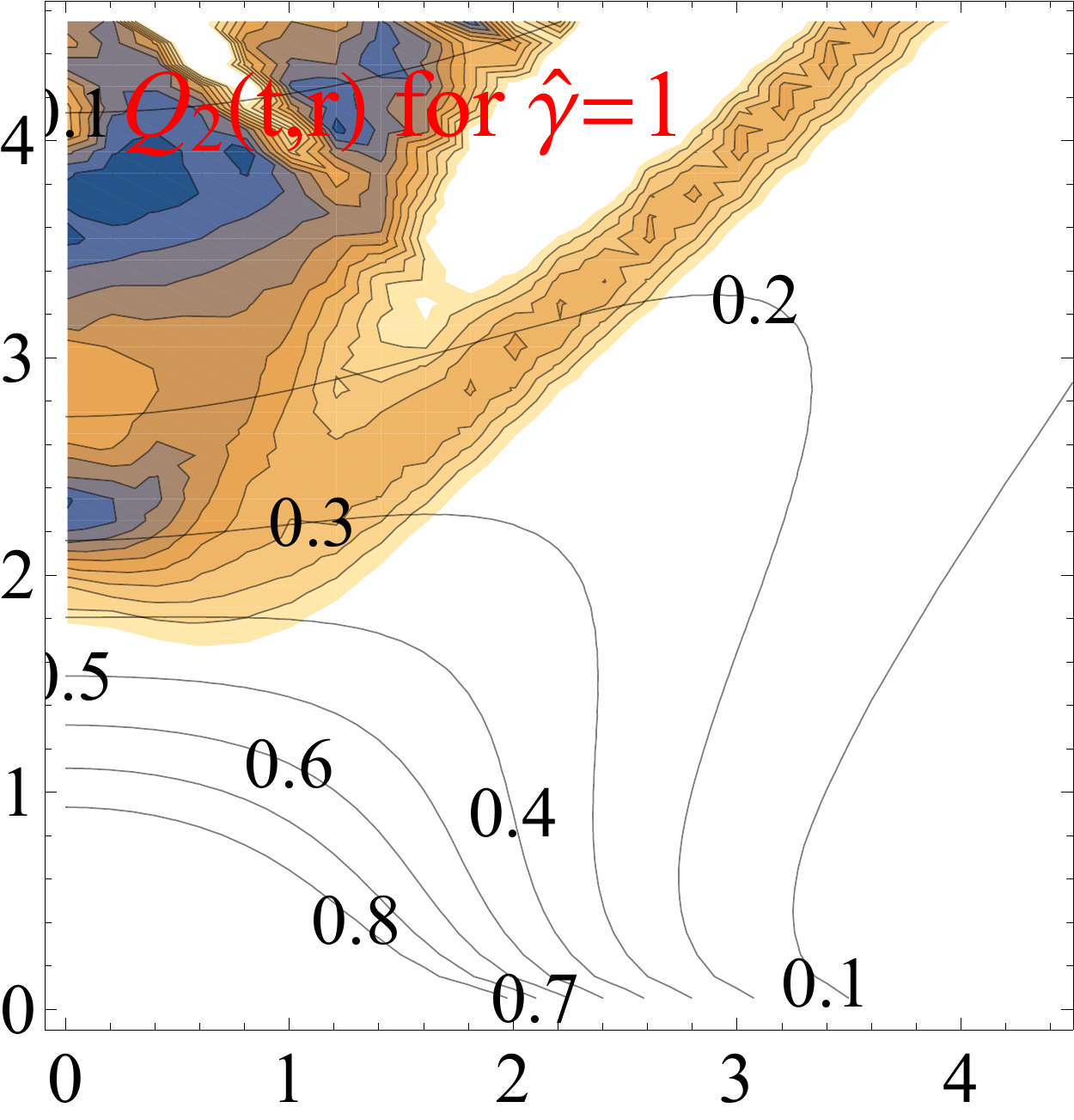}}
\\
\\
$\hat\gamma=2 $\hspace{0.5cm} 
&
\noindent\parbox[c]{0.25\hsize}{\includegraphics[width=0.25\textwidth]{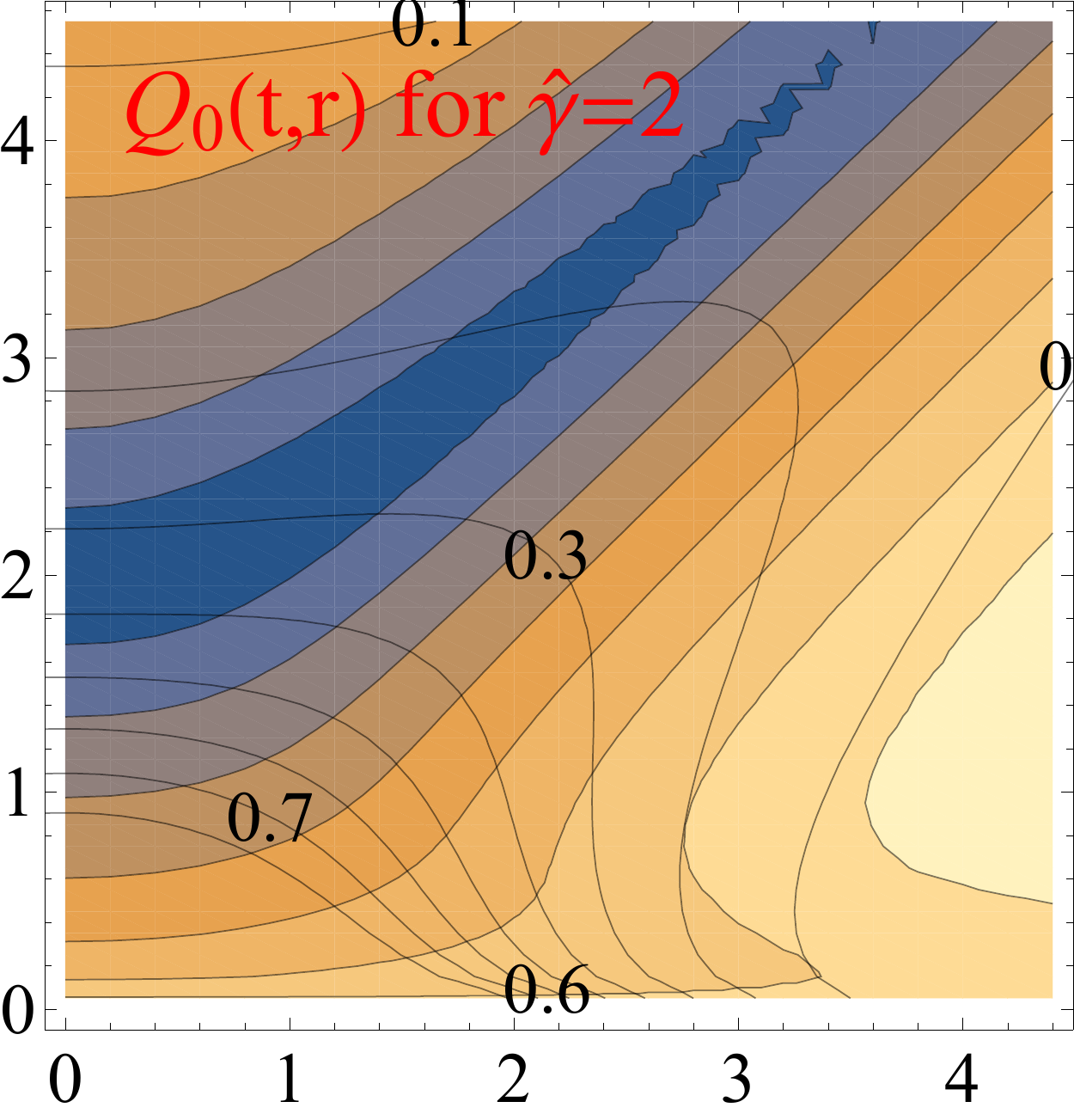}}
&
\noindent\parbox[c]{0.25\hsize}{\includegraphics[width=0.25\textwidth]{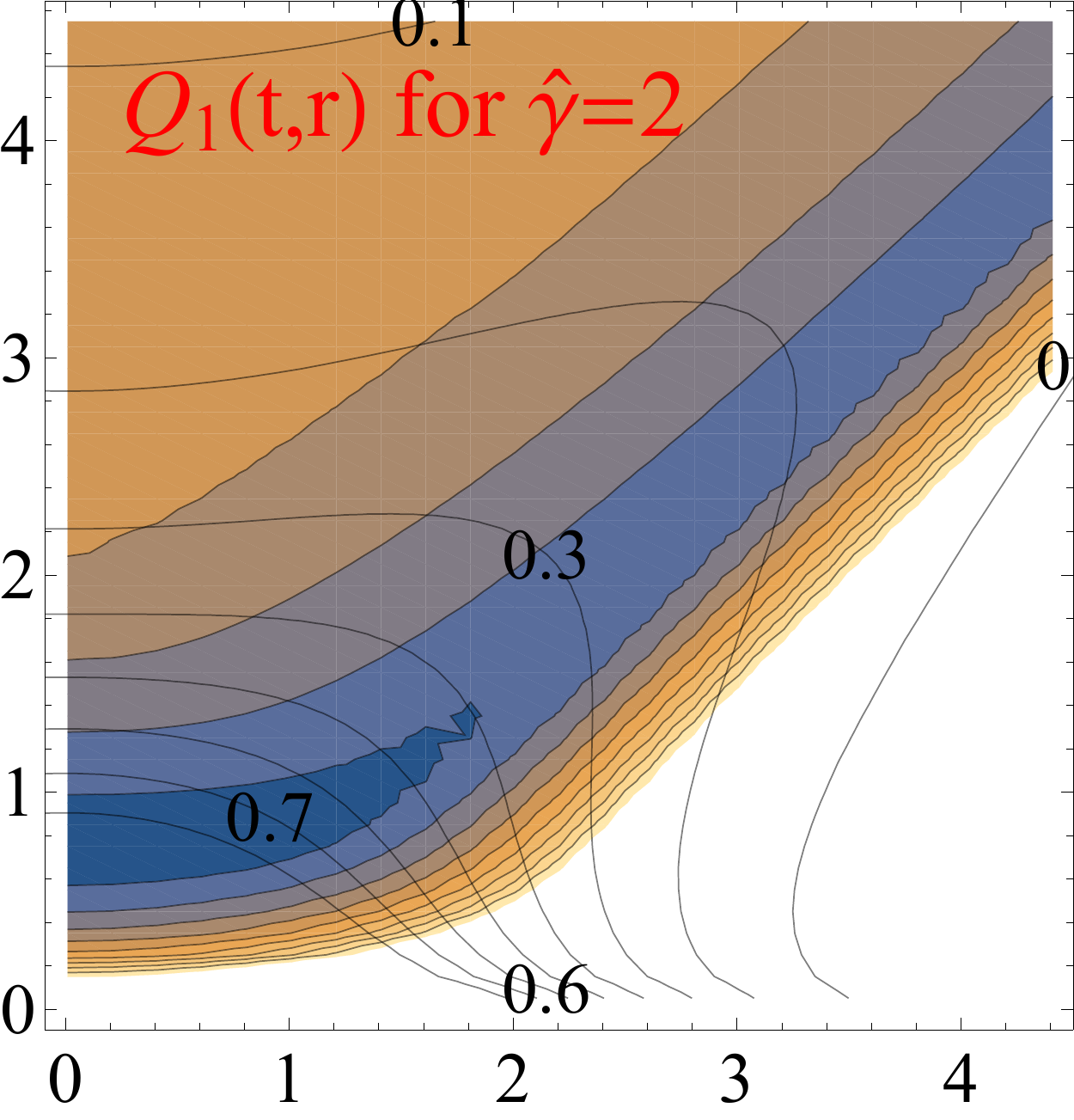}}
&
\noindent\parbox[c]{0.25\hsize}{\includegraphics[width=0.25\textwidth]{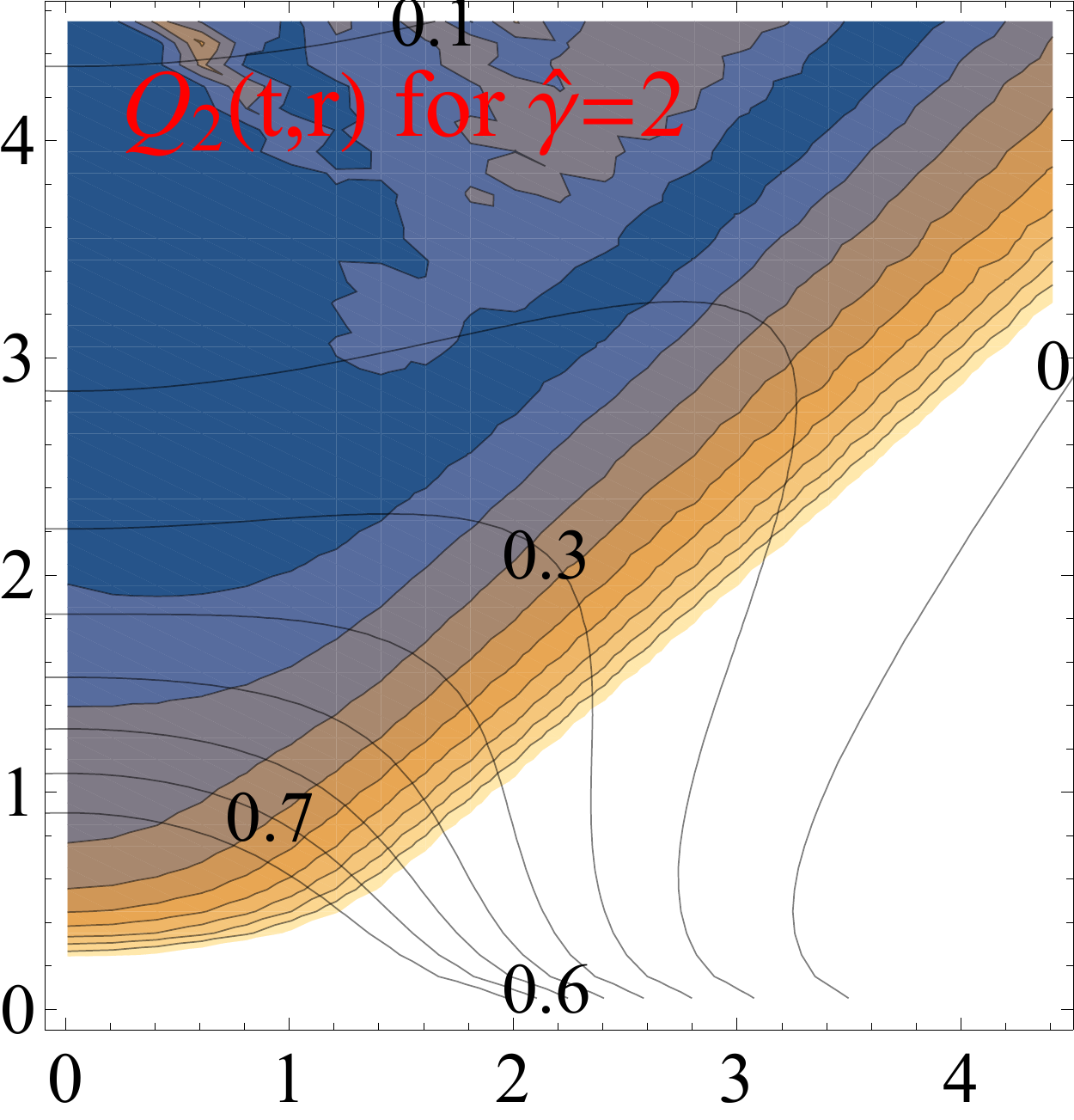}}
\\
\\
$\hat\gamma=4 $\hspace{0.5cm} 
&
\noindent\parbox[c]{0.25\hsize}{\includegraphics[width=0.25\textwidth]{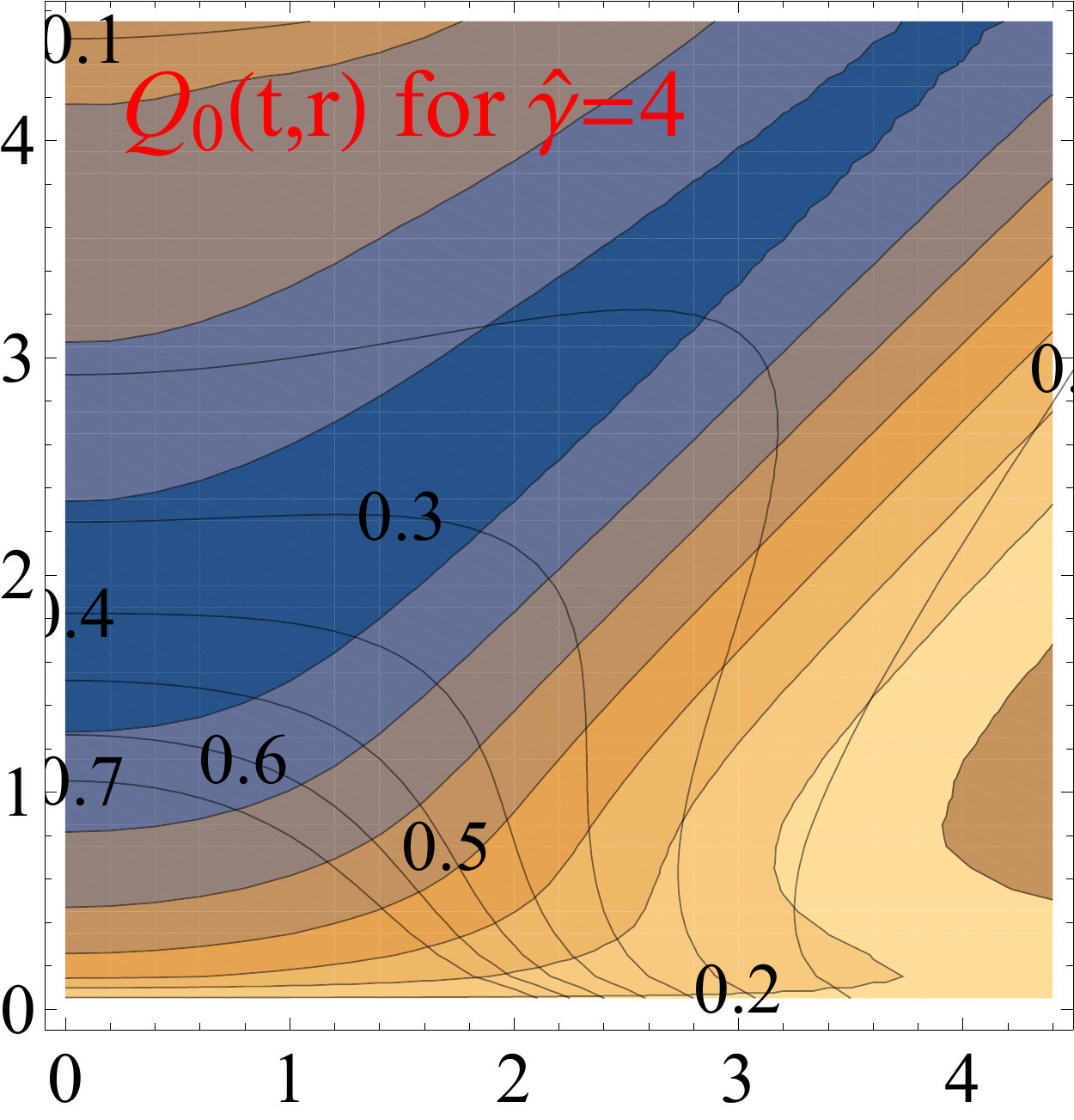}}
&
\noindent\parbox[c]{0.25\hsize}{\includegraphics[width=0.25\textwidth]{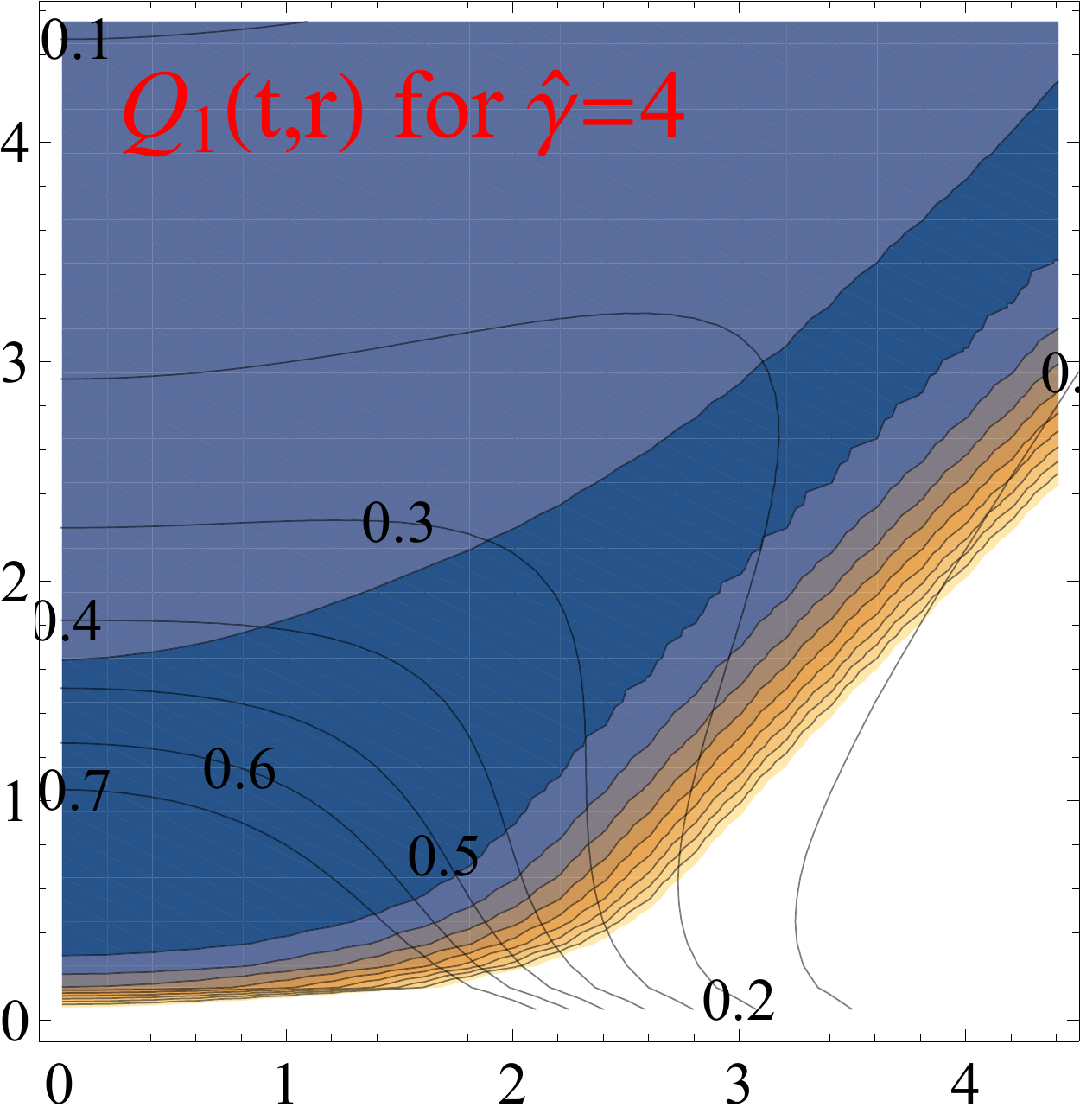} }
&
\noindent\parbox[c]{0.25\hsize}{\includegraphics[width=0.25\textwidth]{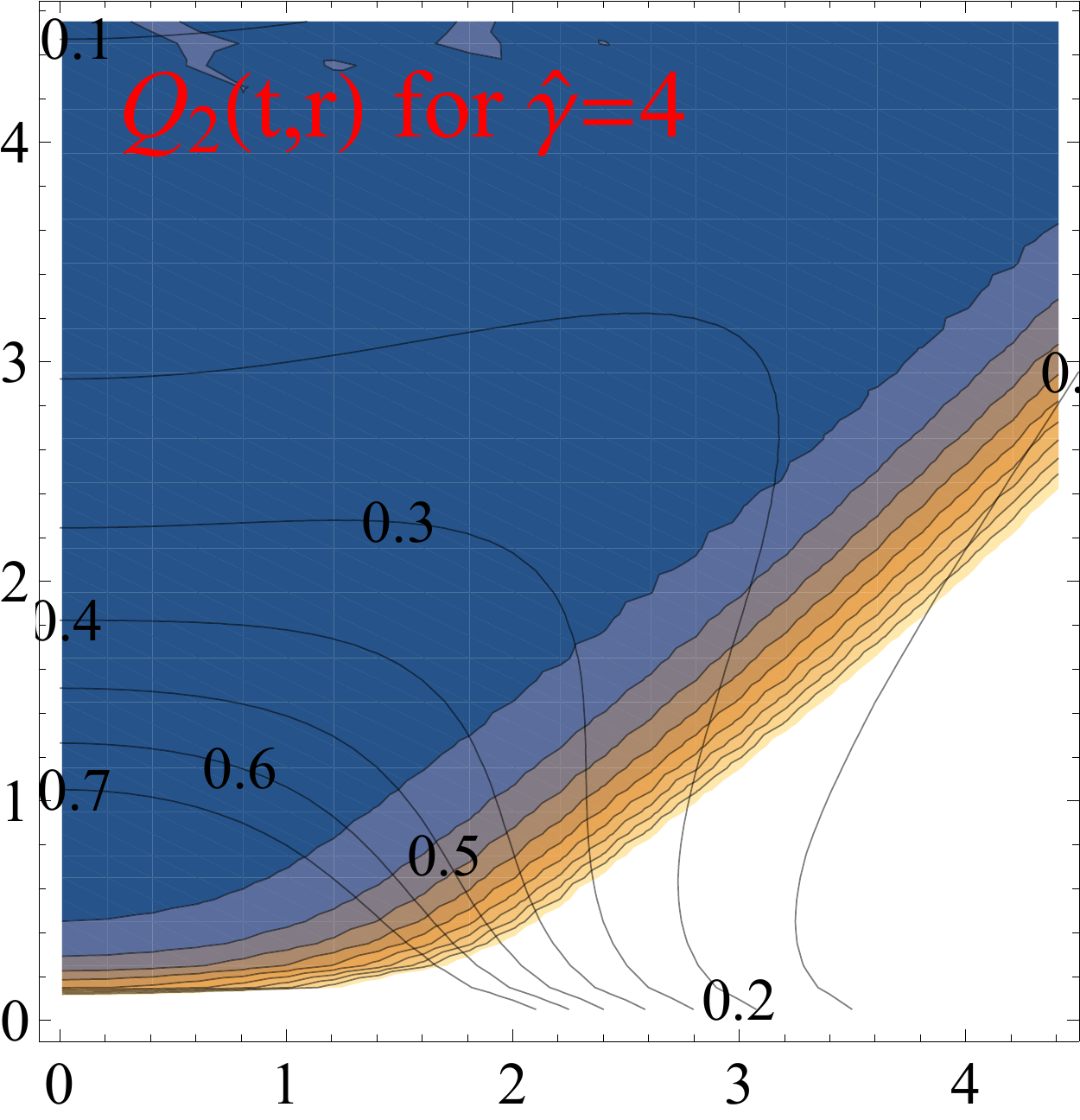}}
\\
\\
$\hat\gamma=8 $\hspace{0.5cm} 
&
\noindent\parbox[c]{0.25\hsize}{\includegraphics[width=0.25\textwidth]{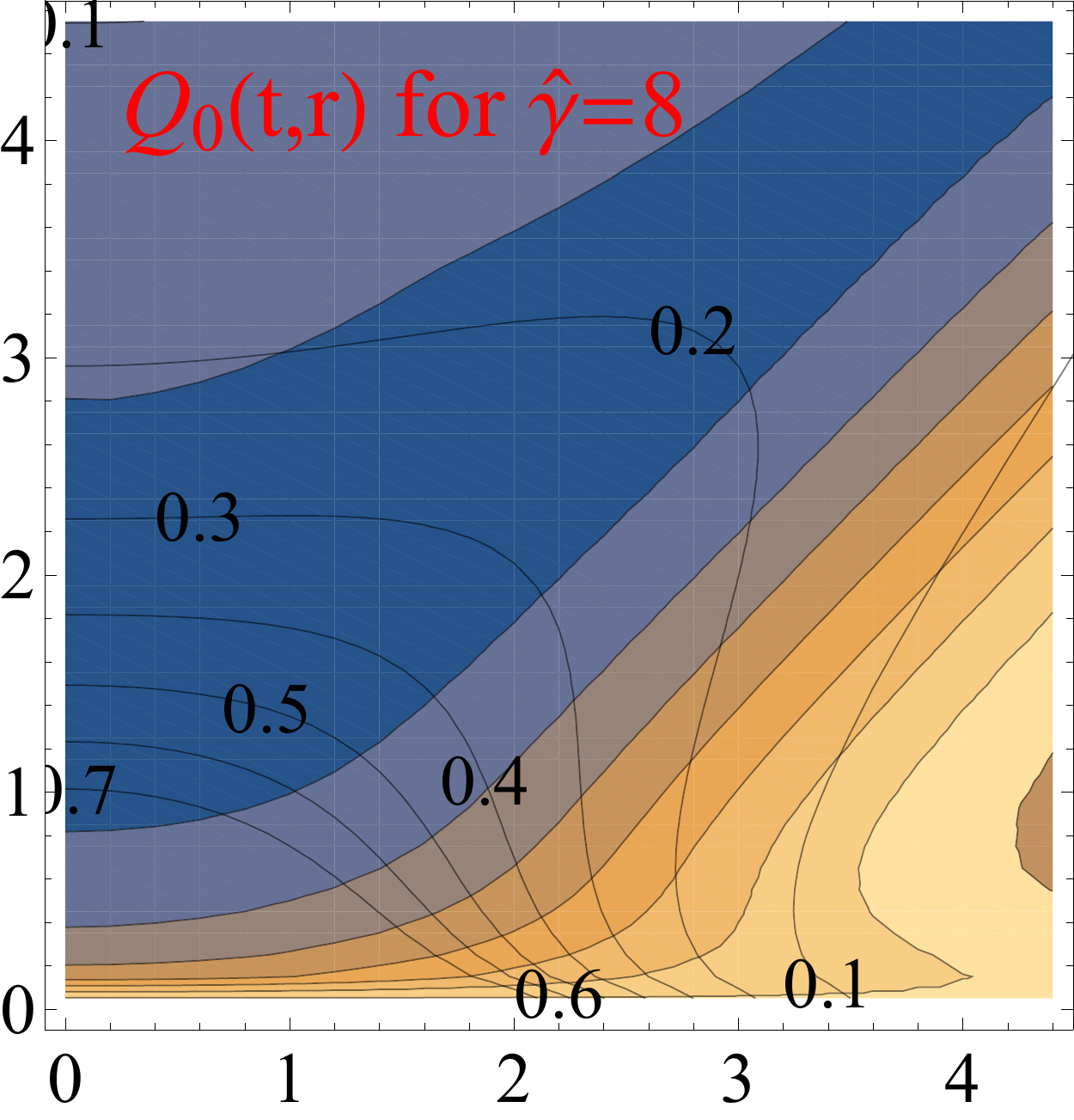}}
&
\noindent\parbox[c]{0.25\hsize}{\includegraphics[width=0.25\textwidth]{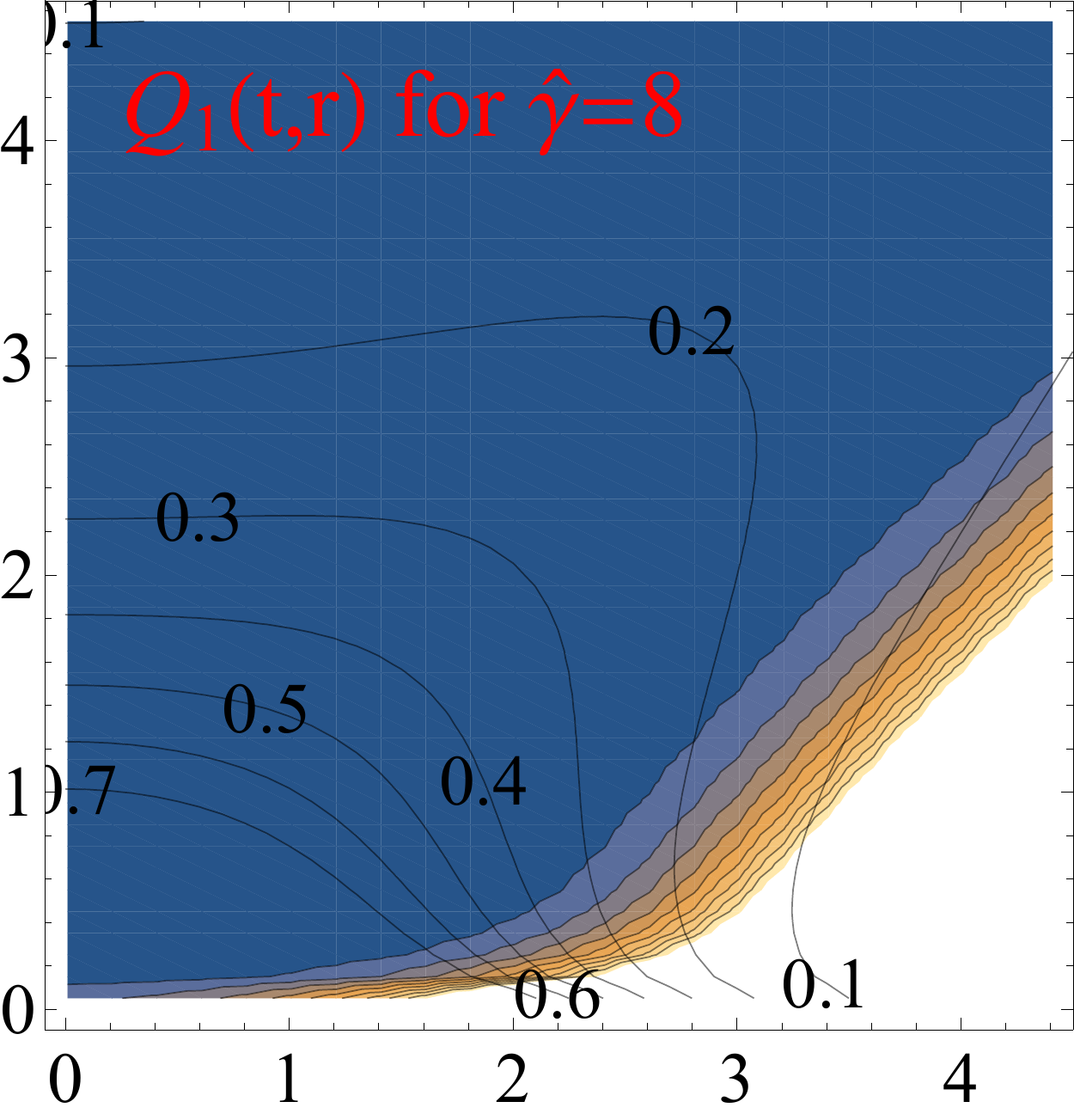} }
&
\noindent\parbox[c]{0.25\hsize}{\includegraphics[width=0.25\textwidth]{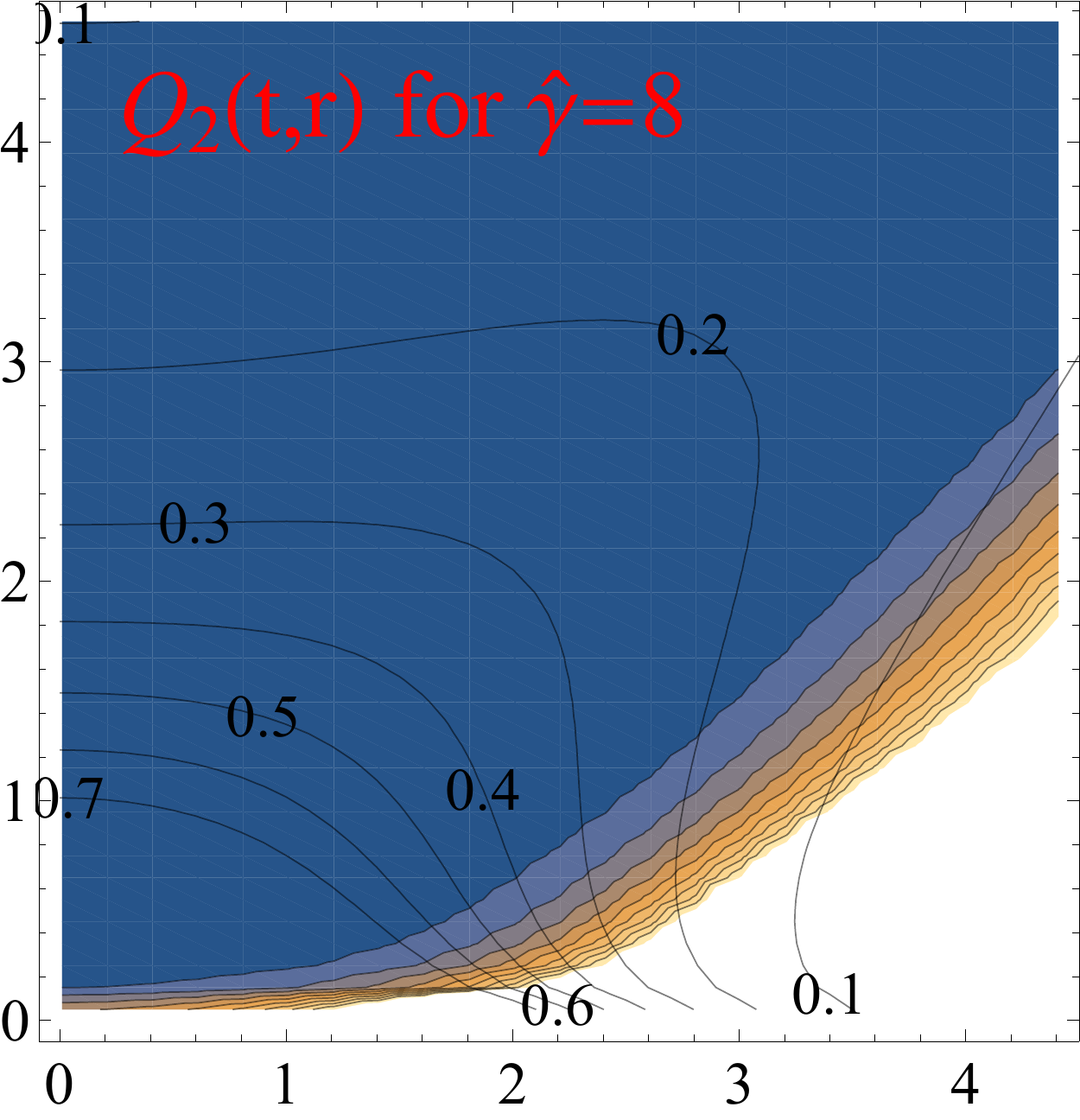}}

\end{tabular}
&
\noindent\parbox[c]{0.1\hsize}{\includegraphics{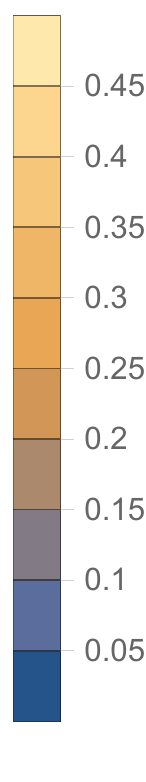} }
\end{tabular}
\caption{The difference between the energy-momentum tensor $T^{\mu\nu}$ calculated in kinetic theory, and $T^{\mu\nu}_{\rm hyd}$ defined via 
the constitutive hydrodynamic equations (\ref{eq52})-(\ref{eq55}), plotted in the $(r,t)$-plane in units of the local energy density $\varepsilon(t,r)$. Different
columns correspond to different values of $\hat\gamma = R/l_{\rm mfp} = 1, 2, 4$ and $8$. Upper (lower) panel: comparison of  $T^{\mu\nu}$ to
 $T^{\mu\nu}_{\rm hyd}$ up to first (second) order in the fluid dynamic gradient expansion, as defined in eqs. (\ref{eq60}), (\ref{eq61}). }
\label{fig3}
\end{figure*}

The subscript 2 indicates that $Q_2(t,r)$ compares the energy-momentum tensor of the kinetic theory with the constitutive hydrodynamic equations of motion to second order
in gradient expansion. It is of interest to compare also to the corresponding first order gradient expansion,
\begin{equation}
	Q_1(t,r) =  \sqrt{ \frac{  \left( \Pi_{\rm kin}- \Pi_{\rm hyd\, 1st} \right)^{\mu\nu}   \left( \Pi_{\rm kin}- \Pi_{\rm hyd\, 1st}  \right)_{\mu\nu} }{  \left(T_{\rm id} \right)^{\mu\nu}    \left(T_{\rm id} \right)_{\mu\nu} } }\, ,
	\label{eq61}
\end{equation}
and one can further consider the quantity obtained from comparing $T_{\rm kin}^{\mu\nu}$ to the zeroth order in fluid dynamic gradients, \emph{i.e.}, to $T_{\rm id}^{\mu\nu}$ 
\begin{equation}
	Q_0(t,r) =  \sqrt{ \frac{  \left( \Pi_{\rm kin} \right)^{\mu\nu}   \left( \Pi_{\rm kin} \right)_{\mu\nu} }{  \left(T_{\rm id} \right)^{\mu\nu}    \left(T_{\rm id} \right)_{\mu\nu} } }\, .
	\label{eq62}
\end{equation}
The quantity $Q_0(t,r)$ is of interest since it informs us about the extent to which the system is locally isotropic at $(t,r)$ in its local rest frame. 
The results shown for $Q_0$ in Fig.~\ref{fig3} indicate that for any non-transparent system with $\hat \gamma>0$, there is always a region of space-time in which the system is almost isotropic. 
For $r=0$, the corresponding blue regions in which $Q_0$ is small are centered around $t/R = 2$. 
The time $t/R=2$ is special in that it is the maximum time up to which particles produced 
initially at large radius can reach the center $r=0$ along free-streaming inward-going trajectories. For small $\hat\gamma$, these particles will cross each other with negligible isotropizing interactions.
In the local rest frame defined by Landau matching, such crossing particle streams lead to transient, locally isotropic systems. This effect is known from other fields in which one 
uses fluid dynamics to describe collective phenomena that is inherently non-fluid dynamical. For instance, this feature is referred to as \emph{shell-crossing} in the literature on 
cosmological Large Scale Structure~\cite{Rampf:2017jan,Pietroni:2018ebj}. There, one refers to a velocity field as \emph{multi-valued} if the velocity of individual particles in the neighborhood of
a space-time point $(x,t)$ is not characterized by the (thermal) dispersion around the unique rest frame velocity $u^\mu(t,x)$, but where it is determined instead by the (multiple) directions  in 
which particles were located in the distant past and from which they were streaming towards $(t,x)$. Kinetic theory realizes this scenario for almost transparent systems with small opacity $\hat\gamma$.

Non-interacting or weakly interacting streams of particles that cross each other can yield transient approximate isotropization in a neighborhood of $(t,x)$. The tell-tale sign of this phenomenon is a negligible $Q_0(t,x)$ 
combined with a sizeable $Q_1(t,x)$  or $Q_2(t,x)$ that indicate that fluid dynamics predicts \emph{larger}
shear viscous components in the energy momentum tensor that is realized in the kinetic theory. 

Fluid dynamics becomes a quantitatively controlled description of the collective dynamics in space-time regions where increasing the order of the fluid dynamic constitutive relation leads to a better description of the full energy-momentum tensor. We therefore refer
to a system evolved with kinetic equations as behaving fluid-dynamical in a space-time region around $(t,r)$ if $Q_1(t,r)$ is sufficiently small and if this smallness
persists to higher order in gradients, \emph{i.e.}, for $Q_2(r,t)$. To be specific, we use the criterion $Q_i(t,r)<0.1$ which corresponds to a situation in which the summed 
deviations of the kinetic shear viscous tensor from its constitutive hydrodynamical form is less than 10 \% of the total enthalpy $\varepsilon + p$ of the system. This
condition corresponds to the blue regions in Fig.~\ref{fig3}. 

 For $\hat\gamma=1$, even though a small band with $Q_0(r,t) < 0.1$ indicates that the
 system passes locally through approximately isotropic distributions, the region in which $Q_1(r,t) < 0.1$ is vastly different from the region in which  $Q_0(r,t) < 0.1$ 
 or $Q_2(r,t) < 0.1$. This  indicates that the gradient expansion does not converge and that the system displays characteristics that are very different
from those of a fluid. We draw a similar conclusion for systems characterized by $\hat\gamma = 2$, since there is no extended region in $(t,r)$ for which the difference $Q_1(t,r)$
is small {\it and} for which this smallness persists in $Q_2(t,r)$. For $\hat \gamma > 4$, however, our calculations indicate that the fluid dynamic gradient expansion is reliable, since the differences between $T^{\mu\nu}_{\rm kin}$
and the fluid dynamic ansatz for the energy-momentum tensor improves order by order in the gradient expansion. Further increasing $\hat\gamma $ to $8$, we find that the
regions of small $Q_1(r,t)$ and $Q_2(r,t)$ widen, and that they extend in particular to earlier time. In the phenomenological practice, such behavior would allow one to switch
at an earlier time from a full kinetic theory description to a fluid-dynamic one. For a discussion in the present framework, see \cite{Kurkela:2018qeb}.

In summary, 
 for the kinetic theory considered here, we know analytically the depth of the particle-like cut $\omega_{\rm cut} = -i/\tau_{\rm iso} = -i\, \gamma\, \varepsilon^{1/4}$ and the position of
 the (shear mode) pole $\omega_{\rm pole} = -i \textstyle\frac{3\eta}{4 \varepsilon} k^2$  in the complex plane of Fig.~\ref{fig1}. Parametrically, an excitation 
 of wavelength $1/k$ is expected to  propagate more fluid-like or more particle-like, depending on whether $\omega_{\rm cut}$ or $\omega_{\rm pole}$ is closer to the real axis. Since
 cut- and pole-terms depend inversely on $\gamma$, it is clear that for decreasing $\gamma$, particle-like excitations dominate. Also, since we consider the evolution of the components
 of the energy-momentum tensor which receives contributions from all wavelengths $1/k$, it is clear that  the transition from a system dominated by particle-like excitations to a  system 
 dominated by fluid dynamic excitations  proceeds gradually with increasing opacity $\hat\gamma$. 
 Quoting a precise value of $\hat\gamma$ for this transition would rely on agreeing on a criterion (such as $Q_i < 0.1$) of what is meant by {\it small}. Qualitatively, however, we observe
from Fig.~\ref{fig3} that systems with $\hat\gamma < 2$ display characteristics in the evolution of their energy-momentum tensor that show marked differences from a fluid dynamic evolution,
while systems with $\hat\gamma > 4$ evolve fluid-like from early times $t /R\ll 1$ onwards. In the first case, the fluid-dynamic gradient expansion does not converge in any extended
space-time region relevant for building up elliptic flow, while it does in the second case. 
This indicates that as the system becomes more opaque, particle-like excitations cease to dominate the
collective dynamics  in the range  $2 \lesssim  \hat\gamma \lesssim 4$.  
 
 %
%%%%%%%%%%%%%%%%%%%%%%%%%%%%%%%%%%%%%%%%%%%%%%%%%%%%%%%%%%
\begin{figure}[t]
\includegraphics[width=0.45\textwidth]{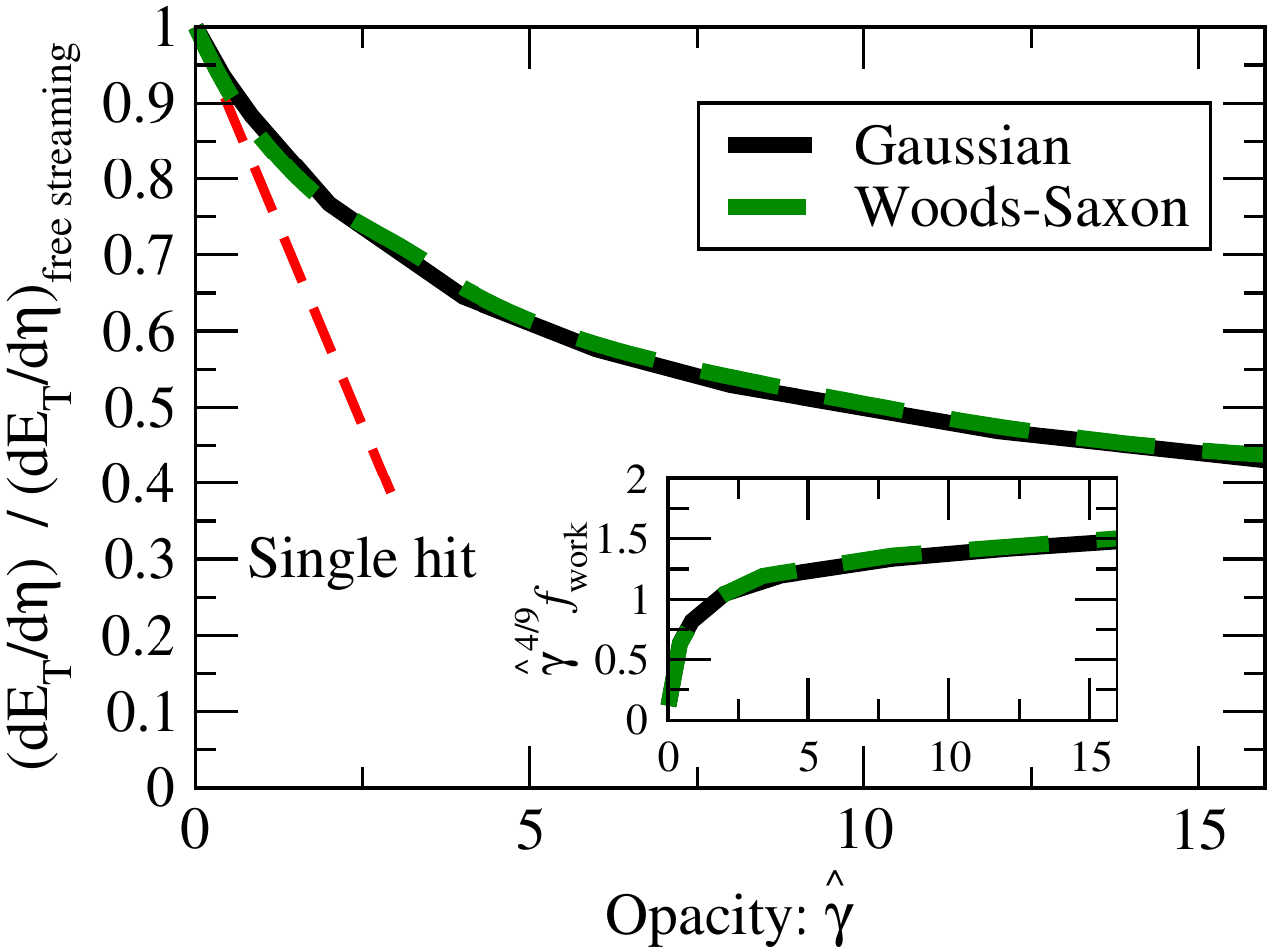}
\caption{The fractional work (\ref{eq65}) done during the kinetic theory evolution as a function of $\hat\gamma$. Numerical results for two different initial transverse profiles are compared
with the perturbative $O(\hat\gamma)$-result of ~\cite{Kurkela:2018ygx} (red line). }
\label{fig4}
\end{figure} 
%%%%%%%%%%%%%%%%%%%%%%%%%%%%%%%%%%%%%%%%%%%%%%%%%%%%%%%%%%%
\subsection{Work in kinetic theory}

The time-evolution of the transverse energy per unit rapidity and its azimuthal distribution $\frac{dE_\perp(t)}{d\eta_s d\phi}$ can be calculated
from kinetic theory according to (\ref{eq14}). The late time limit of these quantities corresponds to the experimentally accessible transverse energy distribution. In the phenomenological discussion
in section~\ref{sec4}, we need to know to what extent the azimuthal average  $\frac{dE_\perp(t\to \infty)}{d\eta_s}$ at late time differs from that at early times which is tantamount to asking how
much work the system does as a function of opacity $\hat\gamma$. 

To address this question, we determine $\frac{dE_{\perp, {\rm free}} }{d\eta_s} $ by free streaming the 
initial conditions to late times, and by comparing to the late time behavior of an opaque system $\frac{dE_\perp(t)}{d\eta_s}$ at finite $\hat\gamma$. The free-streamed transverse energy $\frac{dE_{\perp, {\rm free}} }{d\eta_s }$ follows from inserting the initial condition for $F(\tau_0,r,\theta;\phi, v_z)$ into (\ref{eq14}), and in the case of very anisotropic initial conditions $F \propto \delta(v_z)$ the transverse energy remains unchanged by the free-streaming evolution. In particular, for the
Gaussian initial condition  $F(\tau_0,r,\theta;\phi, v_z) = 2 \varepsilon_0 \delta(v_z)\, e^{-r^2/R^2}$ introduced in eqs. (\ref{eq41text}), one finds
\begin{equation}
	\frac{dE_{\perp, {\rm free}} }{d\eta_s} = \tau_0\, \varepsilon_0\, \pi\ R^2\, .
	\label{eq63}
\end{equation}
This is equivalent to Bjorken's estimate of the initial energy density,
\begin{equation}
	 \varepsilon_0 = \frac{\frac{dE_{\perp, {\rm free}} }{d\eta_s}}{\tau_0\, \pi\ R^2 }\, .
	 \label{eq64}
\end{equation}
Analogously, after having calculated for a system with given $\hat\gamma$ the time dependence of $F(\tau,r,\theta;\phi, v_z)$, we can determine $\frac{dE_\perp(t)}{d\eta_s}$
from (\ref{eq14}). In Fig.~\ref{fig4}, we plot the resulting ratio
\begin{equation}
	f_{\rm work}(\hat\gamma) = \frac{ \frac{dE_\perp(t\to\infty)}{d\eta_s} }{ \frac{dE_{\perp, {\rm free}} }{d\eta_s} }\, .
	\label{eq65}
\end{equation} 
To first order in a perturbative expansion of $\hat\gamma$ and for Gaussian initial conditions, we know that $f_{\rm work}(\hat\gamma) = -0.210\, \hat\gamma$ (red dashed line)~\cite{Kurkela:2018ygx}.
From the numerical results in Fig~\ref{fig4}, we see that rare occasional collisions in almost transparent systems are most efficient in increasing the work. As collisions between particle-like
excitations become more frequent with increasing opacity $\hat\gamma$ and as fluid dynamic excitations become gradually more important for the collective dynamics and the system does more work. 

At large opacity $\hat\gamma \gg 1$, the qualitative $\hat \gamma$-dependence of $f_{\rm work}$ can be understood in 
a simple toy model that neglects radial dynamics: We assume that the system is free-streaming until the time $\tau_{*}= \tau_{\rm iso}(\tau_*,R=0)$ which is the first time the system has had time to isotropize.  The isotropized system then evolves hydrodynamically until it disintegrates at time $\tau \sim R$. The free-streaming evolution---with vanishing longitudinal pressure $P_L$ and with fixed $\varepsilon(\tau) \tau$---does not do longitudinal work and all the work is done during the hydrodynamical phase between $\tau_* < \tau < R$, during which the energy density evolves with constant $\varepsilon(\tau) \tau^{4/3}$. The $f_{\rm work}$ is then determined by the length of the hydrodynamical phase. Solving self-consistently for the isotropization time gives 
\begin{equation}
\tau_{\rm *} \sim \frac{1}{\gamma \varepsilon(\tau_{\rm *})^{1/4}} \sim \frac{\tau_{\rm *}^{1/4}}{\gamma (\varepsilon_0\tau_0)^{1/4}}  \sim \frac{1}{\gamma^{4/3}(\varepsilon_0 \tau_0)^{1/3}},
\end{equation}
which in part gives
\begin{align}
\label{eq:scaling}
f_{\rm work } \sim \frac{\varepsilon(\tau = R)R}{\varepsilon_0 \tau_0} \sim \left( \frac{\tau_{\rm iso}}{R}\right)^{1/3} \sim \hat \gamma^{-4/9}.
\end{align}
This approximate scaling for $f_{\rm work}$ is  observed for opacities $\hat \gamma \gg 1$ in the inset of Fig.~\ref{fig4}, which shows the flattening of the rescaled quantity $\hat \gamma^{4/9} f_{\rm work}(\hat \gamma)$ at large  $\hat \gamma$-values.

Finally, we note the remarkable insensitivity of $f_{\rm work}$ to the initial profile illustrated by the numerical similarity between the results in obtained using either gaussian (Black line) or Woods-Saxon (Green dashed line) initial profile in Fig.~\ref{fig4}. 

\subsection{Results for the linear response $v_n/\epsilon_n$}
\label{sec3d}
The produced transverse energy is of particular interest as it is insensitive to the modeling of hadronization. It is defined by the 
first $p_\perp$-moment of the measured particle distributions $\textstyle\frac{dN}{dy dp_\perp^2 d\phi}$,
\begin{eqnarray}
	\frac{dE_\perp}{d\eta_s d\phi} &\equiv& \int dp_\perp^2\, p_\perp\, \frac{dN}{dy\, dp_\perp^2\, d\phi}
	\label{eq13} \\
	&\underset{t\rightarrow \infty}{=}&\frac{1}{2\pi} \frac{dE_\perp}{d\eta_s}\left(1+2 \sum_{n=1}^\infty v_n \cos\left( n(\phi-\psi_n)\right) \right).
	\nonumber 
\end{eqnarray}
The Fourier coefficients $v_n$ and reaction plane orientations $\psi_n$
parametrize the azimuthal dependence of the transverse {\it energy} flow.  Here we have used the fact that particles with 
momentum rapidity $y$ after the last interaction will end up in a spacetime rapidity $\eta_s = y$. See Appendix \ref{sec2c} for details.

 Spatial deformations $\delta_{n,m}$ are introduced in the initial conditions according to (\ref{eq40text}).
For the kinetic theory formulation explored in this work, the first results for the linear response of the elliptic energy flow $v_2$ to initial elliptic perturbations $\epsilon_2$ were reported in Ref.~ \cite{Kurkela:2018qeb}. The purpose of the present subsection is to complete this information with results on higher harmonics, and to demonstrate that the results depend negligibly on the transverse 
profile function. These results are complete to all orders in $\hat\gamma$, but they are first order in $\epsilon_n$. The formulation in section~\ref{sec2} allows for an extension to all orders in $\epsilon_n$ 
in future work.

\begin{figure}[t]
\includegraphics[width=0.45\textwidth]{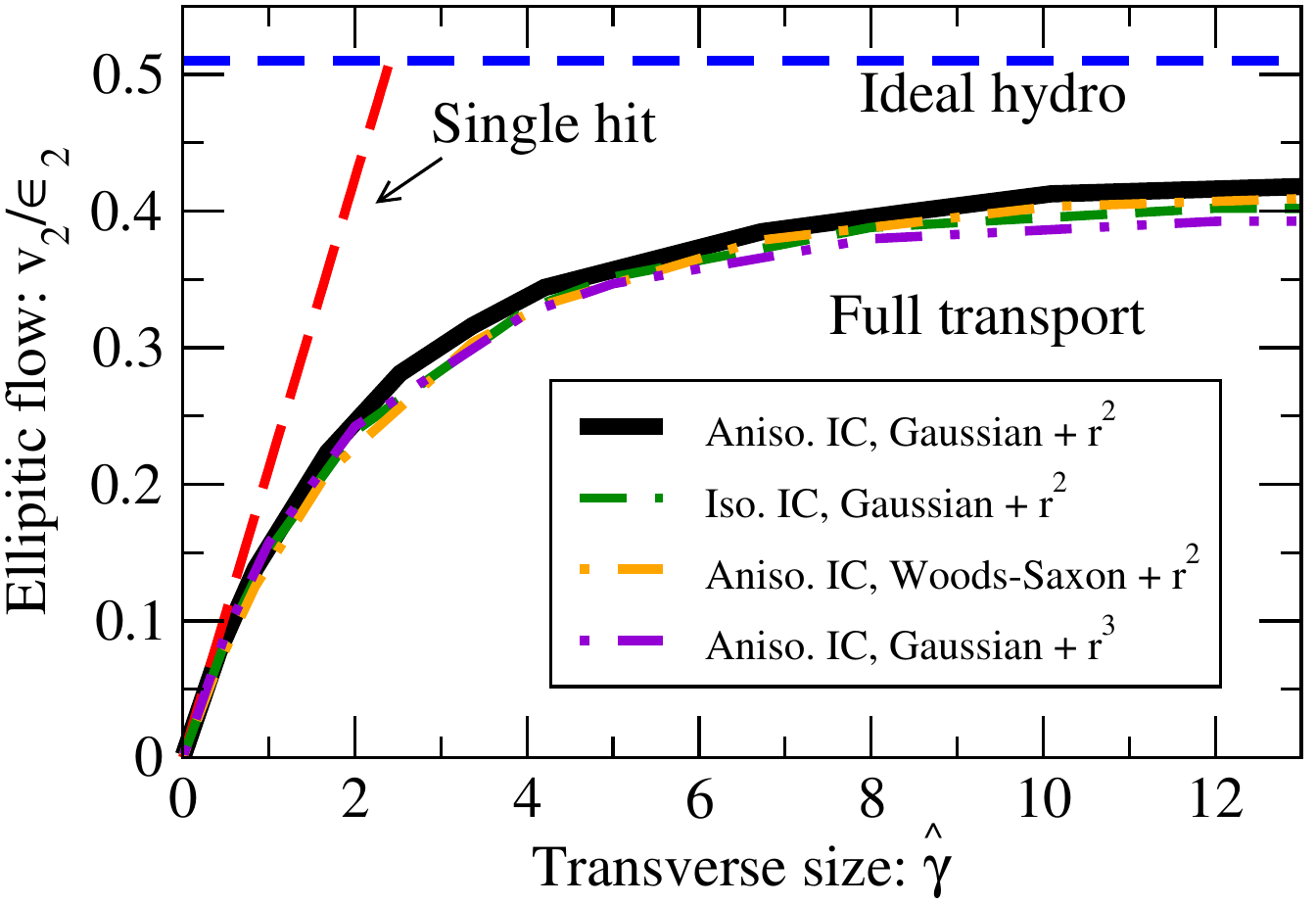}
\includegraphics[width=0.45\textwidth]{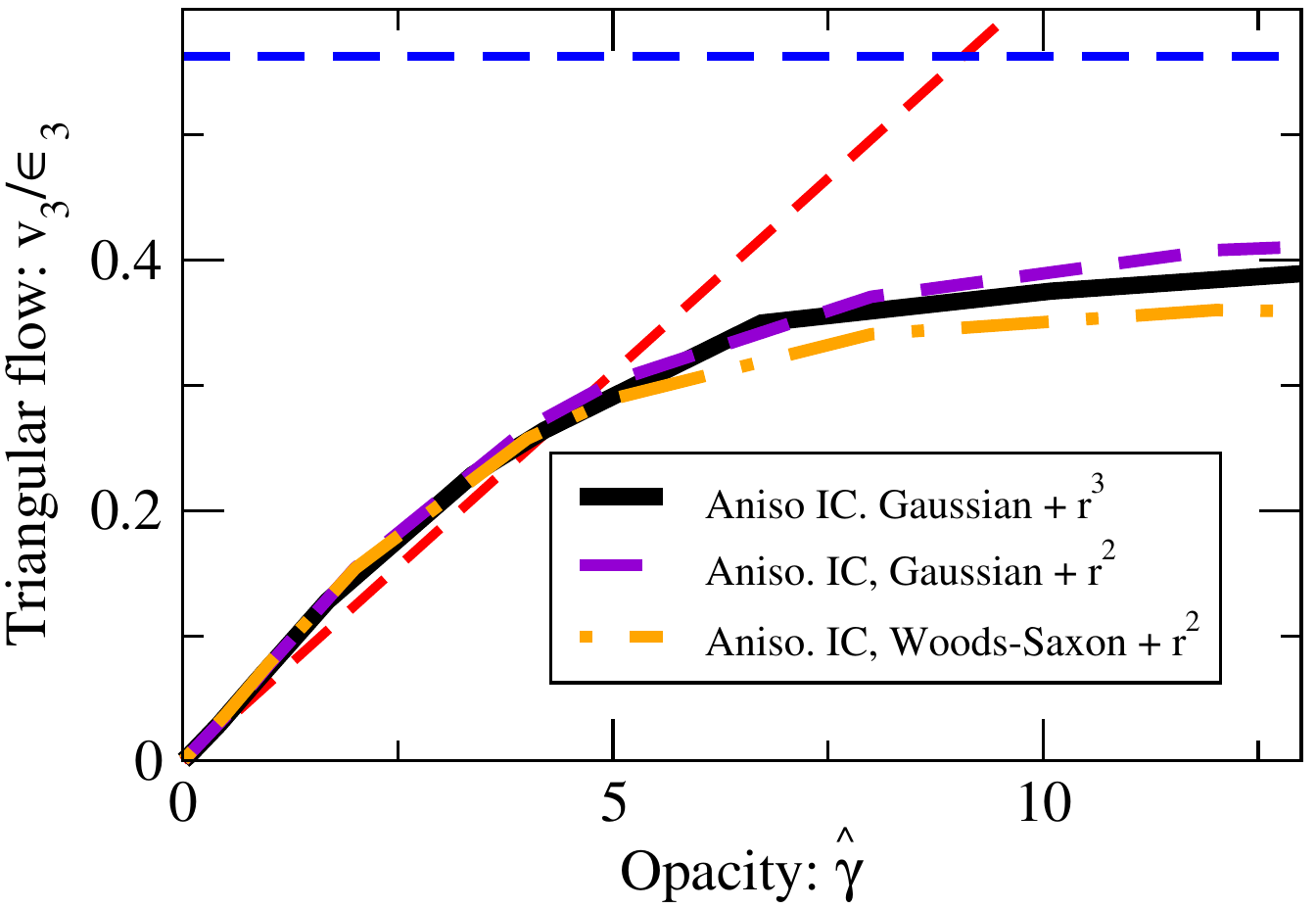}
\caption{The linear response $v_n/\epsilon_n$ of energy flow coefficients $v_n$ to for elliptic (upper panel) and triangular (lower panel) eccentricities $\epsilon_n$. Different radial profiles yield similar results,
curves discussed in section~\ref{sec3d}.}
\label{fig5}
\end{figure}

Fig.~\ref{fig5} shows kinetic theory results for elliptic and triangular flow normalized by eccentricity (for $n\geq 2$)
\begin{align}
\epsilon_n = \frac{\int d^2 x_\perp r^{n} \cos(n \theta) \varepsilon(r,\theta)}{\int d^2 x_\perp r^{n}  \varepsilon(r,\theta)}\, .
\label{eq:epsilon}
\end{align}
This standard definition of $\epsilon_n$ is adopted for all initial conditions irrespective of their radial dependence ($\propto \delta_{n,m}r^m$) in (\ref{eq40text}) and their azimuthally averaged transverse profile.

The black and the purple lines in Fig.~\ref{fig5} correspond to Gaussian background initial conditions of (\ref{eq42text}) and (\ref{eq40text}) with $\delta_{2,2}\neq 0$ (black line) or $\delta_{2,3}\neq 0$ (purple line). 
This illustrates the insensitivity of the 
response to details of the radial profile of the perturbation. Moreover, there is a remarkable insensitivity to the background profile, since Wood-Saxon (yellow line) and Gaussian (black line) initial conditions yield numerically similar linear responses. Furthermore, the green line in Fig.~\ref{fig5} corresponds to an isotropic initial momentum distribution $F(r,\theta,v_z) = \varepsilon_0 e^{-r^2/R^2}$ with a $\delta_{2,2}$ perturbation. Again no significant sensitivity to the initial condition is visible. Similar observations can be made for $v_3/\epsilon_3$ from the bottom panel of Fig.~\ref{fig5}.

Fig.~\ref{fig5} illustrates for an important class of observables how kinetic theory interpolates between the limits of free-streaming and perfect fluidity. In the free streaming limit
($\hat\gamma = 0$) momentum asymmetries $v_n$ are not built up from initial spatial asymmetries $\epsilon_n$. As a consequence, $v_n/\epsilon_n$ starts from zero, and its $\hat\gamma$-slope
(red lines in Fig.~\ref{fig5}) 
close to $\hat\gamma = 0$ can be understood as the consequence of perturbatively rare final state interactions in a small and/or dilute system. In the opposite limit of ideal fluid
dynamics  ($\hat\gamma \to \infty$), momentum asymmetries are built up from spatial eccentricities with vanishing dissipation and thus most efficiently. As a consequence, $v_n/\epsilon_n$ increases monotonously
with increasing opacity $\hat\gamma$ and it approaches at very large $\hat\gamma$ the ideal fluid limit (dashed blue lines in Fig.~\ref{fig5}) from below. For $\hat\gamma < 16$, the difference
between this ideal fluid limit and the result of full kinetic theory indicate dissipative effects. As discussed in the context of Fig.~\ref{fig3},
these dissipative effects are accounted for by viscous fluid dynamics or, for smaller $\hat\gamma$, by particle-like excitations outside the fluid dynamic description. 

\section{Data comparison}
\label{sec4}

The kinetic theory defined in section~\ref{sec2} and analyzed in section~\ref{sec3} is a one-parameter model. 
By comparing it to experimental data, we do not intend to claim that a complete phenomenological understanding of data could be based on this one-parameter model. It is obvious that it 
cannot, and by stating at the beginning of this section the obvious, we hope to preempt any possible misunderstanding of this point. Rather, by comparing to data, we want to 
understand to what extent the one generic physics effect implemented in this model could account for one central qualitative feature in the data, namely for the remarkable system size dependence
of measures of collectivity. Is the centrality dependence of flow measurements consistent with a dynamics that interpolates between free-streaming and viscous fluid dynamics, or is
it more consistent with a collective dynamics that is always fluid-like in the sense that it depends negligibly on particle-like excitations even for the smallest collision systems studied at the LHC?

\subsection{The centrality dependence of opacity $\hat\gamma$ }
We recall that in kinetic theory, the retarded two-point correlation functions of the energy-momentum tensor display fluid-dynamic poles and particle-like cuts. The position of the
hydrodyamic poles is unambigously determined by the parameters of the kinetic theory. In particular, for the kinetic theory studied here, the relation 
\begin{equation}
	\gamma \approx \frac{0.11}{\frac{\eta}{s}}
	\label{eq66}
\end{equation}
follows from eq.(\ref{eq57}) if one choses $\varepsilon = 13\, T^4$  consistent with results from lattice QCD calculations \cite{Borsanyi:2013bia, Bazavov:2014pvz}. 
As discussed in section~\ref{sec2}, the numerical solutions of this kinetic theory depend only on the dimensionless opacity
 $\hat\gamma = \gamma R^{3/4} \left(\varepsilon_0 \tau_0\right)^{1/4}$,
which we rewrite with the help of eqs.~(\ref{eq63}), (\ref{eq65}) and (\ref{eq66}) in the form
\begin{equation}
	\hat\gamma = \frac{0.11}{\frac{\eta}{s}}  \left( \frac{1}{\pi f_{\rm work}(\hat\gamma)} \right)^{1/4}  \left(R\, \frac{dE_\perp}{d\eta_s} \right)^{1/4}\, .
	\label{eq67}
\end{equation}
Here, the factor $ f_{\rm work}(\hat\gamma) $ appears since we have traded the initial energy density 
$\varepsilon_0$  for an expression in terms of the experimentally measured final transverse energy $\frac{dE_\perp}{d\eta_s}$. The function $ f_{\rm work}(\hat\gamma) $  is a 
prediction of kinetic theory. 

\begin{figure}[t]
\includegraphics[width=0.49\textwidth]{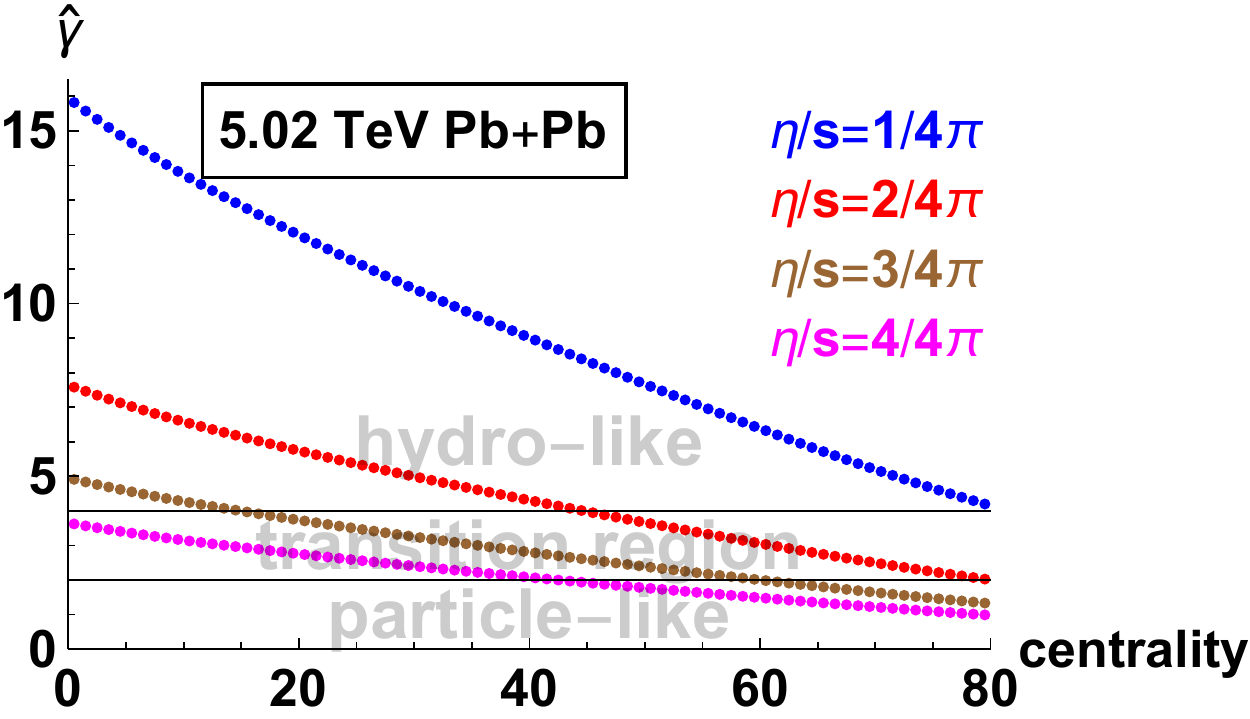}
\caption{The centrality dependence of $\hat\gamma$, calculated from eq.(\ref{eq67}) for different values of $\eta/s$. Horizontal lines denote the values $\hat\gamma =2$
below which the kinetic transport evolution is far from fluid dynamics and $\hat\gamma = 4$ above which the system evolves fluid-like.}
\label{fig6}
\end{figure}

By solving (\ref{eq67}) for $\hat\gamma$, we can determine the centrality-dependence of $\hat\gamma$ from the centrality dependence of the transverse energy $\frac{dE_\perp}{d\eta_s}$
and the centrality dependence of the rms radius $R$ of the initial transverse energy profile. In the following, we take the  centrality dependence of $\frac{dE_\perp}{d\eta_s}$ from data, and 
we use for simplicity for the rms radius $R$  the centrality dependence obtained in an optical Glauber model~\cite{Kharzeev:2000ph}. Since both quantities enter (\ref{eq67}) only as the 
fourth root, experimental and theoretical uncertainties resulting from this procedure are negligible 
(\emph{e.g.}, varying $R$ by 10 \% amounts only to doubling the line width in Fig.~\ref{fig6}).

\subsubsection{Opacity $\hat\gamma$ in PbPb at the LHC}

The centrality dependence of $\hat\gamma$ shown in Fig.~\ref{fig6} for $\sqrt{s_{\rm NN}} = 5.02$ TeV PbPb collisions at the LHC was obtained with $\frac{dE_\perp}{d\eta_s}$ -values 
determined from the $p_\perp$-spectrum $\textstyle\frac{dN}{dp_\perp d\eta_s}$ published by ALICE in Ref.\cite{Acharya:2018qsh} for 
9 centrality bins between 0-5 \% and 70-80\%, corrected for the presence of neutrals and interpolated to finer intermediate centrality bins. 
Combined with the conclusions drawn from Fig.~\ref{fig3}, Fig.~\ref{fig6} informs us for which values of $\eta/s$ fluid dynamic behavior can be expected as a function of system size.

For instance, for $\eta/s=2/4\pi$, the value $\hat\gamma \approx 8$ reached in the most central PbPb collisions (see Fig.~\ref{fig6}) signals according to Fig~\ref{fig3}
that for matter of this viscosity the collective dynamics realized in central PbPb collisions is fluid-like. But for semi-peripheral PbPb collisions with centrality $> 50 \%$, 
the $\eta/s=2$-curve in Fig.~\ref{fig6} shows values $\hat\gamma < 4$ for which a fluid dynamic picture of the evolution becomes questionable, and for $> 80 \%$ centrality, 
values $\hat\gamma < 2$ signal a collective dynamics in peripheral collisions which is qualitatively different from that of a fluid.
As seen from Fig.~\ref{fig6}, the centrality range for which a fluid dynamic gradient expansion is expected to be quantitatively reliable ($\hat\gamma > 4$) or
qualitatively indicative ($\hat\gamma > 2$) changes rapidly with $\eta/s$. %  For $\eta/s = 3/4\pi$, only the top 15 \% central collisions correspond to values $\hat\gamma >4$ for which we have evidence of a good convergence of the fluid dynamic gradient expansion. 

According to Fig.~\ref{fig6}, a gradual numerical change form $\eta/s =1/4\pi$ to $\eta/s =4/4\pi$ translates into a qualitatively different picture of the physics at work,
since it translates into scenarios in which particle-like excitations make either the negligible or the dominant contribution to the collective dynamics in semi-peripheral
and peripheral PbPb collisions at the LHC. Irrespective of the specific kinetic theory studied in the present work, this provides a strong physics argument for a 
precise determination of $\eta/s$ from data. 

\begin{figure}[t]
\includegraphics[width=0.49\textwidth]{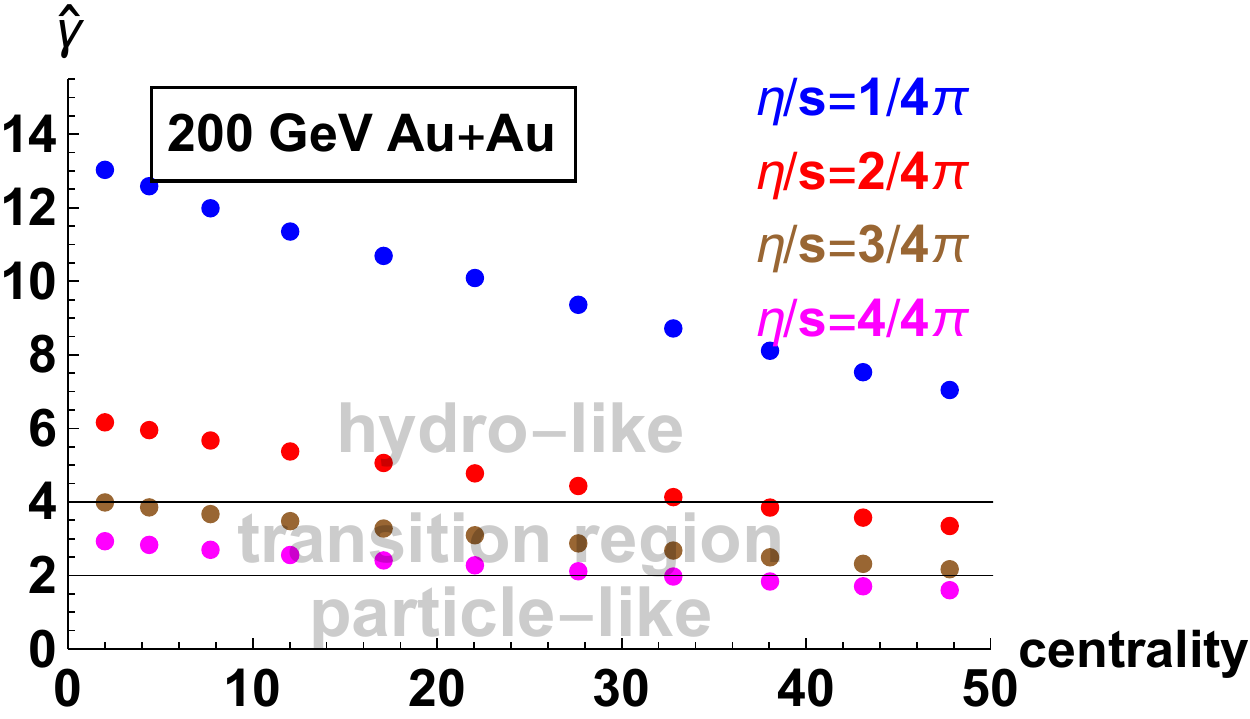}
\caption{Same as Fig.~\ref{fig6}, but for Au+Au collisions at RHIC where smaller $\frac{dE_\perp}{d\eta_s}$ values lead to reduced $\hat\gamma$ values in the same centrality bin.}
\label{fig7}
\end{figure} 

\subsubsection{Opacity $\hat\gamma$ in AuAu at RHIC}
In close analogy to the calculation of $\hat\gamma$ for PbPb collisions at the LHC, we calculate $\hat\gamma$ 
for $\sqrt{s_{\rm NN}} = 200$ GeV AuAu collisions based on STAR data of the $N_{\rm part}$-dependence of $dE_\perp/d\eta_s$~\cite{Adams:2004cb}.
For the same centrality and the same assumed value of $\eta/s$, $\hat\gamma$ is approximately
$20 \%$ smaller at RHIC than at the LHC, and the parameter regions in which fluid-like and particle-like excitations dominate collectivity shift accordingly, see Figs.~\ref{fig6} and~\ref{fig7}. 
For nucleus-nucleus collisions at RHIC to show a more fluid-like behavior  (i.e. correspond to a larger
$\hat\gamma$) than at the LHC, the matter produced at RHIC would thus have to display a significantly smaller $\eta/s$, since at comparable values of $\eta/s$,  
the higher center of mass energy and the higher transverse energy density attained at LHC facilitates fluid-like behavior.

\subsubsection{Opacity $\hat\gamma$ in pPb at the LHC}
To calculate the opacity $\hat\gamma$ for pPb collisions at the LHC, we make the simplifying assymption that the rms-width $R$ of the transverse area initially active in pPb collisions does
{\it not} change with event multiplicity. This assumption has a physical and a pragmatic motivation. Physically, it amounts to stating that the dispersion in the multiplicity distribution
at fixed impact parameter is so broad that the correlation between event multiplicity and event centrality (i.e., impact parameter) is weak. From a pragmatic point, choosing $R$ to
be multiplicity-independent is reasonable, since we do not have firm knowledge about how the initial geometry of pPb collisions changes with multiplicity. Moreover, for the purpose
of calculating $\hat\gamma$, this lack of knowledge is partially alleviated by the fact that information about the multiplicity-dependence of $R$ enters $\hat\gamma$ only as a fourth
root. We therefore choose in the following $R = 1$ fm as default value for pPb collisions, noting that even extreme values of $R= 2$ fm would affect results for $\hat\gamma$ only
by 20\%.  

\begin{figure}[t]
\includegraphics[width=0.49\textwidth]{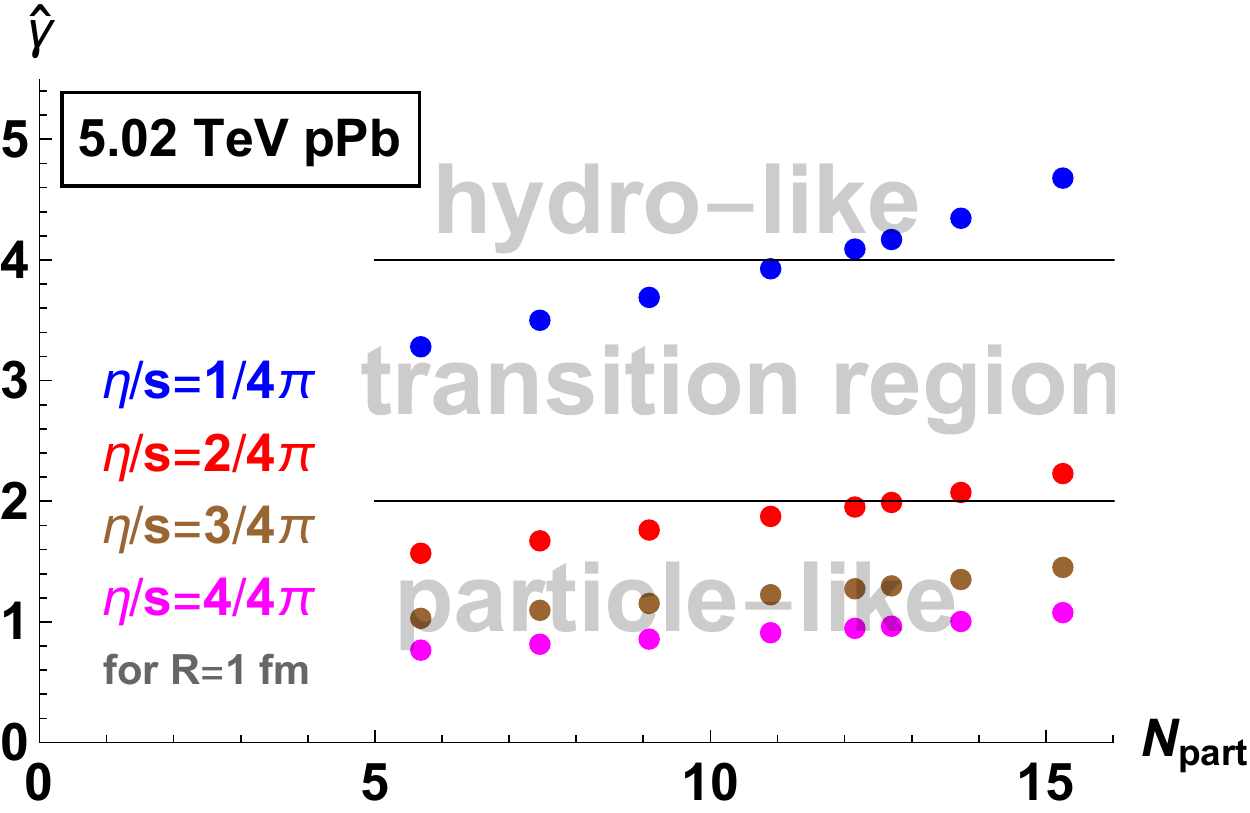}
\caption{The $N_{\rm part}$-dependenct of $\hat\gamma$ in pPb collisions, determined from (\ref{eq67}) for different values of $\eta/s$. }
\label{fig8}
\end{figure}

With the assumption of a multiplicity-independent geometry in pPb, the entire centrality dependence of $ \hat\gamma$ in eq.(\ref{eq67}) originates now from the experimentally inferred
$N_{\rm part}$-dependence  of $dE_\perp/d\eta_s$ that we take from measurements of the CMS collaboration~\cite{Sirunyan:2018nqr}.
The results of Fig.~\ref{fig8} show a substantial reduction of $\hat\gamma$ in the smaller pPb collision systems if compared to heavy ion collisions. 
In pPb collisions, a transport coefficient $\eta/s = 1/4\pi$ indicating maximal fluidity needs to be assumed to reach at least in the highest multiplicity-bins values of $\hat\gamma > 4$. In summary,
under essentially all conditions explored in Fig.~\ref{fig8}, value of $\hat\gamma$ is so small in pPb collisions that collective dynamics cannot develop fluid-like behavior.

\subsection{Comparing kinetic theory to $v_n/\epsilon_n (\hat\gamma)$: assumptions and uncertainties }
\label{sec4b}
Before presenting in the following subsections comparisons of kinetic theory to data on $v_n/\epsilon_n$, we list here the main elements of this comparison and we discuss the underlying assumptions and resulting uncertainties.
\subsubsection{Transverse energy $ \frac{dE_\perp}{d\eta_s} $ and energy flow $v_2$ and $v_3$}
\label{sec4b1}
While data on $\textstyle \frac{dE_\perp}{d\eta_s} $ are published for all collision systems, data on the harmonic decomposion (\ref{eq13}) of 
$\textstyle \frac{dE_\perp}{d\eta_s\, d\phi} $ in terms of energy flow coefficients $v_2$ and $v_3$ are measurable but have not been published yet. 
Therefore, we have to infer them from the measured particle flows $v_m^{\rm particle}(p_\perp)$, $m=2,3$ and from the published single inclusive charged
hadron spectra $\textstyle\frac{dN}{dp_\perp\, d\eta_s}$ by calculating
\begin{equation}
	v_m\equiv v_m^{\rm energy} = \frac{\int_0^\infty dp_\perp\, p_\perp\, v_m^{\rm particle}(p_\perp) \frac{dN}{dp_\perp\, d\eta_s}}{\int_0^\infty dp_\perp\, p_\perp\, \frac{dN}{dp_\perp\, d\eta_s}}\, ,
	\label{eq*2}
\end{equation}
where $\textstyle\frac{dE_\perp}{d\eta_s} = \textstyle\frac{3}{2} \int_0^\infty dp_\perp\, p_\perp\, \frac{dN}{dp_\perp\, d\eta_s}$ (the factor $\textstyle\frac{3}{2}$ corrects approximately for neutrals). 
The integrands in (\ref{eq*2}) differ by one power of $p_\perp$ from the ones
used to determine the $p_\perp$-integrated particle flow
\begin{equation}
	v_m^{\rm particle} = \frac{\int_{p_\perp^{\rm min}}^{p_\perp^{\rm max}} dp_\perp\, v_m^{\rm particle}(p_\perp) \frac{dN}{dp_\perp\, d\eta_s}}{
			\int_{p_\perp^{\rm min}}^{p_\perp^{\rm max}} dp_\perp\,  \frac{dN}{dp_\perp\, d\eta_s}}\, .
			\label{eq*1}
\end{equation}
Data on $v_m^{\rm particle}$ are usually quoted as $p_\perp$-integrated within a finite $p_\perp$-range that varies between experiments. In all collision systems discussed in the following,
we have determined $v_{2,3}^{\rm energy}$ from  $v_{2,3}^{\rm particle}$ and $ \textstyle\frac{dN}{dp_\perp\, d\eta_s}$ by first checking consistency of the published $p_\perp$-integrated 
particle flows with (\ref{eq*1}), and then determining the energy flow coefficients according to (\ref{eq*2}). One can then determine the correction factor $c_{\rm e-corr}$
between measured particle flow and extracted energy flow,
\begin{equation}
	v_{m}^{\rm energy} = c_{\rm e-corr,m} v_{m}^{\rm particle}\Big\vert_{{\rm 0.2 GeV} < p_\perp < {\rm 3 GeV}}\, .
	\label{eq*3}
\end{equation}
Both particle flow and momentum flow coefficients (\ref{eq*2}) and (\ref{eq*1}), respectively, are centrality dependent and therefore, in principle, $c_{\rm e-corr,m}$ could be
centrality-dependent as well. We anticipate that the centrality-dependence of $c_{\rm e-corr,m}$ turns out to be negligible in all collision systems considered. 

Particle flow coefficients $v_{m}^{\rm particle}$ are measured with different methods that are designed to include different sources of particle correlations and that therefore yield 
numerically different results. In particular, the 2nd and higher order cumulant flows $v_{m}^{\rm particle}\lbrace 2 \rbrace$, $v_{m}^{\rm particle}\lbrace 4\rbrace$, $v_{m}^{\rm particle}\lbrace 6\rbrace$, ...,
are designed such that they include only particle correlations that persist in connected 2-, 4-, 6-, ... particle correlation functions. Moreover, to eliminate contributions from jet-like structures,
data for $v_{m}^{\rm particle}$ are often designed to include only the correlation of particles separated by a sizeable rapidity gap $\vert \Delta\eta_s\vert$. 

To choose the variant of particle flow measurements from which we infer $v_m^{\rm energy}$ via (\ref{eq*2}), we start from the following two conisiderations. 
First, one of the simplest perturbative corrections to free-streaming that is accounted for by 
kinetic theory may be thought of as a single rare scattering that leads to correlations amongst very few of the final state particles and that leads, a fortiori, to an azimuthal correlation of 
$\textstyle \frac{dE_\perp}{d\eta_s\, d\phi} $ carried by a small fraction of the entire $ \frac{dE_\perp}{d\eta_s} $. As such a correlation is not shared by all particles in the event, it would 
not fully persist in higher order cumulants, while it is fully accounted for in kinetic theory.  Second, the formulation of
the kinetic theory presented here is not sufficiently detailed to include the characteristic dynamics of jet-like fragmentation patterns, and a meaningful comparison of kinetic theory to data
should thus start from an energy flow $v_m^{\rm energy} $ from which jet-like correlations are eliminated to the extent possible. The combination of both considerations prompts us to calculate
$v_m^{\rm energy} $ from $2$-particle cumulant measurements with rapidity gap.

Based on these considerations, the following data sets enter our analyses:
\begin{enumerate}
\item
{\it Data for PbPb collisions at the LHC}\\
$p_\perp$-differential particle flow $v_{2,3}\lbrace 2, \vert \Delta\eta_s\vert \rbrace (p_\perp)$~\cite{Acharya:2018lmh} published in nine centrality
classes between 0-5 \% and 70-80\%, and single inclusive charged hadron spectra~\cite{Acharya:2018qsh} in the same centrality classes and the same mid-rapidity range are used to determine 
$c_{\rm e-corr,2} = 1.34 (1.33)$ and $c_{\rm e-corr,3} = 1.52 (1.51)$ for $\sqrt{s_{NN}} = 5.02$ ($\sqrt{s_{NN}} = 2.76$ ) TeV PbPb collisions.
Fluctuations of these numbers with centrality are small, do not reveal any systematic trend, and may be attributed to
uncertainties of $p_\perp$-differential flow data at higher $p_\perp$. The $p_\perp$-integrated particle flow calculated from these data according to (\ref{eq*1}) is checked to be consistent with 
the $p_\perp$-integrated $v_2 \lbrace {2, \vert\Delta\eta_s\vert > 1} \rbrace$ ~\cite{Acharya:2018lmh} restricted to the experimentally chosen range $0.2 {\rm GeV} < p_\perp < 3 {\rm GeV} $.
The factors $c_{\rm e-corr,2}$ and $c_{\rm e-corr,3}$ are then used
to infer the energy flow via (\ref{eq*3}) for the 80 different $1 \%$-wide centrality bins published in ~\cite{Acharya:2018lmh} for $\sqrt{s_{NN}} = 5.02$ TeV PbPb, and for the somewhat fewer
bins between 0\% and 80\% centrality published for $\sqrt{s_{NN}} = 2.76$ TeV PbPb.\\
The data on single inclusive charged hadron spectra~\cite{Acharya:2018qsh} are further used to determine $dE_\perp/d\eta_s$ for the nine bins between 0\% and 80\% centrality, and
are then interpolated to obtain centrality dependent values for $dE_\perp/d\eta_s$ in 80 1\%-wide centrality bins.  This is checked to be consistent with published results on the centrality dependence
of $dE_\perp/d\eta_s$. \\
In addition to these data of the ALICE collaboration, there are CMS-data on $v_2\lbrace {2, \vert\Delta\eta_s\vert > 2}\rbrace$ in $\sqrt{s_{NN}} = 5.02$ PbPb~\cite{Sirunyan:2017uyl} that could allow one to
extend particle flow measurements to $\approx 93\% $ centrality (Ref.~\cite{Sirunyan:2017uyl} reports data on pp and pPb, but its HEPDATA-repository documents the
latest CMS data on PbPb collisions). As these data are presented as a function of off-line tracks $N^{\rm offline}_{\rm trk}$, as they are integrated over a slightly different $p_\perp$-range,
and as they are taken in a much wider rapidity range, there are subtelties in consistently infering energy flow as a function of centrality with CMS and ALICE-data on the same plot. We are in 
contact with both collaborations to arrive at such a combination, but this lies outside the scope of the present manuscript. 
\item
{\it Data on pPb collisions at the LHC}\\
CMS data on the pseudorapidity dependence of the transverse energy $dE_\perp/d\eta_s$ in pPb collisions at $\sqrt{s_{\rm NN}} = 5.02$ were presented in~\cite{Sirunyan:2018nqr}
as a function of $N_{\rm part}$, and pPb particle flow measurements $v_2\lbrace {2, \vert\Delta\eta_s\vert > 2}\rbrace$ were presented in~\cite{Sirunyan:2017uyl} against 
the experiment-specific number of offline tracks  $N^{\rm offline}_{\rm trk}$.  Information tabulated in Ref.~\cite{Chatrchyan:2013nka} relates $N^{\rm offline}_{\rm trk}$
to the fraction of the total pPb cross section (i.e. centrality), and this average centrality can be converted into an average $N_{\rm part}$, as done for instance in Ref.~\cite{Adam:2014qja}.
However,   CMS data points for $N_{\rm offline}^{\rm track} \gtrsim 150$ that correspond to the $< 1\%$ most active events of the total cross section cannot be related
meaningfully to $N_{\rm part}$. As $v_2\lbrace {2, \vert\Delta\eta_s\vert > 2}\rbrace$ for $N_{\rm offline}^{\rm track} \gtrsim 150$ is approximately constant~\cite{Sirunyan:2018nqr}
information from all data points at larger $N^{\rm offline}_{\rm trk}$ can be regarded as
grouped into the data point of largest $N_{\rm part}$. In this way, both particle flow and transverse energy in pPb at the LHC is known as a function of $N_{\rm part}$. \\
 The following discussion uses the $N_{\rm part}$-dependence of $v_2\lbrace {2, \vert\Delta\eta_s\vert > 2}\rbrace$ and $dE_\perp/d\eta_s$
thus obtained {\it without} relating $N_{\rm part}$ to a geometrical picture of the collision. The same factor $c_{\rm e-corr,2}=1.34$ determined for $\sqrt{s_{\rm NN}} = 5.02$ PbPb collisions is
used to determine energy flow, since the absence of sufficiently fine-binned $p_\perp$-differential $v_2$ data does not allow one to perform the same determination of $c_{\rm e-corr,2}=1.34$ 
as for PbPb. We  anticipate that none of our conclusions will depend on the precise numerical value of $c_{\rm e-corr,2}=1.34$ used in pPb.  
\item
{\it Data for AuAu collisions at RHIC}\\
The centrality dependence of $dE_\perp/d\eta_s$ is taken from Ref.~\cite{Adams:2004cb}. 
We infer energy flow $v_2$ for $\sqrt{s_{\rm NN}} = 200$ GeV Au+Au collisions at RHIC by paralleling closely the procedure described for PbPb collisions at the LHC. 
STAR $p_\perp$-differential charged hadron $v_2(p_\perp)$ data~\cite{Abelev:2008ae} in the wide 10-40\% centrality bin is weighted with charged single inclusive hadron spectra from~\cite{Adams:2003kv}
according to eqs.~(\ref{eq*2}) and (\ref{eq*1}) to infer a correction factor $c_{e-corr,2} = 1.5$ for $p_\perp^{\rm min} = 0.1$ GeV. This factor is applied to particle flow data of 
$p_\perp$-integrated $v_2(\vert\Delta \eta_s\vert)$ presented in 11 $N_{\rm part}$-bins by PHOBOS~\cite{Back:2004mh}. To allow for a better comparison with other calculations 
presented in this paper, the $N_{\rm part}$-dependence is then converted into the impact parameter dependence obtained from an optical Glauber calculation~\cite{Kharzeev:2000ph}.
\end{enumerate}
\subsubsection{Non-ideal equation of state}
\label{sec4b2}
The kinetic transport theory studied here is for a conformal system with $c_s^2= 1/3$. While this assumption may be
relaxed in future work, we have not calculated yet $v_n/\epsilon_n (\hat\gamma)$ for more realistic equations of state in kinetic theory. Recent  lattice QCD results~\cite{Borsanyi:2013bia,Bazavov:2014pvz} indicate values $c_s^2(T)$ which decrease gradually with decreasing temperature, reaching values 0.23, 0.25 and 0.29 at
temperatures $T \approx 200$, $225$ and $300$ MeV, respectively. A more realistic formulation of the expansion would dynamically average over this temperature-dependence. To estimate 
the difference that such a formulation could make, we calculate first from ideal fluid dynamics the correction factor $c_{\rm eos}$ with which the 
ideal hydro baselines (blue dashed lines in Fig.~\ref{fig5}) must be multiplied to reflect a reduced sound velocity $c_s^2$. We find that in ideal fluid dynamics
\begin{equation}
\left( \frac{v_n}{\epsilon_n}\right)_{c_s^2} = c_{\rm eos} \left( \frac{v_n}{\epsilon_n}\right)_{c_s^2=1/3}
\end{equation}
with $c_{\rm eos} = 0.86$ (0.93) for  $c_s^2 = 0.25$ (=0.29), respectively. For sufficiently large $\hat\gamma$, the kinetic theory result for $v_n/\epsilon_n$ approaches 
this ideal hydro baseline from below, and must be affected by the same factor $c_{\rm eos}$ in case of a non-ideal equation of state. To estimate uncertainties, we 
now assume that this correction factor $c_{\rm eos}$ can be applied to all values of $\hat\gamma$. We choose in the following $c_{\rm eos} = 0.9$ as a default (corresponding to  $c_s^2 = 0.27$),
and we compare to $c_{\rm eos} = 0.8$, which---according to these numerical estimates---may be viewed as a conservative overestimation of the expected correction.
\begin{figure}[t]
\includegraphics[width=0.48\textwidth]{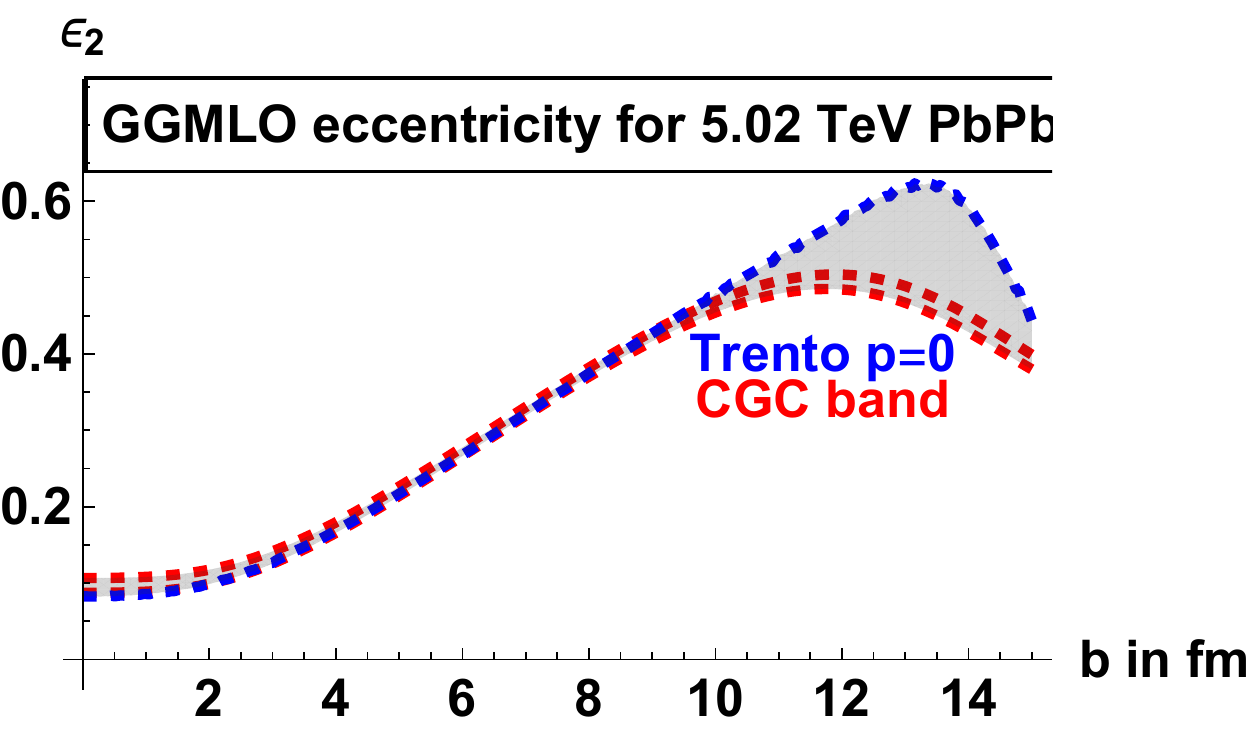}
\includegraphics[width=0.48\textwidth]{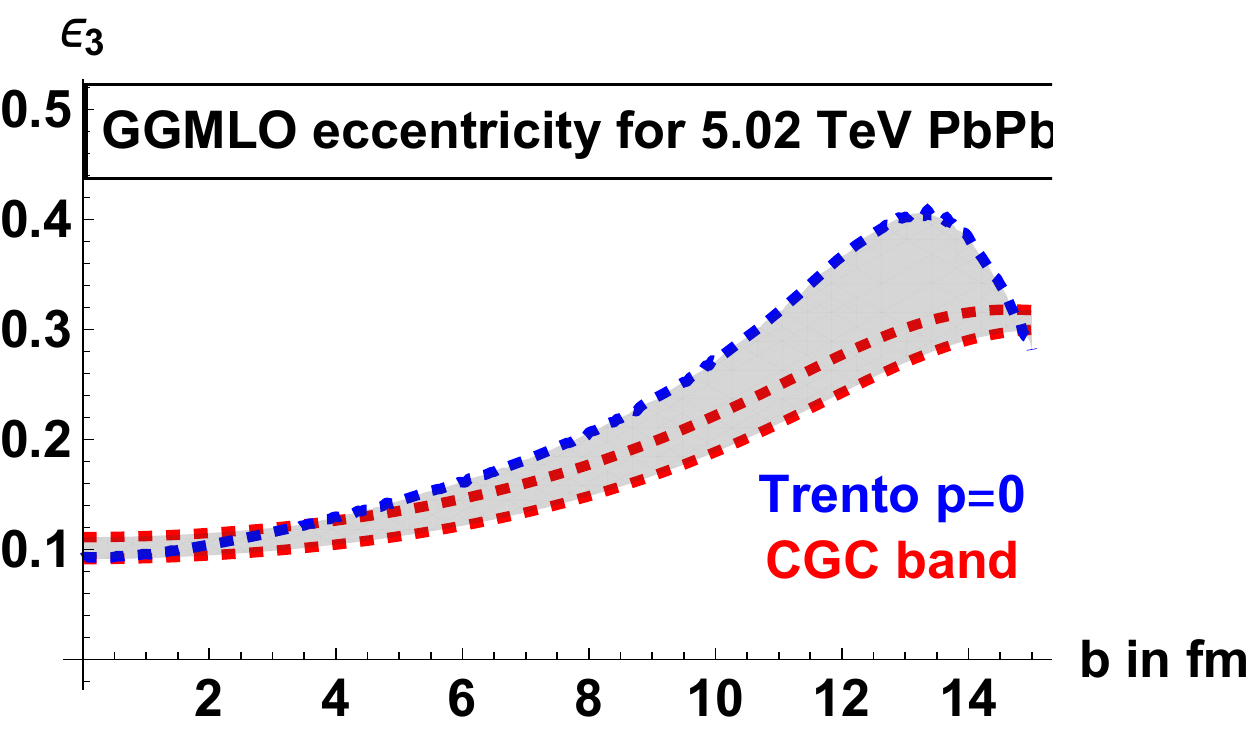}
\caption{The centrality dependence of the elliptic eccentricity $\epsilon_2$ (upper plot) and triangular ecdentricity (lower plot) for two models discussed in the text. Data taken from Ref.~\cite{Giacalone:2019kgg} and displayed over a wider range of impact parameter. The grey band is used to characterize the model-uncertainty of these eccentricities in the following study. 
}
\label{fig9}
\end{figure} 
\subsubsection{Non-linear response to spatial eccentricities}
\label{sec4b3}
The kinetic theory calculations of $v_n/\epsilon_n (\hat\gamma)$ in Fig.~\ref{fig5} determine only the linear response  $v_n$ to an initial spatial eccentricity $\epsilon_n$
\begin{equation}
\frac{\partial v_n(\epsilon_n)}{\partial \epsilon_n}\Bigg |_{\epsilon_n = 0}.
\end{equation}
Both theoretical and experimental evidence indicates that the true response has a subleading non-linear component that is non-negligible 
in particular for $v_2$ in mid-central collisions where eccentricities are large. Future calculations may allow us to determine these non-linear corrections. Here we estimate
them by assuming that they have the same size in kinetic theory as in fluid dynamic simulations in which they have been quantified. To this end, we recall first that symmetry arguments 
dictate that the leading non-linear corrections to $v_2$ is $O\left( \epsilon_2^3\right)$. From fitting the visible non-linearities in the Fig. 17 of of the fluid dynamic study~\cite{Niemi:2015qia}, we find
\begin{eqnarray}
	\frac{v^{\rm particle}_2(\epsilon_2)}{\epsilon_2} &=& \frac{\partial v_n}{\partial \epsilon_n}\Bigg |_{\epsilon_n = 0} f_{\rm corr}(\epsilon_2)\, , \nonumber \\
	%f_{\rm nl}(\epsilon_2) &=& 1 + 0.75 (\epsilon_2)^2\, .
	f_{\rm corr}(\epsilon_2) &=& 1 +c_{\rm nl} \, \epsilon_2^2\, , \nonumber \\
	c_{\rm nl}&=&0.75\, .
\end{eqnarray}
We adopt this as our default value.
The same correction is also applied to $v_3$, where non-linear corrections are much less important since $\epsilon_3$ is smaller. 
\subsubsection{Centrality dependence of initial eccentricities $\epsilon_2$ and $\epsilon_3$}
In comparing data to the kinetic theory prediction of $v_n/\epsilon_n (\hat\gamma)$, the model-dependence of the initial eccentricities $\epsilon_n$ needs to be accounted for. Here,
we consider two recent initial state models that were studied in Ref.~\cite{Giacalone:2019kgg} (referred to as GGMLO in the following) and that are based on different physics assumptions.  
They compare one of the current phenomenological standards, the  so-called Trento model for $p=0$~\cite{Moreland:2014oya}, to a model based on saturation physics for which 
the variation of the saturation scale $Q_s$ translates into a band in the centrality dependence of initial eccentricities. 
Ref.~\cite{Giacalone:2019kgg} shows the corresponding eccentricities for impact parameter $b < 10$ fm,
since saturation physics is thought to lose predictive power for more peripheral collisions. As we study in the following systems up to 80 \% centrality, we require the impact parameter
dependence of $\epsilon_n$ for $b \lesssim 13.9$ fm.  We thank the authors of Ref.~\cite{Giacalone:2019kgg} for providing their data for this extended regime which we replot in 
Fig.~\ref{fig9}. We have further obtained corresponding data sets for 2.76 TeV PbPb collisions that take for the saturation model slightly larger values of the eccentricity (data not shown)
and that we shall use. The gray band in these figures will enter as model-dependent uncertainty in the following study. 
%

%%%%%%%%%%%%%%%%%%%%%%%%%%%%%%%%%%%%%%%%%%%%%%%%%%%%%%%%%%%%%%%%%
\subsection{Confronting kinetic transport to data on $v_n/\epsilon_n(\hat\gamma)$ in PbPb at the LHC}
\label{sec4c}
We can determine now for each centrality bin in each collision system
\begin{enumerate}[i.]
\item
 the energy flow $v_n$ from data,
 \item the transverse energy per unit rapidity $\frac{dE_\perp}{d\eta_s}$ from data,
 \item the rms radius $R$ of the initial distribution as a function of centrality from a Glauber model,
 \item the eccentricity $\epsilon_n$.
\end{enumerate}

\begin{figure}[t]
\includegraphics[width=0.45\textwidth]{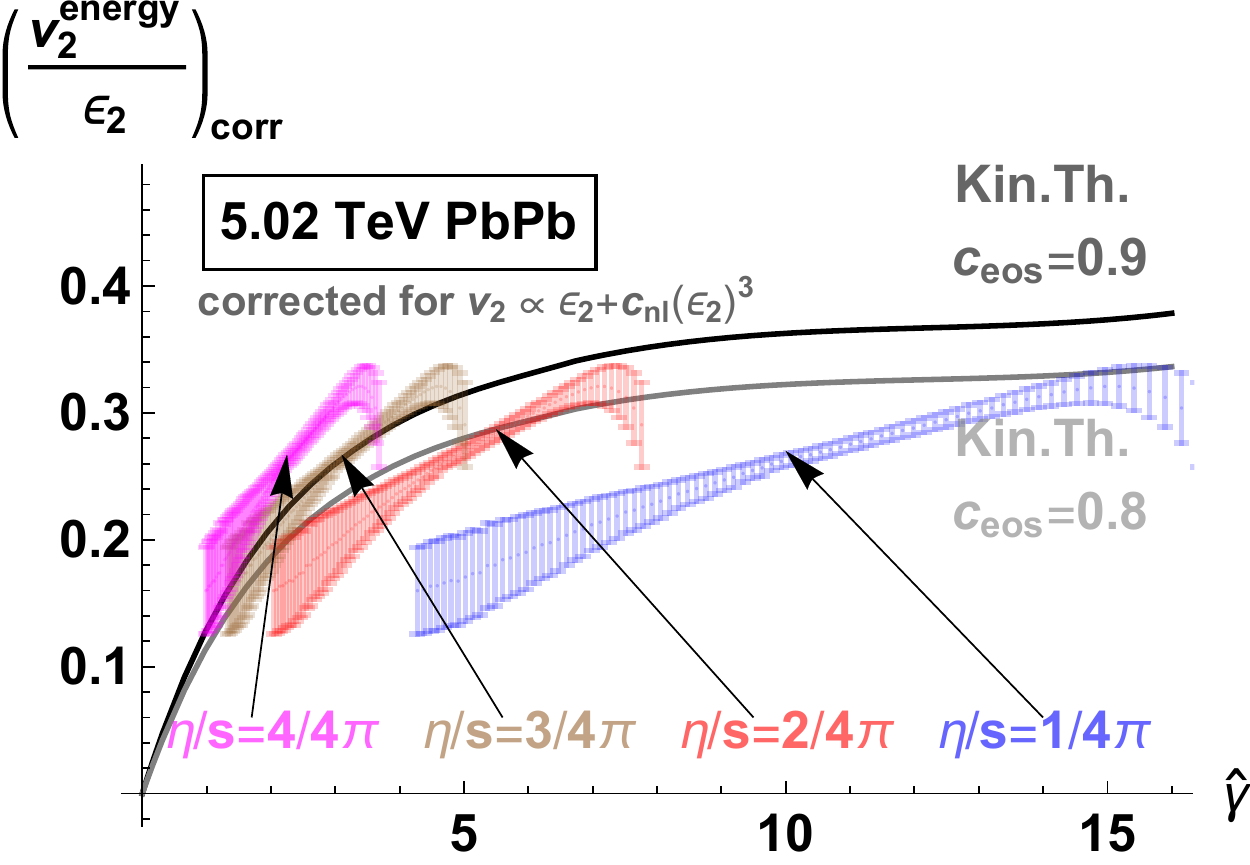}
\includegraphics[width=0.45\textwidth]{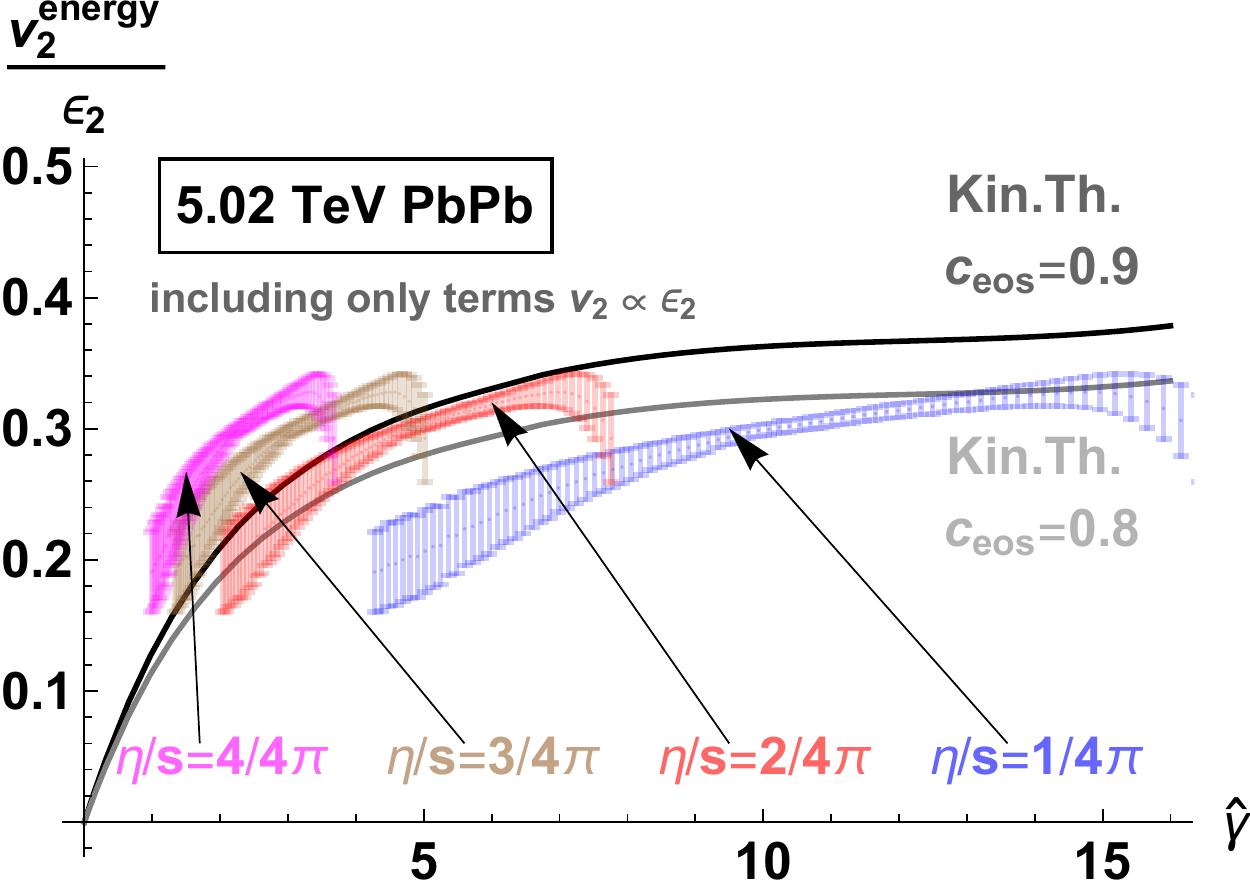}
\caption{The ratio $v_2/\epsilon_2$ as a function of $\hat\gamma$. Results for kinetic theory, supplemented by a factor $c_{\rm eos} = 0.8$ and $0.9$ to correct for a non-ideal equation of state, are compared to ratios  $v_2/\epsilon_2(\hat\gamma)$ constructed from ALICE data. Uncertainty bands arise from uncertainties in the centrality-dependent eccentricity only. Non-linearities $f_{\rm corr}(\epsilon_2)$ in
the $\epsilon_2$-dependence of $v_2$ are corrected for according to eq. ~(\ref{eq7X}) (upper plot). To illustrate the size of this necessary correction, 
the lower plot shows results for the case that it is neglected ($f_{\rm corr}(\epsilon_2)=1$). }
\label{fig20}
\end{figure} 

How to arrive at these quantities has been detailed in the previous subsection. The resulting basis for comparing data to the kinetic transport formulated in section~\ref{sec2} can be 
summarized in the simple expression
\begin{eqnarray}
\left( \frac{v^{\rm energy}_n}{\epsilon_n}\right)_{\rm corr} &\equiv& \frac{c_{\rm e-corr}}{f_{\rm corr}(\epsilon_n)}\, \left( \frac{v_n^{\rm particle}}{\epsilon_n} \right) \label{eq7X} \\
&=& c_{\rm eos} \left( \frac{\partial v_n }{\partial \epsilon_n} \right)\Bigg | _{\substack{\epsilon_n=0 \\c_s^2 = 1/3}}\, . \label{eq7Y}
\end{eqnarray}
Here, the quantities $c_{\rm e-corr}$, $c_{\rm eos}$ and $f_{\rm corr}$ 
are fixed as discussed in sections~\ref{sec4b1}, ~\ref{sec4b2} and ~\ref{sec4b3}, respectively, and the model-dependent eccentricities $\epsilon_2$, $\epsilon_3$ are
varied according to Fig.~\ref{fig9}. 
The first line (\ref{eq7X}) represents the centrality-dependent experimental  input. The factor $\textstyle\left( \frac{\partial v_n }{\partial \epsilon_n} \right)$ in the second line (\ref{eq7Y}) 
represents the theoretical expectation. It is calculated from kinetic transport and its centrality dependence arises solely from its $\hat\gamma$-dependence. 

\begin{figure}[t]
\includegraphics[width=0.45\textwidth]{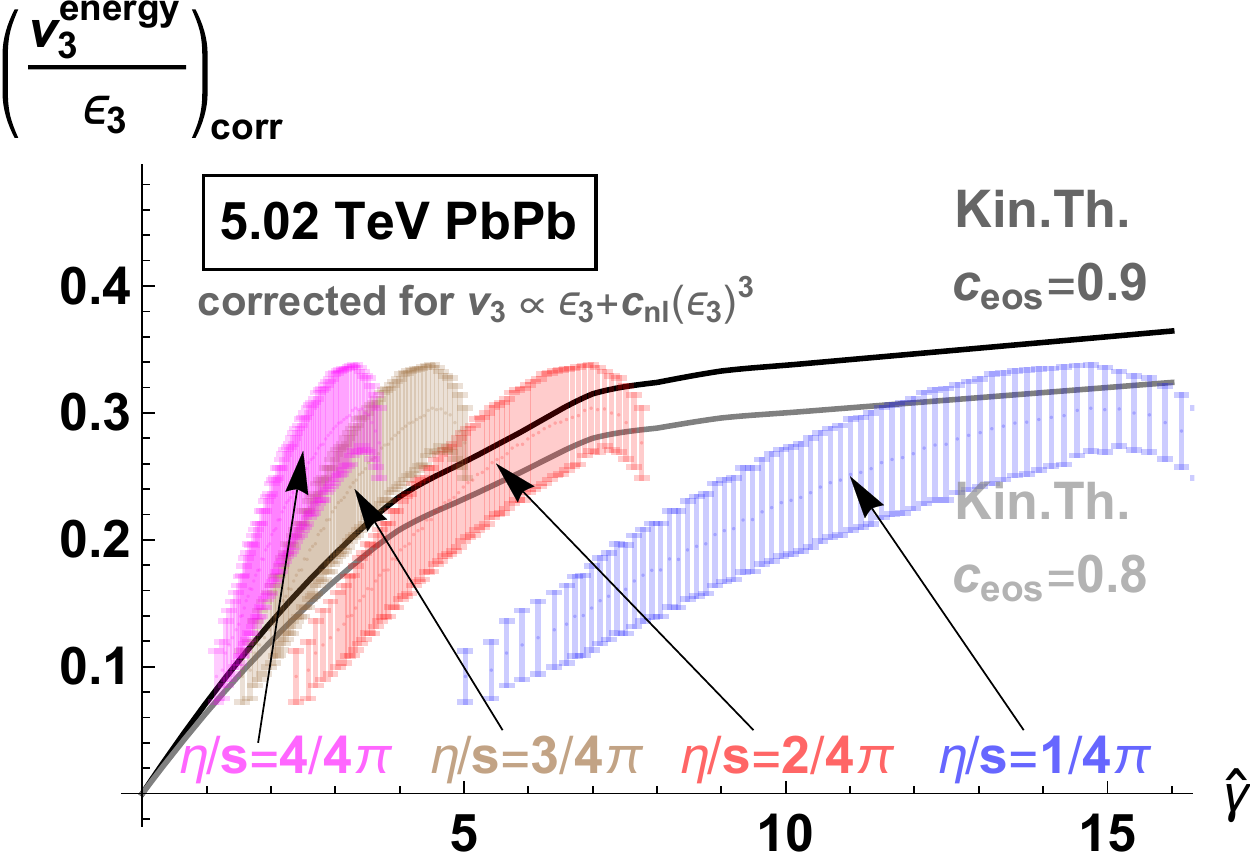}
\caption{Same as Fig.~\ref{fig20}, but for the ratio $v_3/\epsilon_3$ as a function of $\hat\gamma$. Since $\epsilon_3$ is significantly smaller than $\epsilon_2$, correcting or not correcting with eq. (\ref{eq7X}) for the non-linear $\epsilon_3$-dependence of $v_3$ makes a negligible effect.}
\label{fig21}
\end{figure} 

We use data for 80 1\%-wide centrality bins in 5.02 TeV PbPb collisions to evaluate (\ref{eq7X}) and to plot the results against the 
80 $\hat\gamma$-values of the corresponding centrality bins (see eq.(\ref{eq67}) and Fig.~\ref{fig6}). Figs.~\ref{fig20} and ~\ref{fig21} 
show the comparison of these data to kinetic theory results for $\textstyle\left( \frac{v_2}{\epsilon_2} \right)_{\rm corr}(\hat\gamma)$  and $\textstyle\left( \frac{v_3}{\epsilon_3} \right)_{\rm corr} (\hat\gamma)$.

As seen from Fig.~\ref{fig6}, the 80 bins between 1\% and 80\% centrality correspond to ranges $
4\lesssim \hat\gamma \lesssim 16$ ( $2\lesssim \hat\gamma \lesssim 7.5$,
$1.3\lesssim \hat\gamma \lesssim 5.0$) for $\eta/s = 1/4\pi$ ($\eta/s = 2/4\pi$, $\eta/s = 3/4\pi$), respectively. Accordingly, the blue, red and brown bands for 
$\eta/s = 1/4\pi$ ($\eta/s = 2/4\pi$, $\eta/s = 3/4\pi$) in Fig.~\ref{fig20} scan over the corresponding $\hat\gamma$-ranges, with the most peripheral centrality bin determining
the left-most data of the corresponding band, and the most central bin determining the right-most data point. For fixed value of $\hat\gamma$, the sizes of the plotted error bars
reflect only the modelling uncertainties in the initial eccentricities which are depicted by the grey band in Fig.~\ref{fig9}. These uncertainties get larger for more peripheral 
centralities, when the uncertainties in the models
(see Fig.~\ref{fig9}) are larger. The uncertainties are also somewhat larger in the most central bins, when $\textstyle\left( \frac{v_2}{\epsilon_2} \right)_{\rm corr}$ is obtained by dividing $v_2$ with a small
eccentricity $\epsilon_2$ of non-negligible uncertainty. 

To illustrate the size of the correction terms in eqs.~(\ref{eq7X}), (\ref{eq7Y}), we display the same data  in the lower panel of Fig.~\ref{fig20}
without correcting for nonlinearities, that is, by setting $f_{\rm corr} = 1$. Similarly we expose the theoretical uncertainty arising from the equation-of-state correction factor $c_{\rm eos}$ by displaying the kinetic theory result of Fig.~\ref{fig5} corrected with $c_{\rm eos} = 0.8$ and 0.9 in both panels. 

We find that the centrality dependence of $\textstyle\left( \frac{v_2}{\epsilon_2} \right)_{\rm corr} $ extracted from data coincides well with that from kinetic theory for $\eta/s \approx 3/4\pi$ for the default estimate $c_{\rm eos} \approx 0.9$. For the conservative overestimate of this correction, $c_{\rm eos} = 0.8$, a value close to $\eta/s= 2/4\pi$ would be favoured, while neglecting this correction ($c_{\rm eos} = 1.$) would
translate to a favoured value  $\eta/s \approx 4/4\pi$. 

\begin{figure}[t]
\includegraphics[width=0.48\textwidth]{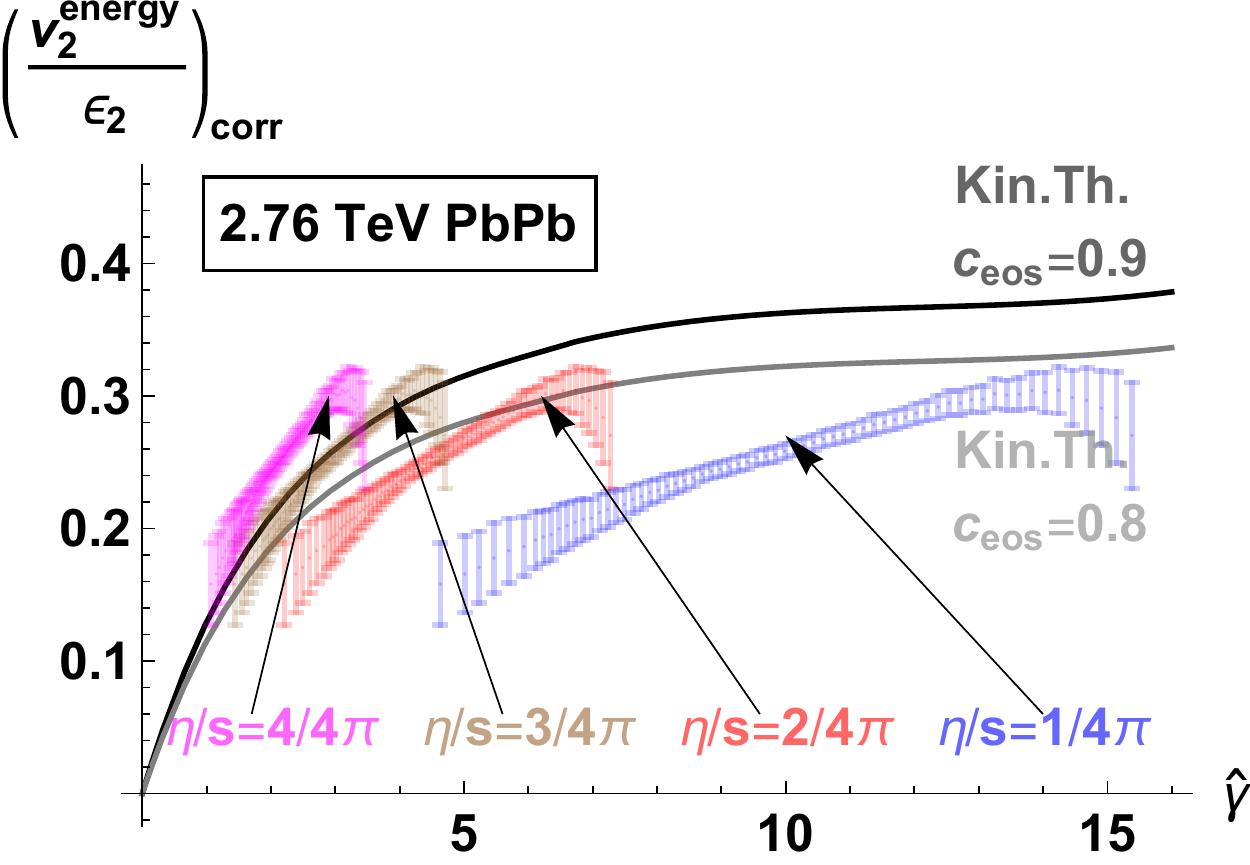}
\includegraphics[width=0.48\textwidth]{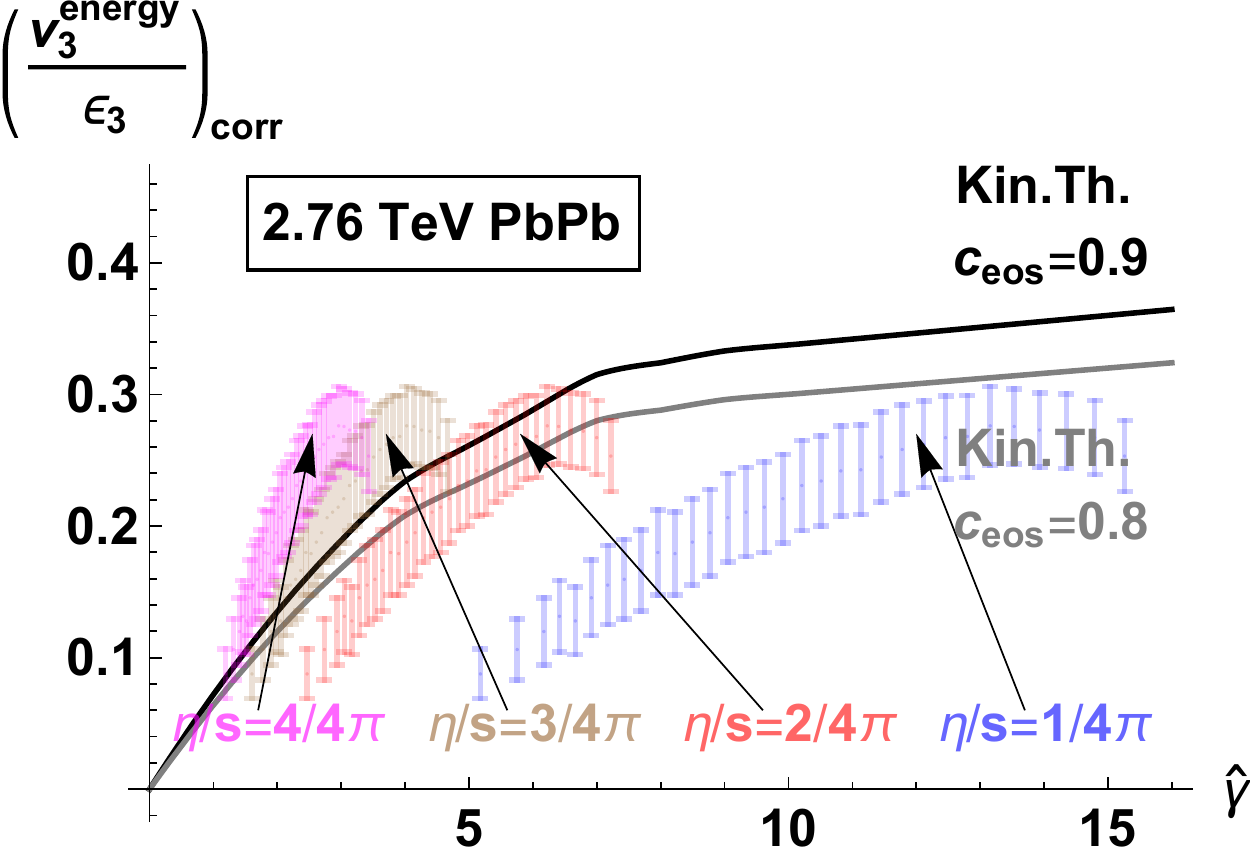}
\caption{Selected data from Figs.~\ref{fig20} and \ref{fig21} with an indication of the centrality range corresponding to different values of $\hat\gamma$.}
\label{fig22}
\end{figure} 

This analysis is repeated for triangular flow in Fig.~\ref{fig21}. Since the maximal $\epsilon_3$-values are significantly smaller than those for
$\epsilon_2$ (see Fig.~\ref{fig9}), a correction for a non-linear dependence on eccentricity $f_{\rm corr}(\epsilon_3)$ turns out to be negligible (data not shown). Also, since the gray uncertainty band shown for
the centrality dependence of eccentricity in Fig.~\ref{fig9} is in semi-central collisions much broader for $\epsilon_3$ than for $\epsilon_2$, the corresponding uncertainties in plotting 
$\textstyle\left( \frac{v_3}{\epsilon_3} \right)_{\rm corr} $ in Fig.~\ref{fig21} are more pronounced for data from semi-central events. 
We see that kinetic theory accounts for the magnitude and centrality-dependence of $\textstyle\left( \frac{v_2}{\epsilon_2} \right)_{\rm corr} $ and $\textstyle\left( \frac{v_3}{\epsilon_3} \right)_{\rm corr} $
with values of $\eta/s$ that lie within the same ball-park, although the triangular flow data prefer slightly lower $\eta/s$ or --- alternatively --- prefer a slightly smaller $\epsilon_3$ than the
one displayed in Fig.~\ref{fig9}.

The analysis of the elliptic and triangular flow are repeated for  $\sqrt{s_{\rm{NN}}}=2.76$ TeV PbPb collisions in Fig.~\ref{fig22}. Compared to the same centrality bin at the higher center of mass 
energy of 5.02 TeV, the smaller produced transverse energy results for the same centrality in slightly smaller values of $\hat\gamma$. The conclusion about the preferred parameter range for $\eta/s$
remains approximately the same for $\sqrt{s_{\rm{NN}}}=2.76$ and $\sqrt{s_{\rm{NN}}}=5.02$ TeV PbPb collisions.

\begin{figure}[t]
\includegraphics[width=0.47\textwidth]{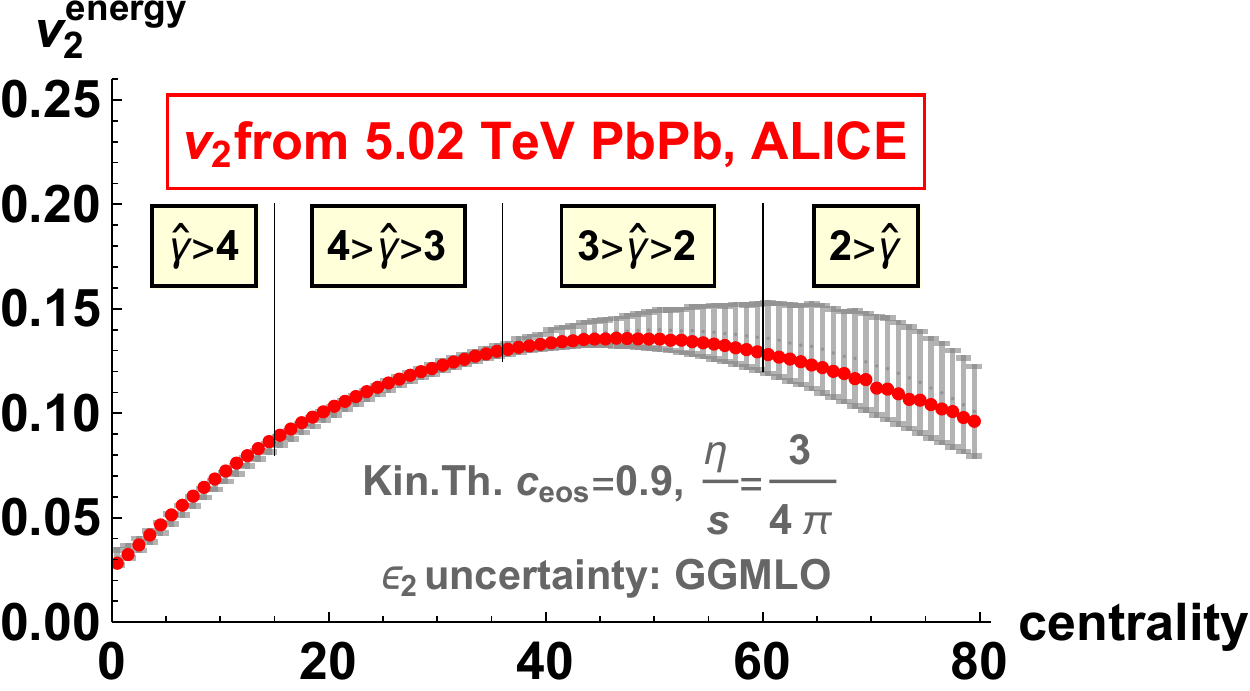}
\includegraphics[width=0.47\textwidth]{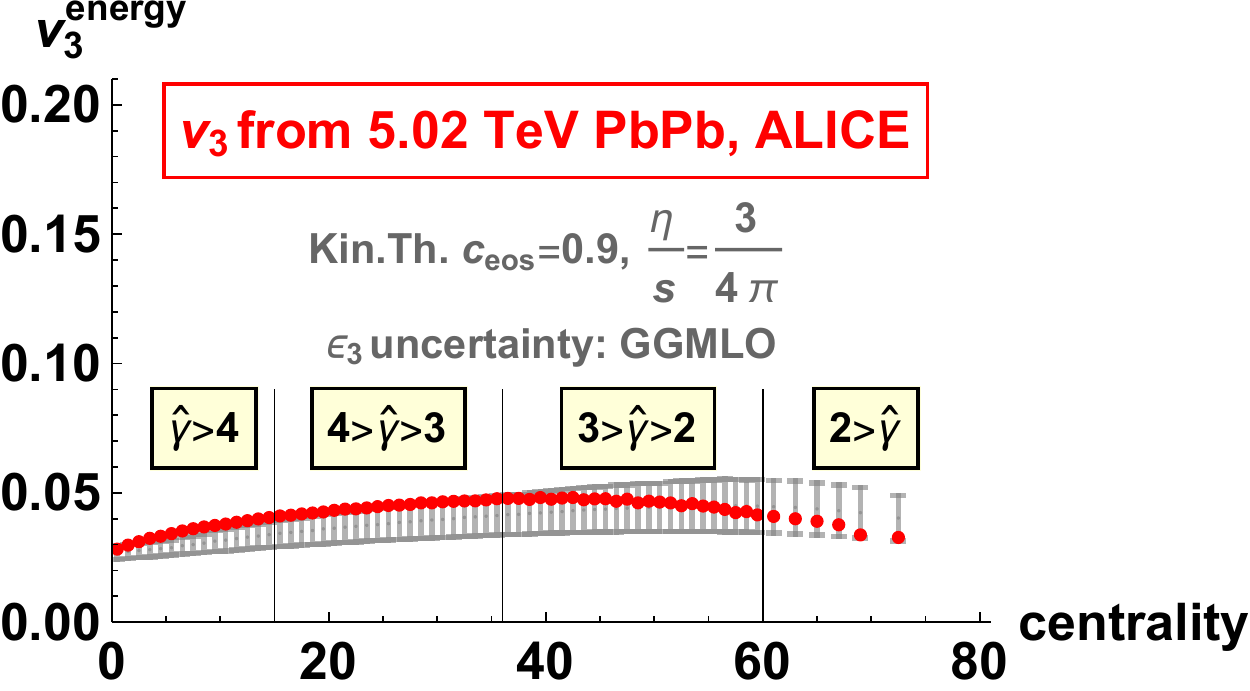}
\caption{The energy flows $v_2$ and $v_3$, calculated from kinetic theory according to eqs.~(\ref{eq7X}),(\ref{eq7Y}) as response to the GGMLO eccentricities plotted in Fig.~\ref{fig9}.
Correcting for a QCD-like equation of state with $c_{\rm eos}=0.9$, a value $\eta/s= 3/4\pi$ is favored. }
\label{fig23}
\end{figure} 

Given the simplicity of the kinetic theory studied here and the uncertainties enumerated in section~\ref{sec4b}, we do not draw too tight numerical conclusions about 
$\eta/s$ from the present study. We solely note that the parameter range $2/4\pi < \eta/s < 4/4\pi$ is favored by this analysis. The constraint $\eta/s < 4/4\pi$ is significantly lower than the value 
favored in our earlier analysis~\cite{Kurkela:2018ygx}. We note that this earlier analysis was based on the single hit line in Fig.~\ref{fig5}, and that the difference compared to the present study
arises from a combination of several improvements.
In particular, the present analysis includes non-perturbative control over the $\hat\gamma$-dependence beyond the linear approximation, it takes into account the correction factors in (\ref{eq7X}), (\ref{eq7Y}), and it exploits the full experimentally available centrality dependence. 

In combination with the $\eta/s$-dependence of $\hat\gamma$, the loose constraint $2/4\pi < \eta/s < 4/4\pi$ implies an interesting qualitative statement. 
Irrespective of where the favored value of $\eta/s$ is located in the range $2/4\pi < \eta/s < 4/4\pi$, semi-peripheral or peripheral PbPb collisions 
reach down to values $\hat\gamma < 2$ for which a fluid dynamic picture does not correspond to the physical reality, while the most central PbPb collisions reach up to values
$\hat\gamma > 4$ for which a fluid dynamic picture is a suitable interpretation of the physical reality.  

To further emphasize this point, Fig.~\ref{fig23} compares kinetic theory with $\eta/s = 3/4\pi$ to elliptic and triangular energy flow data
without deviding them by eccentricity. Centrality ranges of PbPb collision correspond---provided that the chosen values of $\eta/s$ is physically realized---to ranges of $\hat\gamma$ for which a physical picture of the collision based on fluid dynamics is ($\hat\gamma > 4$)  or is not ($\hat\gamma < 2$) suitable. The uncertainties in favoring $\eta/s = 3/4\pi$ can be judged from the parameter scans in Figs.~\ref{fig20} and \ref{fig21}. 
We note as an aside that Figs.~\ref{fig20} and \ref{fig21} for $\eta/s = 3/4\pi$ and Fig.~\ref{fig23} contain identical information; any visual impression that the agreement between theory and experiment
is better in Fig.~\ref{fig23} is erroneous. 
\begin{figure}[t]
\includegraphics[width=0.48\textwidth]{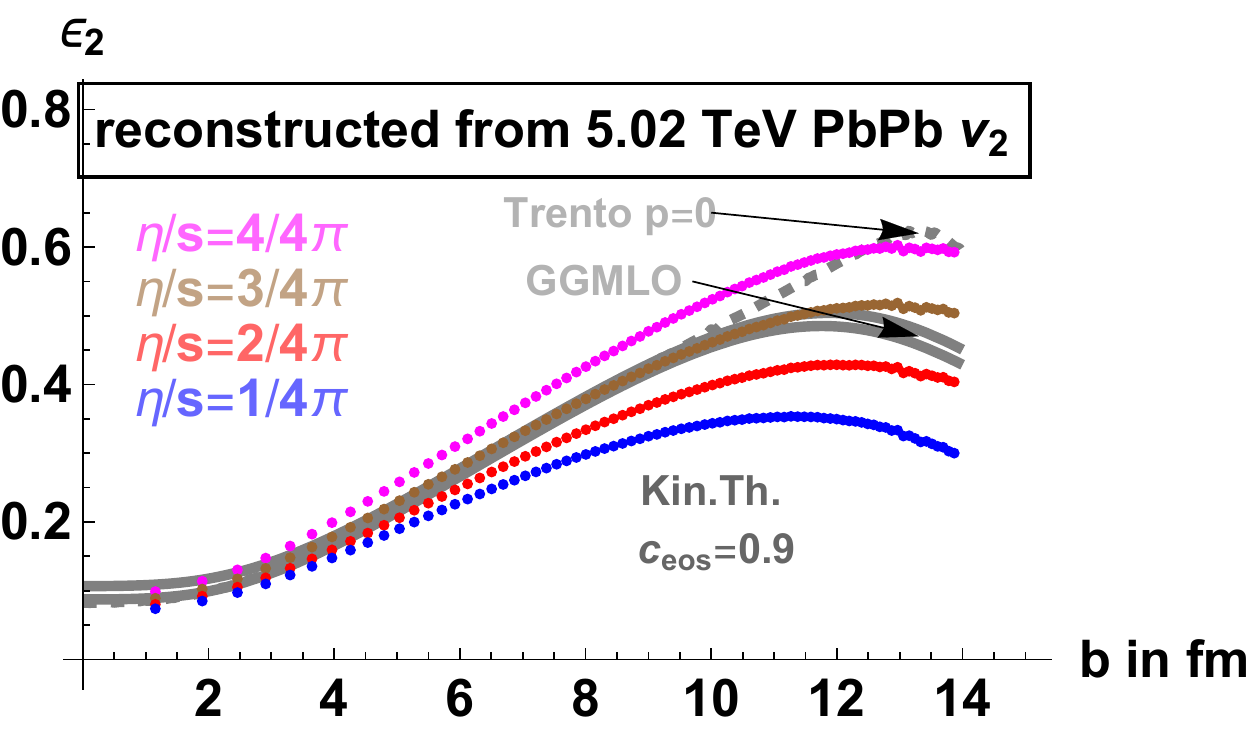}
\includegraphics[width=0.48\textwidth]{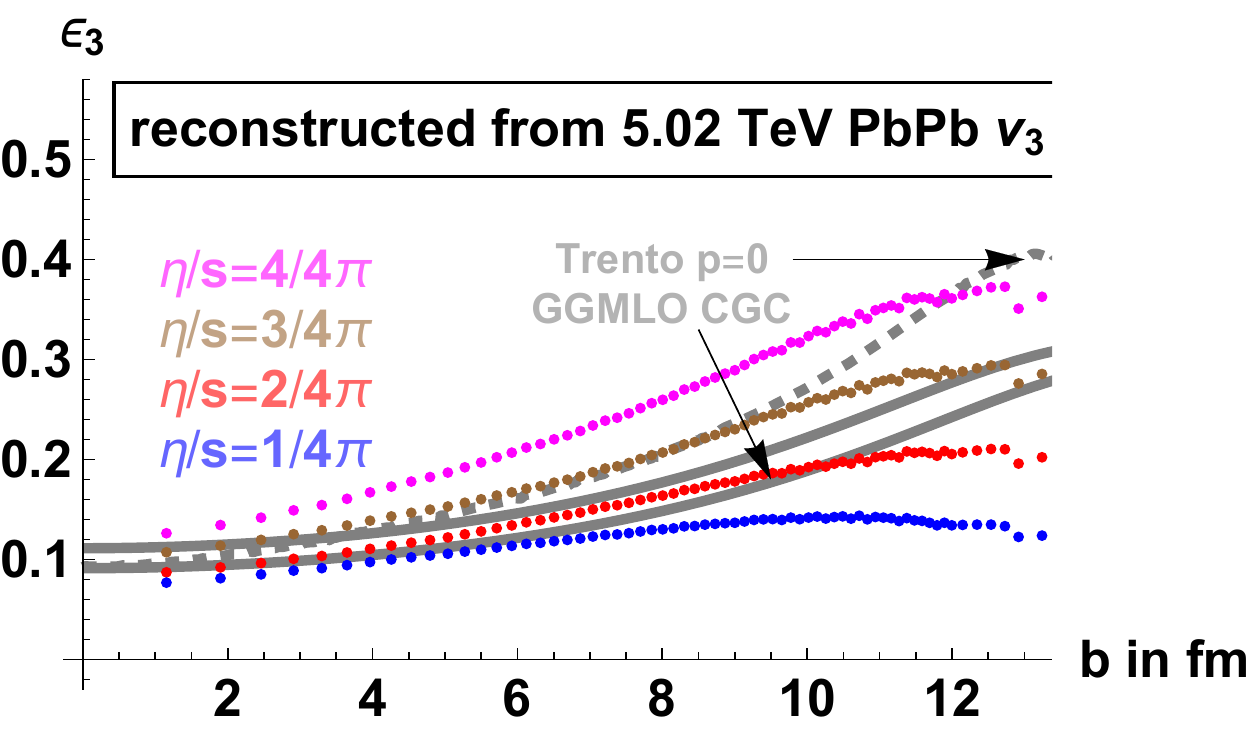}
\caption{The eccentricities $\epsilon_2$ and $\epsilon_3$ preferred by kinetic theory with $c_{\rm eos}$ according to (\ref{eq72}) and compared to two eccentricity models
of Ref.~\cite{Giacalone:2019kgg}.  }
\label{fig14}
\end{figure} 

\subsection{The centrality dependence of eccentricities $\epsilon_2$ and $\epsilon_3$ preferred by kinetic theory.} 
Rather than testing a
combination of kinetic theory and model-dependent initial eccentricity versus flow data,  as done in the previous subsection, we invert now the logic of the analysis.
We assume that the kinetic theory reflects physical reality and 
we use it to extract the eccentricity of the collision system from data. To this end, we rewrite (\ref{eq7X}),
\begin{equation}
	\epsilon_n\, f_{\rm corr}(\epsilon_n) = \frac{c_{\rm e-corr}\, v_n^{\rm particle} }{ c_{\rm eos}  } \left( \frac{\partial v_n }{\partial \epsilon_n} \right)\Bigg | _{\substack{\epsilon_n=0 \\c_s^2 = 1/3}}^{-1}\, ,
	\label{eq72}
\end{equation}
and we solve for $\epsilon_n$ for each centrality bin. As this procedure can be followed without making 
assumptions about $\epsilon_n$, it allows us in the following to compare different collision systems (PbPb and pPb at the LHC, AuAu at RHIC) without discussing the relative uncertainties 
with which the initial eccentricities of these systems are thought to be known. 
The centrality dependence of $\epsilon_n$ thus determined depends on $\eta/s$, and the physical value of $\gamma = 0.11/(\eta/s)$ can
be constrained from it to the extent to which information about the size and centrality dependence of $\epsilon_n$ is assumed to be known. 
\begin{figure}[t]
\includegraphics[width=0.48\textwidth]{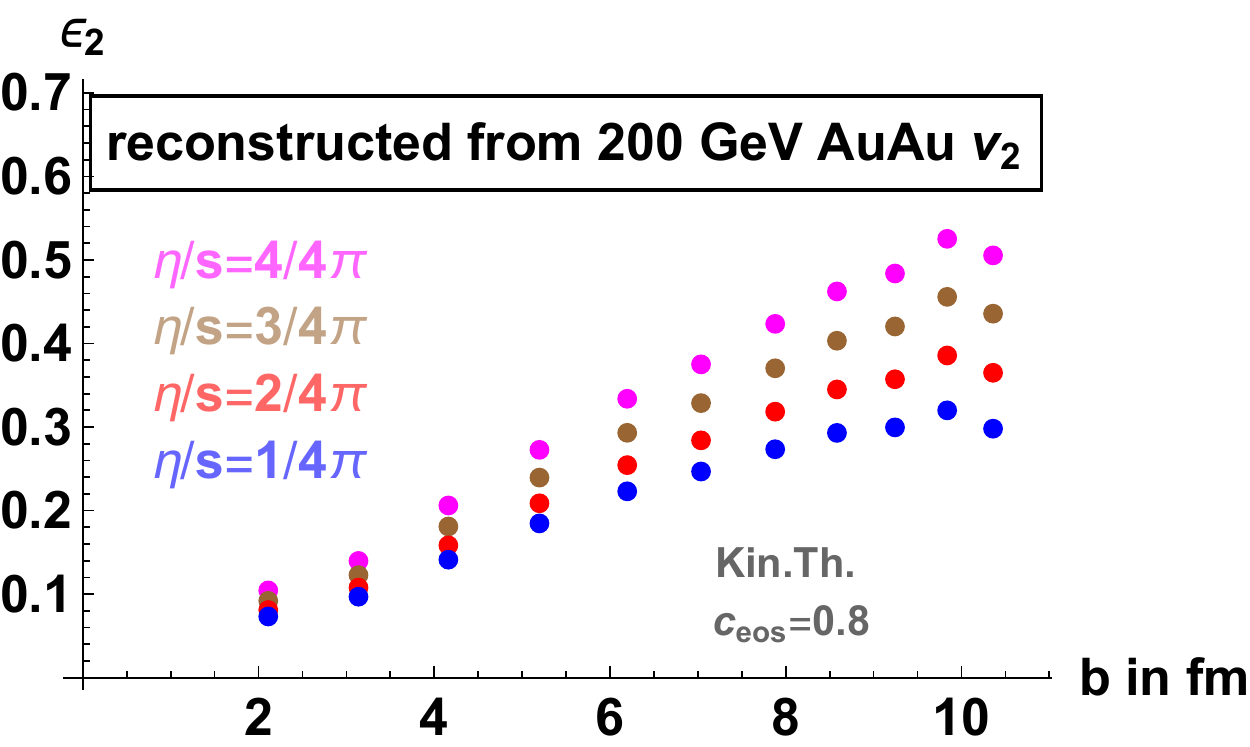}
\includegraphics[width=0.48\textwidth]{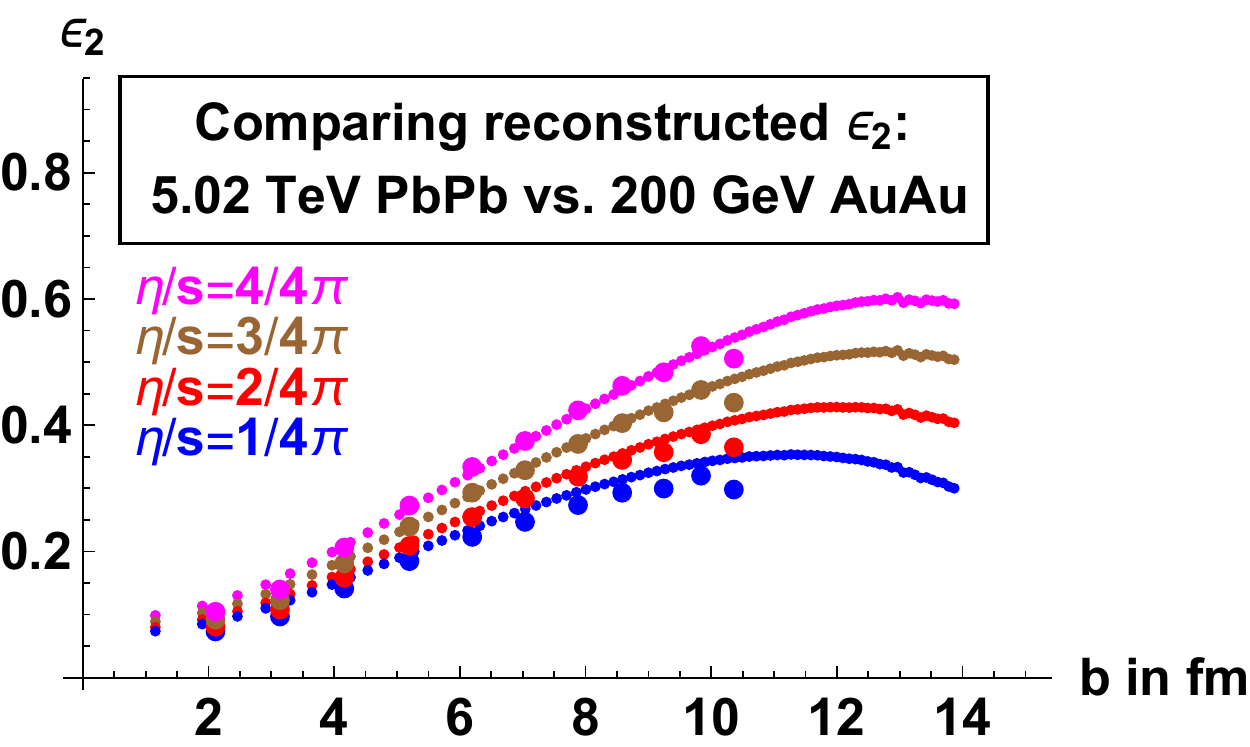}
\caption{The eccentricity $\epsilon_2$ reconstructed for 200 GeV AuAu STAR data from (\ref{eq72}) (upper panel) and compared to the eccentricity reconstructed from 5.02 TeV PbPb data in Fig.~\ref{fig14} (lower panel).   }
\label{fig15}
\end{figure} 

We start this discussion with the 5.02 TeV PbPb data, for which Fig.\ref{fig14} shows the centrality dependence of eccentricities  $\epsilon_2$, $\epsilon_3$ reconstructed from (\ref{eq72})
and overlaid with the eccentricity models depicted in Fig.~\ref{fig9}. The conclusion is obviously the same as the one reached in section~\ref{sec4c}: combining LHC data and the
eccentricity model  of Fig.~\ref{fig9} favors a value of $\eta/s$ close to $3/4\pi$. In addition, (\ref{eq72}) informs us by how much one would have to distort the model of initial eccentricity
to favor a value outside the range $2/4\pi < \eta/s < 4/4\pi$. 

We have reconstructed the eccentricity $\epsilon_2$ for 200 GeV AuAu collisions from data sets discussed in section~\ref{sec4b}. The initial geometrical distribution in AuAu collisions at RHIC
and in PbPb collisions at the LHC can differ not only because of the slighly different nuclear size, but also because particle production processes and the ensuing event-wise fluctuations are
expected to evolve over an order of magnitude in $\sqrt{s_{\rm NN}}$. Because of the uncertainties discussed in section~\ref{sec4b}, the reconstructed eccentricities displayed in
Fig.~\ref{fig15}  do not exclude small difference in the eccentricity of both collision systems. Such differences could arise within the present analysis for instance if
 the value of $\eta/s$ varies significantly with $\sqrt{s_{\rm NN}}$, or if the factor $c_{\rm eos}$ in (\ref{eq72}) takes a slighly different value for both collision systems. Also, the slightly different
 size of the Au and Pb nucleus implies that larger differences would certainly become visible if RHIC data were extended to more peripheral collisions. Despite these caveats, 
 Fig.~\ref{fig15} seems to supports an approximately $\sqrt{s_{\rm NN}}$-independent elliptic eccentricity.

\subsection{Reconstructing eccentricity for pPb collisions}
The eccentricity in pPb collisions is particularly difficult to infer from multiplicity distributions close to mid-rapidity, since the correlation between event multiplicity and transverse geometry 
weakens when the dispersion of the multiplicity produced at fixed 
impact parameter approaches the width of the entire multiplicity distribution. In plotting $\hat\gamma$ for pPb collisions in Fig.~\ref{fig8}, we had assumed already
the limiting case that this correlation is negligible so that events of different multiplicity (classified experimentally as a function of $N_{\rm part}$) correspond to
systems of initially similar transverse width $R$.

\begin{figure}[t]
\includegraphics[width=0.48\textwidth]{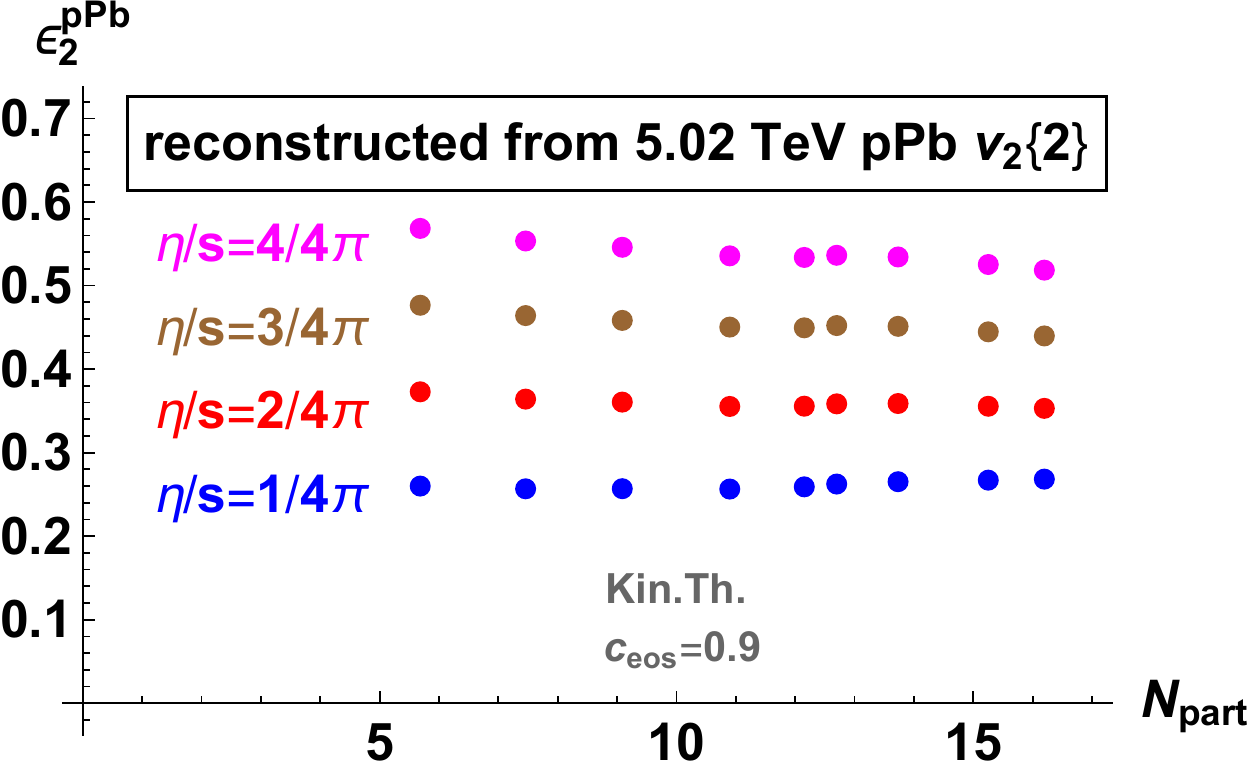}
\caption{The eccentricities $\epsilon_2$ reconstructed for pPb CMS data }
\label{fig16}
\end{figure}

In reconstructing eccentricity according to (\ref{eq72}) from the 5.02 TeV pPb data discussed in section~\ref{sec4b}, Fig.~\ref{fig16} gives independent support to
this assumption of a multiplicity-independent initial geometry of the active area in pPb collisions. While the value of the measured
eccentricity $\epsilon_{2}^{\rm pPb}$ depends on the assumed value of $\eta/s$, the qualitative feature that it
does not depend on the final state multiplicity remains unchanged for all values of $\eta/s$. This suggests that the 
multiplicity in a pPb collision is not determined by
geometric overlap but rather by fluctuations which are independent of the geometrical fluctuations leading to the eccentricity. 

The flat eccentricity in Fig.~\ref{fig16} is qualitatively different from the one predicted by Glauber-type models for which $\epsilon^{\rm pPb}_2$ 
decreases with increasing $N_{\rm part}$~\cite{Bozek:2011if}. These Glauber-type models have been used to initialize phenomenologically successful fluid dynamic simulations of pPb 
collisions~\cite{Bozek:2011if,Bozek:2013uha}. In this way, these studies support the view that pPb collisions can be viewed as creating initially a small fluid droplet, in which fluid-dynamic
expansion translates an impact parameter-dependent spatial anisotropy into a centrality-dependent momentum anisotropy. 
However, if we try to translate the geometrical picture of pPb collisions documented in~\cite{Bozek:2011if,Bozek:2013uha} into the parameters
of the present study, we systematically arrive at values of $\hat\gamma$ that are too small to support a fluid dynamic picture. Rather, insensitive to initial geometry, the kinetic theory
studied here builds up flow under spatio-temporal constraints in which fluid-dynamic modes cannot dominate collectivity, and the multiplicity-dependence of $v_2/\epsilon_2$ can be
viewed as arising solely from the multiplicity-dependent scattering probability that enhances collectivity within a system of fixed eccentricity.

 \subsection{Comment on measures of system size in small systems}
 \label{sec4f}
As explained in Section \ref{sec2},  the opacity $\hat \gamma$ is the only dimensionless variable that observables may
depend on, and hence it is the unique variable in which the system size can be measured. It is natural to 
think of system size in units of mean free path. In the present model, this quantity at the relevant time when
flow is built up ($\tau =R$) at $r=0$ reads
\begin{align}
\frac{R}{\tau_R(\tau = R)}  = \gamma R  \varepsilon^{1/4}(\tau = R).
\end{align}
This quantity can be calculated exactly in the current kinetic theory setup. Denoting $\frac{\varepsilon(\tau = R)R}{\varepsilon_0 \tau_0} = f_{0\rightarrow R}(\hat \gamma)$, we have%
\footnote{The factor $\pi$ on the right hand side of (\ref{eq80}) arises from a Gaussian density profile, and slightly different factors arise different profiles, \emph{e.g.}, $0.95\pi$ for Woods-Saxon distribution of (\ref{eq43}). }
\begin{align}
\frac{R}{\tau_R(\tau = R)} & = \hat \gamma [ f_{0\rightarrow R}(\hat \gamma)]^{1/4} \nonumber  \\
& = \gamma \left[ \frac{ f_{0\rightarrow R}} {f_{\rm work}}\right]^{1/4} \left[ \frac{R}{\pi} \frac{d E_\perp}{d \eta_s}\right]^{1/4}\, .
\label{eq80}
\end{align}
The fraction $ f_{\rm work}(\hat \gamma)$ is closely related to $f_{0\rightarrow R}$, and in the $\hat\gamma$-range studied here, we find $e^{-1}f_{\rm work}(\hat \gamma)= f_{0\rightarrow R}(\hat \gamma)$ up to $15\%$  corrections.  Therefore, $\frac{R}{\tau_R} \approx \hat \gamma$ up to small corrections in large systems, for which the scaling of eq.~(\ref{eq:scaling}) suggests that this quantity is proportional to 
\begin{equation}
\frac{R}{\tau_R} \sim \hat \gamma^{8/9}, \qquad \hbox{ \rm  [for $\hat\gamma \gg 1$].}
\end{equation} Like the opacity $\hat\gamma$ used in the plots and defined in eq.~(\ref{eq67}), 
the ratio $\frac{R}{\tau_R(\tau = R)}$ is a suitable scaling variable in the sense that it depends only on $\hat \gamma$ and that it increases monotonically with $\hat \gamma$.

The energy density at the origin at time $\tau=R$ may be estimated also in other ways.  For instance, following Bjorken, one may 
assume that the system decouples at time $\tau = R$ and that it evolves free streaming from this time onwards. A simple geometrical picture yields then
\begin{align}
\varepsilon_\perp(R) R \sim \frac{1}{A_\perp} \frac{d E_\perp} {d \eta_s}\, ,
\end{align}
where $\varepsilon_\perp(R)$ is the transverse energy density constructed from the total transverse energy $dE_\perp/d\eta_s$
within an appropriately chosen transverse area $A_\perp \sim \pi R^2$. Using this estimate to evaluate $\tau_R = \textstyle\frac{1}{\gamma \varepsilon^{1/4}}$, 
one finds 
\begin{align}
\frac{R}{\tau_R} \sim  \gamma \left[ \frac{R}{\pi a} \frac{d E_\perp}{d \eta_s}\right]^{1/4}\, ,
\label{eq82}
\end{align}
where $a = \varepsilon_\perp / \varepsilon = [\frac{2}{\pi},1] $. Here, the value $a = 1$ is realized for a maximally anisotropic system ($p_z = 0$) at time $\tau = R$ and 
$a = \frac{2}{\pi}$ for an isotropic one. The parametric estimate (\ref{eq82}) is consistent with the full calculation (\ref{eq80}).
  
In general, the scaling variable (\ref{eq80}) is proportional to the fourth root of the transverse energy. It is noteworthy that for the specific case of large systems ($\hat \gamma \gg 1$)
which had time to reach local thermal equilibrium, scaling in opacity $\hat\gamma$ becomes interchangeable with a scaling in $dN/d\eta_s$. To see this, let us specify the 
assumption that the system has reached thermal equilibirium such that $\langle p_\perp \rangle =  f_{\varepsilon}\,  \varepsilon^{1/4}$, where $f_{\varepsilon}$ is a constant for a conformal system. Inserting this into the expression for $\hat\gamma$ in terms of $\frac{dN}{d\eta_s}$, 
\begin{equation}
\hat\gamma = \gamma \left(\frac{1}{\pi f_{\rm work}(\hat\gamma)}\right)^{1/4} \left(R\, \langle p_\perp \rangle\right)^{1/4} \,
\left(\frac{dN}{d\eta_s} \right)^{1/4} \, ,
\label{eq87}
\end{equation}
and solving
self-consistently for $\hat\gamma$, one finds
\begin{equation}
  \hat\gamma =   c_{\rm hyd} \left[     \frac{dN}{d\eta_s}  \right]^{1/3}  \qquad \hbox{ \rm  [for $\hat\gamma \gg 1$]}\, ,
   \label{eq83}
\end{equation}
where $c_{\rm hyd} = \gamma      \left( \frac{ f_{\varepsilon}}{\pi } \frac{f_{0\rightarrow R}}{f_{\rm work}}  \right)^{1/3} $. 
Eq.~(\ref{eq87}) holds for systems of any size, but   it is only in the limit of large opacity that the scaling variable $\hat\gamma$ seizes to depend on the geometrical extent 
and depends solely on event multiplicity~\cite{Kurkela:2018wud}. 
This is consistent with the wide-spread experimental practice of using $\frac{dN}{d\eta_s}$ as a proxy of system size and as a scaling variable for comparing different
collision systems.  So, according to this kinetic theory, for small systems, measurements will take system-independent values
if plotted against $\hat\gamma$, while they take in general system-dependent values if plotted against  $\frac{dN}{d\eta_s}$. 

We finally comment on differences between our formulation of kinetic theory and formulations of massless kinetic theory with fixed cross section $\sigma$. 
In these latter models, the mean free path $\tau_R(\tau) = \textstyle\frac{1}{n(\tau)\, \sigma}$
at time $\tau$ may be expressed in terms of the particle density 
$n(\tau) \sim \frac{1}{\tau A_\perp} \frac{dN}{d\eta_s}$. In a small, low-density system, the first correction to free-streaming is
expected to be proportional to the inverse Knudsen number that may be written as 
\begin{equation}
	 \frac{R}{\tau_R(\tau = R)} \sim \frac{\sigma}{A_\perp} \frac{dN}{d\eta_s}\, , \qquad  \hbox{\rm [for kin. th. with fixed $\sigma$]}.
\end{equation}
The elliptic flow may therefore be expected to increase for small systems linearly with $\textstyle\frac{1}{A_\perp} \frac{dN}{d\eta_s}$,~\cite{Heiselberg:1998es,Voloshin:1999gs,Gyulassy:2004zy,Drescher:2007cd}
\begin{equation}
	v_2 \propto \frac{\sigma}{A_\perp} \frac{dN}{d\eta_s}\, ,\qquad  \hbox{\rm [for kin. th. with fixed $\sigma$]}.
	\label{eq86}
\end{equation}
For small systems, the parametric dependences of this $v_2$ on $R$ and on $\frac{dN}{d\eta_s}$ are characteristically different from the ones obtained for the kinetic theory studied here, see eq.~(\ref{eq87}), where we find 
\begin{equation}
	v_2 \propto \hat\gamma \propto \left(R\, \frac{dN}{d\eta_s} \right)^{1/4} \, ,\quad  \hbox{\rm [for a small conformal system]}.
	\label{eq88}
\end{equation}
The scaling of eq.~(\ref{eq88}) is that of a system with conformal symmetry, while (\ref{eq86}) arises from a model of explicitly broken conformal symmetry. 
Up to corrections due to renormalization group running,
conformal symmetry is realized in high energy QCD in small systems, as well as in high-temperature QCD. There are additional scales that arise at
lower energies, including e.g. particle masses. The introduction of a fixed cross section may be viewed as a model of non-conformal matter. 

It is important to understand whether conformal scaling (\ref{eq88}) of anisotropic flow with multiplicity can be established experimentally in the smallest 
systems. Whether or not corrections to this scaling could be established, would inform us about important elements of the microscopic dynamics underlying collectivity.

\section{Conclusions}

To ask what is the microscopic structure of quark-gluon plasma, is to ask 
how the plasma behaves away from the hydrodynamic limit. The study of 
the onset of signs of collectivity as a function of system size offers one 
setup where this question can be addressed. 
The effective description of the plasma in terms of fluid dynamics must 
eventually break, giving way to non-hydrodynamic modes to start dominating
the dynamics, and the precise way how this happens reflects the microscopic structure of the plasma.
Here, we have studied the consequences of assuming that the plasma has an 
underlying quasiparticle description. We have chosen this as our starting point
because it is consistent with free-streaming particles in small systems.

Many microscopic models may appear to be consistent with the data on small systems
and given the large unknowns in the initial condition of the collision it may be difficult to discriminate amongst them.
One qualitative feature that quasiparticle models have in common is
that the location of the hydrodynamic pole (the value of $\eta/s$) determines also the location of the 
non-hydrodynamic sector in the complex frequency plane. While the details of this correspondence may vary between different
specific quasiparticle models, the scale at which non-hydrodynamic structures become dominant $\Delta R$ is given by the mean free path $\Delta R \sim 1/(\gamma \varepsilon^{1/4})$ which also determines the specific shear viscosity $\eta/s \sim 1/\gamma$. Therefore the knowledge of transport properties of the plasma also uniquely 
determines how the free-streaming behaviour is reached in systems that have transverse sizes smaller than the mean free path. 
In Section \ref{sec4} we have seen that this prediction is not in contradiction with the available data. 

Based on this exercise, to what extent can we then conclude that the plasma does have a quasiparticle description?
In quantum field theory at weak coupling, there are additional structures in the complex plane to the hydrodynamic poles and the quasiparticle
cut. These appear at the scale of the first Matsubara mode ${\rm Im}\left[{\omega}\right] \approx - 4 \pi  T$. They reflect the quantum mechanical nature and the de Broglie wavelength of the quasiparticles. When the mean free path becomes $\sim 1/T \sim 1/\varepsilon^{1/4}$, the quantum field theory becomes strongly coupled and there is no clear separation between the quasiparticle cut and the quantum mechanical Matsubara modes. We have no firm knowledge about what happens in quantum field theory at these couplings, but we do know that in the limit of infinite coupling the nonhydrodynamics structures still lie at the scale of ${\rm Im}\left[{\omega}\right] \sim  - T$. Therefore one could expect that in a model that goes beyond quasi-particles, the scale at which non-hydrodynamical modes appear saturates such that $\Delta R \sim \min(\frac{1}{\gamma \varepsilon^{1/4}}, \frac{1}{\varepsilon^{1/4}})$. This is qualitatively in contrast to the quasiparticle models for which $\Delta R \sim \frac{1}{\gamma \varepsilon^{1/4}}$. 
It may be that such models also describe the data well, but given the good agreement with the data to the quasiparticle model, we must
acknowledge that we do not have evidence that the quark-gluon plasma does not have quasiparticle excitations. 

\begin{appendix}
\section{Solution of the kinetic theory using free-streaming coordinate system}
 \label{appendA}
 
 This appendix gives details of the formulation and solution of massless boost-invariant kinetic transport in the isotropization time approximation. This approach, summarized in section~\ref{sec2}, was first presented 
 in Ref.~\cite{Kurkela:2018qeb}. The starting point is the massless kinetic transport equation
\begin{align}
\partial_t f + \vec{v}_\perp \cdot \partial_{\vec{x}_\perp} f +v_z \partial_{z}f = -C[f] \, .
\label{eq1}
\end{align}
We consider longitudinally boost-invariant systems for which the physics at time $t$ and longitudinal position $z$ is identical to the physics at $z=0$ and time $\tau=\sqrt{t^2-z^2}$.
The invariance of the distribution  $f$ under a boost with velocity $u_z=z/t$ implies 
 $f(t,\vec{x}_\perp,z;\vec{p}_\perp,p_z) = f(\tau,\vec{x}_\perp;\vec{p}_\perp,p'_z)$ where $p'_z=(p_z-p u_z)t/\tau$. Using $\partial  p'_z/\partial z\vert_{z=0} = -p/t$,
 one can therefore rewrite eq.~(\ref{eq1})  in the central region $z=0$ as~\cite{Baym:1984np}
\begin{align}
\partial_t f + \vec{v}_\perp \cdot \partial_{\vec{x}_\perp} f -\frac{p_z}{t}\partial_{p_z}f = -C[f] \, .
\label{eq2}
\end{align}

\subsection{Free-streaming variables}
\label{sec2b}
It is technically advantageous to introduce free-streaming variables $\vec{\tilde{x}}_\perp$, $\tilde{p}_z$, such that $f$---if understood as a function of $\vec{\tilde{x}}_\perp$, $\tilde{p}_z$ 
rather $\vec{x}_\perp$, $p_z$---satisfies
\begin{align}
\partial_t f(t,\vec{\tilde{x}}_\perp;\vec{p}_\perp,\tilde{p}_z)  = -C[f] \, .
\label{eq3}
\end{align}
To find these variables, we recall first from Ref.~\cite{Baym:1984np} that the $\partial_{p_z}$-derivative 
in (\ref{eq2}) can be absorbed in taking the time derivative at constant $p_z t $,
\begin{equation}
	\partial_t f\vert_{p_z t } = \partial_t f\vert_{p_z} -\frac{p_z}{t}\partial_{p_z}f\, .
	\label{eq4}
\end{equation}
We therefore introduce the free-streaming variable 
\begin{equation}
	\tilde{p}_z \equiv p_z \frac{t}{t_0}\, ,
	\label{eq5}
\end{equation}
which equals $p_z$ at initial time $t_0$. In the absence of collisions, $C[f]=0$, eq.~(\ref{eq2}) can then be written as
\begin{equation}
	\partial_t f_{\rm free} (t,\vec{k}_\perp;\vec{p}_\perp,\tilde{p}_z) = -i \frac{\vec{p}_\perp\cdot \vec{k}}{\sqrt{p_\perp^2 +\tilde{p}_z^2 \textstyle\frac{t_0^2}{t^2}}} 
	f_{\rm free}(t,\vec{k}_\perp;\vec{p}_\perp,\tilde{p}_z)\, .
	\label{eq6}
\end{equation}
 Here, $\partial_t$ is taken at constant $\tilde{p}_z$ and 
$f_{\rm free}(t,\vec{k}_\perp;\vec{p}_\perp,\tilde{p}_z)$ is the Fourier transform of $f_{\rm free}(t,\vec{x}_\perp;\vec{p}_\perp,\tilde{p}_z)$. Integrating (\ref{eq6}) from $t_0$ to $t$, 
 we find the free-streaming solution
\begin{align}
	f_{\rm free} (t,\vec{x}_\perp;\vec{p}_\perp,\tilde{p}_z)  = \int \frac{d\vec{k}_\perp}{2\pi}  e^{i\vec{k}_\perp\cdot \vec{\tilde{x}}_\perp} 
	f_{\rm free} (t_0,\vec{k}_\perp;\vec{p}_\perp,\tilde{p}_z) \, ,
	\label{eq7}
\end{align} 
where
\begin{eqnarray}
	\vec{\tilde{x}}_\perp&=& \vec{x}_\perp  - \vec{p}_\perp \int_{t_0}^{t}  \frac{d\bar{t}}{\sqrt{p_\perp^2 +\tilde{p}_z^2 \textstyle\frac{t_0^2}{\bar{t}^2}}} 
		\nonumber \\
		&=&  \vec{x}_\perp + \frac{\hat{p}_\perp}{\sqrt{1-v_z^2}} \left(t_0 \sqrt{1-v_z^2 +v_z^2\textstyle\frac{t^2}{t_0^2} } - t \right)
		\label{eq8}
\end{eqnarray}
with $\hat{p}_\perp = \vec{p}_\perp / p_\perp = \left(\cos\phi,\sin\phi \right)$ denoting the unit direction of the transverse momentum.
We refer to   $\vec{\tilde{x}}_\perp$ as free-streaming coordinate, since it does not change with time for a free-streaming particle. Indeed, 
eq.~(\ref{eq7}) shows that the time dependence 
of the free-streaming distribution becomes trivial if $f_{\rm free}$ is written in terms of $\vec{\tilde{x}}_\perp$ and $\tilde{p}_z$,
\begin{equation}
	\partial_t f_{\rm free} (t,\vec{\tilde{x}}_\perp;\vec{p}_\perp,\tilde{p}_z) \vert_{\vec{\tilde{x}}_\perp,\tilde{p}_z} = 0\, .
	\label{eq9}
\end{equation}

Instead of $\tilde{p}_z $, it will be useful to consider the normalized quantity
\begin{equation}
	\tilde{v}_z \equiv \frac{\tilde{p}_z}{\sqrt{p_\perp^2 + \tilde{p}_z^2}} =  
	\frac{v_z \textstyle\frac{t}{t_0} }{\sqrt{1-v_z^2+ (v_z \textstyle\frac{t}{t_0})^2  }}\, .
	\label{eq10}
\end{equation}
The inverses of the free-streaming coordinates (\ref{eq8}) and (\ref{eq10}) are then given by
\begin{align}
	v_z &=  
	\frac{\tilde{v}_z \textstyle\frac{t_0}{t} }{\sqrt{1-\tilde{v}_z^2+ (\tilde{v}_z \textstyle\frac{t_0}{t})^2  }}\, ,
	\label{eq11}\\
	\vec{x}_\perp&= \vec{\tilde{x}}_\perp 
	+ \frac{\hat{p}_\perp}{\sqrt{1-\tilde{v}_z^2 }} \left(t \sqrt{1-\tilde{v}_z^2 +\tilde{v}_z^2\textstyle\frac{t_0^2}{t^2} } - t_0 \right).
	\label{eq12}
\end{align}
In the following, we take $f = f(t,\vec{x}_\perp; p, \phi,v_z)$ with $\vec{v}_\perp = \sqrt{1-v_z^2} \left(\cos\phi,\sin\phi \right)$, and
we change to the free-streaming coordinates (\ref{eq8}) and (\ref{eq10}) where appropriate. We often parametrize transverse positions
in terms of radial coordinates, $ \vec{x}_\perp = r\, \left(\cos\theta,\sin\theta \right)$ and $ \vec{\tilde x}_\perp = \tilde{r}\, \left(\cos\tilde\theta,\sin\tilde\theta \right)$.

As we have focussed on observables constructed from the energy momentum tensor 
 \begin{align}
 	T^{\mu\nu}(t,\vec{x}_\perp) &= \int \frac{d^3p}{(2\pi)^3\, p} p^\mu\, p^\nu\,  f(t,\vec{x}_\perp; p, \phi,v_z)\nonumber \\
		 	&= \int_{-1}^{1} \frac{dv_z}{2} \int \frac{d\phi}{2\pi} v^\mu v^\nu F(t,\vec{x}_\perp; \phi,v_z)\, ,
		 	\label{eq17}
 \end{align}
 we are interested in the time-evolution of the first $p$-integrated momentum moments $F$. To write the 
 equation of motion for $F$ in eq.~(\ref{eq2n}), we used that the last two terms on the left-hand side
 of  eq.~(\ref{eq2n}) arise from the integral 
 $\int \frac{4\pi p^3 dp}{(2\pi)^3} \left(- \frac{p_z}{t}\partial_{p_z}f\right)$ with the help of $\partial_{p_z} f = \textstyle\frac{\partial p}{\partial {p_z}} \partial_p f + 
 \textstyle\frac{\partial {v_z}}{\partial {p_z}} \partial_{v_z} f$.   
Free-streaming coordinates allow one to simplify this equation further. In particular, 
expressing with the help of eqs.~(\ref{eq11}) and (\ref{eq12}) $v_z = v_z(\tau,\tilde{v}_z)$ and $\vec{x}_\perp=\vec{x}_\perp(\tau,\vec{\tilde{x}}_\perp,\tilde{v}_z)$ 
 as functions of free-streaming coordinates, one finds
 \begin{eqnarray}
 	\frac{\partial \vec{x}_\perp(t,\vec{\tilde{x}}_\perp,\tilde{v}_z)}{\partial t} &=& \vec{v}_\perp\, ,
 		\label{eq19}\\
 	\frac{\partial v_z(t,\tilde{v}_z)}{\partial t} &=& -  \frac{v_z(1-v_z^2)}{t}\, ,
 		\label{eq20}
 \end{eqnarray}
 which are the prefactors of the $\partial_{\vec{x}_\perp}$- and $\partial_{v_z}$-derivatives in (\ref{eq18}). Therefore, 
expressing $F$ as a function of $t$,  $\vec{\tilde{x}}_\perp$ and $\tilde{v}_z$ allows one to write  the first three terms in eq.~(\ref{eq18}) 
as a single time derivative $\partial_t F +  \textstyle\frac{\partial \vec{x}_\perp(t,\vec{\tilde{x}}_\perp,\tilde{v}_z)}{\partial t} 
\partial_{\vec{x}_\perp} F +\textstyle \frac{\partial v_z(t,\tilde{v}_z)}{\partial t} \partial_{v_z} F$. To absorb also the fourth term of eq.(\ref{eq18}) in
a time derivative, it is useful to introduce a prefactor $c(v_z,t, t_0)$ that satisfies $c(v_z,t_0, t_0)=1$ and $\textstyle\frac{\partial c}{\partial t}
= \textstyle\frac{4v_z^2}{t} c$. The solution to this equation is $c(v_z,t, t_0) =  \left[1-\tilde{v}_z^2 +\tilde{v}_z^2 \textstyle\frac{t_0^2}{t^2} \right]^{-2}$. We therefore define the function
\begin{equation}
	\tilde{F}(t,\vec{\tilde{x}}_\perp; \phi,\tilde{v}_z) \equiv  \frac{1}{\left[1-\tilde{v}_z^2 +\tilde{v}_z^2 \textstyle\frac{t_0^2}{t^2} \right]^2}\, 
	F(t,\vec{x}_\perp; \phi,v_z)\, ,
	\label{eq21}
\end{equation}
where the arguments of $F$ on the right hand side are understood to be functions of $t$,  $\vec{\tilde{x}}_\perp$ and $\tilde{v}_z$ and $\phi$. 
In terms of $\tilde{F}$, the equation of motion (\ref{eq18}) simplifies to
\begin{equation}
	\partial_t \tilde{F}(t,\vec{\tilde{x}}_\perp; \phi,\tilde{v}_z)= -  \frac{1}{\left[1-\tilde{v}_z^2 +\tilde{v}_z^2 \textstyle\frac{t_0^2}{t^2} \right]^2}\,  
	C[F](t,\vec{x}_\perp; \phi,v_z)\, .
	\label{eq22}
\end{equation}

\subsection{Transverse energy $E_\perp$ and its azimuthal distribution}
\label{sec2c}

To calculate  the transverse energy flow (\ref{eq13}) from kinetic theory, we identify $f(t,r,\theta; p, \phi,v_z) = \textstyle \frac{dN}{d^3x\, d^3p}$. We write 
$d^3x = rdr\, d\theta\, dz$, and we recall that we work in the bin $z=0$ where $z=\tau \sinh\eta_s$ and $dz = t\, d\eta_s$. Therefore 
inserting $\frac{dN}{dy\, dp_\perp^2\, d\phi}
= \frac{dN}{d\eta_s\, dp_\perp^2\, d\phi} = \int dp_z \int rdr\, \int d\theta \, f(t,r,\theta; p, \phi,v_z)$ in eq.(\ref{eq13}), we find that the transverse energy distribution 
$\textstyle\frac{dE_\perp(t)}{d\eta_s d\phi}$ at time $t$ can be written as 
\begin{equation}
	\frac{dE_\perp(t)}{d\eta_s d\phi} = \frac{t}{2\pi} \int_0^\infty\!\!\!\!\!\! rdr \!\! \int_0^{2\pi}\!\!\!\!\!\! d\theta\! \int_{-1}^{1} \frac{dv_z}{2} \sqrt{1-v_z^2}\, F(t,r,\theta; \phi,v_z)
	\label{eq14} \, .
\end{equation}
Using free-streaming coordinates, $\vec{\tilde x}_\perp = \tilde{r} \left(\cos\tilde\theta,\sin\tilde\theta\right)$
 and the Jacobian $\textstyle\frac{\partial v_z}{\partial\tilde{v}_z} = \textstyle\frac{t_0}{t} \left(1 - \tilde{v}_z^2 + \tilde{v}_z^2 \textstyle\frac{t_0^2}{t^2}  \right)^{-2}$,
 we find
\begin{eqnarray}
	\frac{dE_\perp(t)}{d\eta_s d\phi} &=& \frac{t_0}{2\pi} \int_0^\infty \tilde{r} d\tilde{r} \int_0^{2\pi} d\tilde\theta \int_{-1}^{1} \frac{d\tilde{v}_z}{2} 
		\nonumber \\
	&& \frac{\sqrt{1-\tilde{v}_z^2}}{\left(1 - \tilde{v}_z^2 + \tilde{v}_z^2 \textstyle\frac{t_0^2}{t^2}  \right)^2} \, F(t,r,\theta; \phi,v_z)\, ,
	\label{eq16} 
\end{eqnarray}
 where the arguments of the function $F$ are now understood as functions of $\tilde{v}_z$, $\vec{\tilde{x}}_\perp$, see eqs.(\ref{eq11}) and (\ref{eq12}). 
 
\subsubsection{Harmonic decomposition of $F$ and $\textstyle\frac{dE_\perp}{d\eta_sd\phi}$}
\label{sec2d1}
 The azimuthal asymmetries $v_n$ in the Fourier decomposition (\ref{eq13}) of the transverse energy (\ref{eq16})
 arise in response to spatial azimuthal asymmetries in the distribution function $F$ at initial time $t_0$. To parametrize the latter, 
 we Fourier decompose $F$ for all times $t\geq t_0$ 
in the spatial azimuthal orientation $\theta$, 
\begin{equation}
	F(t,r,\theta; \phi,v_z) = \sum_{n=-\infty}^\infty \delta_n\, e^{i n \theta}\, F_{(n)}(t,r; \phi_r,v_z)\, .
	\label{eq23}
\end{equation}
Here, the coefficient functions $F_{(n)}$ are invariant under azimuthal rotation, and they can therefore depend on the azimuthal angles $\phi$ and $\theta$
only via the combination $\phi_r = \phi -\theta$. 
Since the term $\vec{v}_\perp \cdot \partial_{\vec{x}_\perp} F$ 
$= \textstyle\frac{v_\perp}{r} \sin(\phi-\theta) \textstyle\frac{\partial F}{\partial \theta}
+ v_\perp  \cos(\phi-\theta) \frac{\partial F}{\partial r}$ in the evolution equation (\ref{eq18}) includes trigonometric functions with arguments $\phi_r = \phi -\theta$, 
and since we are interested in an observable (\ref{eq16}) that is differential in $\phi$, we shall use $\phi$ and $\phi_r$ as independent variables. 
The kinetic equations of the $F_{(n)}$'s are then obtained from 
expanding eq.~(\ref{eq18}) to linear order in the perturbations $\delta_n$, 
\begin{align}
	&\partial_t F_{(0)} + v_\perp \cos\phi_r \partial_r F_{(0)} - \frac{v_\perp}{r} \sin\phi_r \partial_{\phi_r} F_{(0)} \nonumber \\
	&\quad - \frac{v_z(1-v_z^2)}{t} \partial_{v_z} F_{(0)}  +  \frac{4v_z^2}{t}F_{(0)} = -C[F_{(0)}]\, ,\label{eq24}
\end{align}
\begin{align}
	&\partial_t F_{(n)} + v_\perp \cos\phi_r \partial_r F_{(n)} \nonumber \\
	&\quad + in \frac{v_\perp}{r} \sin\phi_r F_{(n)}
	- \frac{v_\perp}{r} \sin\phi_r \partial_{\phi_r} F_{(n)} \nonumber \\
	&\quad - \frac{v_z(1-v_z^2)}{t} \partial_{v_z} F_{(n)}  +  \frac{4v_z^2}{t}F_{(n)} = -\delta C[F_{(n)}]\, ,\label{eq25}
\end{align}
 where the form of the collision kernels $C[F_{(0)}]$ and $\delta C[F_{(n)}]$ will be given further below.
 The evolution equation (\ref{eq25}) for $F_{(n)}$ has the symmetries
 \begin{eqnarray}
 	{\rm Re} F_{(n)}(t,r; - \phi_r,v_z) = {\rm Re} F_{(n)}(t,r; \phi_r,v_z)\, , \label{eq26}\\
 	{\rm Im} F_{(n)}(t,r; - \phi_r,v_z) = - {\rm Im} F_{(n)}(t,r; \phi_r,v_z)\, . \label{eq27}
 \end{eqnarray}
 Since the initial conditions $F_{(n)}(t_0,r; \phi_r,v_z)$ will be chosen to be $\phi_r$-independent, these symmetries are also satisfied 
 by the final solution. 
 Therefore, inserting (\ref{eq23}) into (\ref{eq14}), changing  from $\theta$- to $\phi_r$-integration (i.e., 
 understanding $F$ in (\ref{eq23}) as a function of $\phi$ and $\phi_r$), and 
 exploiting the symmetries (\ref{eq26}), (\ref{eq27}), one finds
\begin{eqnarray}
	&&\frac{dE_\perp(t)}{d\eta_s d\phi} = \frac{t}{2\pi} \int_0^\infty rdr \int_0^{2\pi} d\phi_r \int_{-1}^{1} \frac{dv_z}{2} \sqrt{1-v_z^2}  
	\nonumber \\
	&& \times \left\{ F_{(0)}(t,r; \phi_r,v_z) \right. \nonumber \\
		&& \qquad + \sum_{n=1}^\infty 2\delta_n\, \cos(n\phi) \left[ \cos(n\phi_r) {\rm Re} F_{(n)}(t,r; \phi_r,v_z) \right. \nonumber \\
		&&	\qquad \qquad \qquad \left. \left.		+ \sin(n\phi_r) {\rm Im} F_{(n)}(t,r; \phi_r,v_z) \right] \right\} \, .
	\label{eq28}
\end{eqnarray}
 In terms of free-streaming variables, the Fourier components $\tilde{F}_{(n)} (t,\vec{\tilde{x}}_\perp; \phi,\tilde{v}_z)$ defined via eqs.~(\ref{eq21}), (\ref{eq23})
 satisfy evolution equations 
\begin{equation}
	\partial_t \tilde{F}_{(0)} (t,\vec{\tilde{x}}_\perp; \phi,\tilde{v}_z)= -  \frac{C[F_{(0)}](t,\vec{x}_\perp; \phi,v_z)}{\left[1-\tilde{v}_z^2 +\tilde{v}_z^2 \textstyle\frac{t_0^2}{t^2} \right]^2}
	\, ,
	\label{eq29}
\end{equation}
and  
\begin{eqnarray}
	&& \partial_t \tilde{F}_{(n)} (t,\vec{\tilde{x}}_\perp; \phi,\tilde{v}_z)  + in \frac{v_\perp}{r} \sin\phi_r \tilde{F}_{(n)}(t,\vec{\tilde{x}}_\perp; \phi,\tilde{v}_z) \nonumber \\
	&& \qquad = -  \frac{\delta C[F_{(n)}](t,\vec{x}_\perp; \phi,v_z)}{\left[1-\tilde{v}_z^2 +\tilde{v}_z^2 \textstyle\frac{t_0^2}{t^2} \right]^2}
	\, ,
	\label{eq30}
\end{eqnarray}
which are consistent with eqs.~(\ref{eq24}), (\ref{eq25}). Since the coefficient functions $ \tilde{F}_{(n)} (t,\vec{\tilde{x}}_\perp; \phi,\tilde{v}_z)$ are invariant
under azimuthal orientations, they can depend on azimuthal angles only via $\tilde{\phi}_r = \phi -\tilde{\theta}$.
Transforming the integrand of (\ref{eq28}) to free-streaming coordinates with the help of eqs.~(\ref{eq16}) and (\ref{eq21}), we write
\begin{align}
	&\frac{dE_\perp(t)}{d\eta_s d\phi} = \frac{t_0}{2\pi}  \int_0^\infty \tilde{r}d\tilde{r} \int_0^{2\pi} d\tilde\phi_r \int_{-1}^{1} \frac{d\tilde{v}_z}{2} \sqrt{1-\tilde{v}_z^2}  
	\nonumber \\
	& \times \left\{ \tilde{F}_{(0)}(t,\tilde{r}; \tilde{\phi}_r,\tilde{v}_z) \right. \nonumber \\
		& \qquad + \sum_{n=1}^\infty 2\delta_n\, \cos(n\phi) \left[ \cos(n\phi_r) {\rm Re} \tilde{F}_{(n)}(t,\tilde{r}; \tilde{\phi}_r,\tilde{v}_z) \right. \nonumber \\
		&	\qquad \qquad \qquad \left. \left.		+ \sin(n\phi_r) {\rm Im} \tilde{F}_{(n)}(t,\tilde{r}; \tilde{\phi}_r,\tilde{v}_z) \right] \right\} \, .
	\label{eq31}
\end{align}
We shall finally determine the linear response coefficients $\textstyle\frac{v_n}{\delta_n}$ by comparing the parametrization (\ref{eq13}) to the late time
limit of (\ref{eq31}). To this end, we solve the kinetic evolution equations (\ref{eq29}), (\ref{eq30}) for $\tilde{F}_{(n)}$ in free-streaming coordinates.

\subsubsection{Isotropization time approximation (ITA)}
\label{sec2d2}
The isotropization time approximation 
\begin{equation}
	 -  C[f] = - \frac{ \left[-v_\mu u^\mu \right]}{\tau_{\rm iso}}\, \left\{f(x_\mu;p_\mu) - f_{\rm iso}(p^\mu u_\mu)
	 \right\}
	 \label{eq32}
\end{equation}
is based solely on the assumption that $f$ evolves towards a distribution $f_{\rm iso}(p^\mu u_\mu)$ which as a consequence of being
isotropic can depend only on the scalar $p^\mu u_\mu$. In general, this assumption is not sufficient to specify the functional form of $f_{\rm iso}$. 
Remarkably, however, this assumption fully specifies the functional form of the first momentum moment
\begin{eqnarray}
	F_{\rm iso}(t,\vec{x}_\perp; \phi,v_z) &=& \int \frac{4\pi p^2 dp}{(2\pi)^3} p\, f_{\rm iso}(p\, v_\mu u^\mu) \nonumber \\
	&=& \frac{\varepsilon(t,\vec{x}_\perp)}{\left(-v_\mu u^\mu \right)^4}\, .
 	\label{eq34}
\end{eqnarray}
Here, the dependence on $1/\left(-v^\mu u_\mu \right)^4$ follows from the dimensionality of the integral. To see that the integration constant is given by
the local energy density, one starts from local energy conservation $\int\frac{d^3p}{(2\pi)^3} p^\mu u_\mu \left( f-f_{\rm iso}\right)=0$, and one rewrites
the first term of this equation with the Landau matching condition $u^\mu T_{\mu}^{\nu} = - \varepsilon\, u^{\nu}$. The collision kernel for the first momentum
moments $F$ takes then the form
\begin{equation}
	 -  C[F] = - \gamma\varepsilon^{1/4} \left(-v_\mu u^\mu \right) F+  \gamma \frac{\varepsilon^{5/4} }{\left(-v_\mu u^\mu \right)^3}\, .
	 \label{eq35}
\end{equation}
To evaluate the kinetic equations (\ref{eq24}), (\ref{eq25}) for the harmonic coefficients $F_{(n)}$, we need to expand the collision kernel $C[F] $
to first order in the perturbation $\delta_n$. To this end, we note that the expansion  (\ref{eq23}) defines a corresponding expansion
\begin{equation}
	T^{\mu\nu}(t,r,\theta) = T_{(0)}^{\mu\nu}(t,r) + \sum_{n\not=0}\delta_n\, e^{i n \theta}\, T_{(n)}^{\mu\nu}(t,r)\, .
	\label{eq36}
\end{equation}
Also, the velocity $u^{\mu}$ of the locally comoving restframe can be written as an expansion in $\delta_n$,
\begin{equation}
	u^{\mu}(t,r,\theta) = u^{\mu}_{(0)}(t,r) + \sum_{n\not=0} \delta_n\, e^{i n \theta}\, u^{\mu}_{(n)}(t,r).
\end{equation}
One can then determine  the energy density $\varepsilon = \varepsilon_{(0)} + \sum_{n\not= 0} \delta_n e^{i n \theta}\,  \varepsilon_{(n)}$ and the flow velocity 
by solving to linear order in $\delta_n$ the Landau matching condition $u^\mu T_{\mu}^{\nu} = - \varepsilon\, u^{\nu}$. 
We proceed as follows:

In the coordinate system $(t,r,\theta,z)$ with mainly 
positive metric $g_{\mu\nu} = {\rm diag}\left(-1,1,r^2,1 \right)$ and with the linearly independent unit vectors
\begin{eqnarray}
	u_{(0)}^\mu &=& \textstyle\frac{1}{\sqrt{1-u^2}} \left(1,u,0,0\right)\, ,\label{eqa1}\\
	 \hat{R}^\mu &=& \textstyle\frac{1}{\sqrt{1-u^2}} \left(u,1,0,0\right)\, ,\label{eqa2}\\
	  \hat{\theta}^\mu &=&  \left(0,0,\textstyle\frac{1}{r},0\right)\, ,\label{eqa3}\\
	  \hat{z}^\mu &=&   \left(0,0,0,1\right)\, ,\label{eqa4}
\end{eqnarray}
we write the unperturbed energy momentum tensor $T_{(0)}^{\mu\nu}$ as 
\begin{eqnarray}
	T_{(0)}^{\mu\nu} &=& \varepsilon_{(0)} u_{(0)}^\mu u_{(0)}^\nu + P_{(0)r} \hat{R}^\mu \hat{R}^\nu \nonumber \\
		&&+ P_{(0)\theta} \hat{\theta}^\mu \hat{\theta}^\nu
		+ P_{(0)z} \hat{z}^\mu \hat{z}^\nu\, .
			\label{eqa5}
\end{eqnarray}
This zeroth harmonic $T_{(0)}^{\mu\nu}$ is defined in terms of the
zeroth harmonic $F_{(0)}$ according to eqs.~(\ref{eq17}) and (\ref{eq23}). The components of $T_{(0)}^{\mu\nu}$ can be expressed 
explicitly in terms of the moments
\begin{align}
	A_{n,m}(t,r) &= \int_{-1}^{1} \frac{dv_z}{2} \left(1-v_z^2\right)^{\textstyle\frac{n}{2}} \nonumber \\
	&\times \int \frac{d\phi_r}{2\pi} \cos\left(m \phi_r\right)  F_{(0)}(t,r;\phi_r,v_z)\, .
	\label{eqa6}
\end{align}
In particular, 
$T_{(0)}^{00} = A_{0,0}$, 
$T_{(0)}^{0r} = A_{1,1}$, $T_{(0)}^{rr} = \textstyle\frac{1}{2} \left( A_{2,0} + A_{2,2} \right)$ and $T_{(0)}^{\theta\theta} = \textstyle\frac{1}{2} \left( A_{2,0} - A_{2,2} \right)$.
By solving the Landau matching condition $u_{(0)}^\mu {T_{(0)\, \mu}}^\nu = - \varepsilon_{(0)}\, u^\nu_{(0)}$, one can then
determine as functions of $T^{00}_{(0)}$, $T_{(0)}^{0r}$ and $T_{(0)}^{rr}$
the eigenvalue $ \varepsilon_{(0)}$ and the variable $u$ that defines the eigenvector $u_{(0)}^\mu$ in (\ref{eqa1}),
\begin{align}
	u &=  \frac{2A_{0,0} + A_{2,0} + A_{2,2} }{4 A_{1,1}} \nonumber \\
	&- \frac{ \sqrt{\left(2A_{0,0} + A_{2,0} + A_{2,2} \right)^2-16 A_{1,1}^2}}{4 A_{1,1}}\, , \label{eqa7}\\
	\varepsilon_{(0)} &= A_{0,0} - u A_{1,1}\, ,\label{eqa8}\\
	P_{(0)r}  &= - A_{0,0} + \frac{A_{1,1}}{u}\, ,\label{eqa9}\\
	P_{(0)\theta} &= \frac{1}{2} \left(A_{2,0} - A_{2,2}  \right)\, .\label{eqa10}
\end{align}
To determine the first order corrections $\varepsilon_{(n)}$ to the unperturbed local energy
density $\varepsilon_{(0)}$ and the corrections $u_{(n)r}$, $u_{(n)\theta}$ to the radial and azimuthal component of the unperturbed flow vector (\ref{eqa1}),
we expand the Landau matching condition to first order in $\delta_n$,
\begin{equation}
	u_{(0)\mu} T_{(n)}^{\mu\nu} + u_{(n)\mu} T_{(0)}^{\mu\nu} = -\varepsilon_{(0)}   u_{(n)}^\nu - \varepsilon_{(n)}   u_{(0)}^\nu\, ,
	\label{eqa11}
\end{equation}
where 
\begin{align}
	T_{(n)}^{\mu\nu}(t,r,\theta) = \int_{-1}^{1} \frac{dv_z}{2} \int \frac{d\phi}{2\pi} v^\mu v^\nu F_{(n)}(t,r; \phi_r,v_z)\, .
	\label{eqa12}
\end{align}
Contracting equation (\ref{eqa11}) with the vectors $u_{(0)}^\mu$, 
and using that $u_{(0)\mu} u_{(n)}^\mu = - \delta_{n0}$ to ensure normalization of the perturbed and unperturbed flow field to $O(\delta_n)$,
one finds
\begin{eqnarray}
	&&\varepsilon_{(n)}(t,r) = u_{(0)\mu}  T_{(n)}^{\mu\nu} u_{(0)\nu} \nonumber \\
	&&\quad =  \int \frac{dv_z\, d\phi_r}{4\pi} \left(v_{\mu} u_{(0)}^\mu\right)^2\, {\rm Re} F_{(n)}(t,r;\phi_r,v_z)\, . \qquad  \quad
	\label{eqa13}
\end{eqnarray}
We recall that the real (imaginary) parts of $F_{(n)}(t,r;\phi_r,v_z)$ are even (odd) under $\phi_r \to - \phi_r$, respectively, see eqs. (\ref{eq26}), (\ref{eq27}).
Since $v_{\mu} u_{(0)}^\mu = -\textstyle\frac{1-v_\perp u \cos(\phi_r)}{\sqrt{1-u^2}}$ is even under $\phi_r\to-\phi_r$, the integrand in eq.(\ref{eqa13}) 
involves only the real part of $F_{(n)}$. 

Similarly, by contracting eq.~(\ref{eqa11}) with $\hat{R}^\mu$ and with $\hat\theta^\mu$, we find
\begin{eqnarray}
	&&u_{(n)r} = \frac{ u_{(0)\mu}  T_{(n)}^{\mu\nu} \hat{R}_\nu }{\varepsilon_{(0)} + P_{(0)r}} \nonumber \\
	&& \quad = \frac{-1}{\varepsilon_{(0)} + P_{(0)r}} 
	\int \frac{dv_z\, d\phi_r}{4\pi} v_{\mu} u_{(0)}^\mu\,  v_{\nu} \hat{R}^\nu  {\rm Re} F_{(n)}\, , \qquad 
	\label{eqa14}
\end{eqnarray}
and 
\begin{eqnarray}
	&&u_{(n)\theta} = \frac{ u_{(0)\mu}  T_{(n)}^{\mu\nu} \hat\theta_\nu }{\varepsilon_{(0)} + P_{(0)\theta}} \nonumber \\
	&& \quad = \frac{-i}{\varepsilon_{(0)} + P_{(0)\theta}} 
	\int \frac{dv_z\, d\phi_r}{4\pi} v_{\mu} u_{(0)}^\mu\,  v_{\nu} \hat\theta^\nu  {\rm Im} F_{(n)}\, , \qquad 
	\label{eqa15}
\end{eqnarray}
where $ v_{\nu} \hat{R}^\nu =-\textstyle\frac{u-v_\perp \cos(\phi_r)}{\sqrt{1-u^2}}$ and $v_{\nu} \hat\theta^\nu = v_\perp \sin\phi_r$.

In this way, we have expressed the zeroth harmonics (\ref{eqa1}),  (\ref{eqa7}),  (\ref{eqa8}), and the higher
harmonics~(\ref{eqa13}), (\ref{eqa14}) and (\ref{eqa15}) in terms of integral moments of the solution $F_{(n)} $ and,
a fortiori, the collision kernel 
\begin{equation}
	C[F_{(0)}] = \gamma \varepsilon^{\textstyle\frac{1}{4}}_{(0)} \left( -v_\mu u^{\mu}_{(0)} \right) F_{(0)} 
	- \gamma \frac{\varepsilon^{\textstyle\frac{5}{4}}_{(0)}}{\left( -v_\mu u^{\mu}_{(0)} \right)^3}\, ,
	\label{eq38}
\end{equation}
and
\begin{eqnarray}
	&&\delta C[F_{(n)}] = - \gamma \frac{5 \varepsilon_{(n)} \varepsilon^{\textstyle\frac{1}{4}}_{(0)}}{4 \left( -v_\mu u^{\mu}_{(0)} \right)^3}
	 + \gamma \frac{3\left( -v_\mu u^{\mu}_{(n)} \right) \varepsilon^{\textstyle\frac{5}{4}}_{(0)}}{\left( -v_\mu u^{\mu}_{(0)} \right)^4}\, ,
	 \nonumber \\
	 &&\quad  + \gamma \frac{ \varepsilon_{(n)} }{4\varepsilon_{(0)}^{\textstyle\frac{3}{4}}}   \left( -v_\mu u^{\mu}_{(0)} \right)F_{(0)}
	 + \gamma \varepsilon^{\textstyle\frac{1}{4}}_{(0)}   \left( -v_\mu u^{\mu}_{(n)} \right)F_{(0)} \nonumber \\
	 &&\quad + \gamma \varepsilon^{\textstyle\frac{1}{4}}_{(0)}   \left( -v_\mu u^{\mu}_{(0)} \right)F_{(n)}. 
	 \label{eq39}
\end{eqnarray}
is now given explicitly in terms of $F$.

\subsection{Scaling of the initial conditions and of the equations of motion}
\label{sec2f}
For numerical evaluation, it is useful to write the initial conditions and equations of motion in dimensionless variables. To this end, 
one notes first that for very early times sufficiently close to $\tau_0$, $F$ can be approximated by the free-streaming solution 
 \begin{align}
 &F(\tau, \vec x_\perp; \phi, v_z) =2 \delta(v_z)\, \varepsilon_0 \frac{\tau_0}{\tau} e^{-\frac{r^2+\tau^2-2 r \tau \cos\phi_r}{R^2}} \nonumber \\
& \quad \left\{1+\frac{\delta_2}{R^2}[(x-\tau\cos\phi)^2-(y-\tau\sin\phi)^2]\right\} .
\label{eq44}
 \end{align}
For notational simplicity, this equation is written restricted to the second harmonic and to a gaussian initial profile, but 
explicit expressions can be given easily for all harmonics and for arbitrary radial profiles. 
One now defines the function ${\cal F}$ from $F$ by rescaling all time and length scales with $R$,
  $\bar \tau \equiv \tau/R$ $,\, \, \vec \bar{\!\! x_\perp} \equiv \vec x_\perp / R$ and dividing out the prefactor $\textstyle \frac{R}{\varepsilon_0 \tau_0 }$, 
 \begin{equation}
 	{\cal F}(\bar \tau,\, \,   \vec \bar{\!\! x_\perp}; \phi, v_z) \equiv \frac{R}{\varepsilon_0 \tau_0 } F\left(\frac{\tau}{R}, \frac{\vec x_\perp}{R}; \phi, v_z\right)\, .
 	\label{eq45}
\end{equation}
 To avoid overloading our notation, we drop the bars on dimensionless time and space coordinates immediately. The arguments of ${\cal F}$ will be understood in the following as 
 being dimensionless, while the arguments of $F$ are understood to be dimensionful. In particular,  
  \begin{align}
&{\cal F}( \tau, \vec x_\perp; \phi, v_z) = \frac{2 \delta(v_z)}{\tau}\, e^{-r^2-\tau^2+2 r \tau \cos\phi_r} \nonumber \\
  & \qquad \qquad  \left\{ 1 + \delta_2 [c_{(2)}\cos(2\theta)  +s_{(2)}\sin(2\theta)]  \right\}\, ,
  \label{eq46}
 \end{align}
 where $c_{(2)}$ and $s_{(2)}$ are functions of the dimensionless $r$, $\tau$ and $\phi_r$, 
\begin{eqnarray}
c_{(2)}(r,\phi_r,\tau) &=& r^2-2 r \tau \cos\phi_r+\tau^2 \cos(2 \phi_r)\, ,  \nonumber \\
s_{(2)}(r,\phi_r,\tau) &=& 2 \tau (r - \tau \cos\phi_r) \sin\phi_r. \nonumber
\end{eqnarray}
When we insert (\ref{eq45}) into the equation of motion (\ref{eq18}), we find
 \begin{eqnarray}
&&\partial_\tau {\cal F} + \vec{v}_\perp \cdot \partial_{\vec{x}_\perp} {\cal F} - \frac{1}{\tau}v_z(1-v_z^2) \partial_{v_z} {\cal F} + \frac{4 v_z^2}{\tau} {\cal F} \nonumber \\
&& = -\hat\gamma\left[\varepsilon_{\cal F}^\frac{1}{4}(-v\cdot u) {\cal F}- \frac{\varepsilon_{\cal F}^\frac{5}{4}}{(-v\cdot u )^3}\right] \,.
\label{eq47}
\end{eqnarray}
Here, all space-time coordinates are understood as being dimensionless and the factors $\varepsilon_{\cal F}$ on the right hand side are understood to be 
calculated from $\cal F$. Since the energy density is defined as a momentum integral over $F$, we have $\varepsilon = \varepsilon_{\cal F}  \frac{\varepsilon_0 \tau_0 }{R}$, 
where $\varepsilon_{\cal F} $ is dimensionless. The prefactor $\hat\gamma$ combines the prefactor $\gamma$ of the ITA in eq.(\ref{eq35}) with a factor 
$\left( \frac{\varepsilon_0 \tau_0 }{R}\right)^{1/4}$ (arising from rescaling the energy density $\varepsilon$ in the collision kernel  (\ref{eq35}) and from changing from $F$ to ${\cal F}$),
and with a factor $R$ (arising from rescaling the space-time derivatives and explicit $\frac{1}{t}$-factors on the left hand side of (\ref{eq18})),
\begin{equation}
	\hat \gamma =R^{3/4}\gamma (\, \varepsilon_0 \tau_0)^{1/4}\, .
	\label{eq48}
\end{equation}
%
%%%%%%%%%%%%%%%%%%%%%%%%%%%%%%%%%%%%%%%%%%%%%%%%%%%%%%%%%%%%%%%%%%%%%%%%%%
\subsection{Numerical solution}
\label{sec2g}
To solve the equation (\ref{eq47}) with initial conditions (\ref{eq46}) numerically, we go with ${\cal F}$ to free-streaming coordinates 
\begin{align}
 & \tilde {\cal F}_{(0)}(\tilde r,\tilde \phi_r,\tilde v_z,\tau_0)=2\frac{\delta(\tilde v_z)}{\tau_0}e^{-\tilde r^2-\tau_0^2+2 \tilde r \tau_0 \cos\tilde\phi_r},\label{eq49}\\
  & \tilde {\cal F}_{(2)}(\tilde r,\tilde \phi_r,\tilde v_z,\tau_0)=\frac{\delta(\tilde v_z)}{\tau_0}
  		e^{-\tilde r^2-\tau_0^2+2 \tilde r \tau_0\cos\tilde\phi_r}   \nonumber \\ 
  & \qquad \qquad  \left[c_{(2)}(\tilde r,\tilde\phi_r,\tau_0)-is_{(2)}(\tilde r,\tilde\phi_r,\tau_0)\right] \label{eq50}\, ,
\end{align}
and we solve for the corresponding simpler evolution equations (\ref{eq29}), (\ref{eq30}) with collision kernels (\ref{eq38}), (\ref{eq39}). 
It is convenient to make 
 a phase redefinition 
\begin{equation}
F_{(n)} \equiv  e^{in\phi_r}\, {\cal F}_{(n)} \, ,
\label{eqb1}
\end{equation}
which we insert in eq.~(\ref{eq25}) to obtain
\begin{align}
& \partial_\tau {\cal F}_{(n)} + v_\perp \cos\phi_r \partial_r {\cal F}_{(n)}-\frac{v_\perp}{r} \sin\phi_r \partial_{\phi_r} {\cal F}_{(n)}\nonumber\\
& -\frac{1}{\tau}v_z(1-v_z^2)\partial_{v_z} {\cal F}_{(n)} +\frac{4v_z^2}{\tau} {\cal F}_{(n)} \nonumber \\
& =  -e^{-in\phi_r}\delta C[e^{in\phi_r} {\cal F}_{(n)}].
\label{eqb2}
\end{align}
In contrast to the evolution eq.(\ref{eq30}) for $\tilde F_{(n)}$, the evolution of   $\tilde {\cal F}_{(n)}$ in free-streaming coordinates is then independent of the $\sin$-term  on the right hand side,
and it reads
\begin{align}% \label{eq:Fredn}
\partial_\tau \tilde {\cal F}_{(n)} &=  -\frac{e^{-in\phi_r}}{\left[1-\tilde{v}_z^2+\left(\frac{\tau_0}{\tau} \tilde{v}_z\right)^2\right]^2}
\nonumber \\
&\delta C\left[e^{in\phi_r}\tilde {\cal F}_{(n)}\left[1-\tilde{v}_z^2+\left(\frac{\tau_0}{\tau} \tilde{v}_z\right)^2\right]^2\right]\, .
\label{eqb3}
\end{align}
This is the starting point for the numerical algorithm that we use to solve  the first momentum moments $\tilde F_{(n)}$  in a three-dimensional phase space spanned in $\tilde r$, $\tilde v_z$ and $\phi_r$. We now give details of the numerical implementation.

\subsubsection{Lattice}
According to eqs.~(\ref{eq8}) and (\ref{eq10}), one always has $\tilde v_z\approx \pm 1$ if $(v_z \tau/\tau_0)^2\gg 1-v_z^2$. If the typical value of $v_z$ at late times is parametrically larger than that due to the free-streaming $\propto\tau_0/\tau$, the corresponding value of $\tilde{v}_z$ is always close to $\pm 1$. In order to capture the detailed evolution of the system near isotropization (with a finite $v_z$), we choose a non-uniform grid for $\tilde v_z$ such that more grid points are assigned to the regions $|\tilde v_z|\sim 1$. Explicitly, in this paper $\tilde v_z$ is discretized as follows
\begin{eqnarray}
  \tilde v_{zm}=\frac{e^{m\Delta \tilde v_z}-1}{e^{m\Delta \tilde v_z}+1}\qquad\text{with $m=0,1,\cdots, n_z$}
  \label{eqb4}
\end{eqnarray}
with a spacing $\Delta\tilde v_z$. The other two variables $\tilde{r}$ and $\phi_r$ are discretized in a uniform grid%spacings with $\Delta \tilde r$ and $\Delta \phi_r$:
\begin{eqnarray}
  &&\tilde r_{m}=m \Delta \tilde r\qquad \text{with $m=0,1,\cdots, n_r$}, \label{eqb6} \\
  &&\tilde \phi_{rm}=m \Delta \phi_r\qquad \text{with $m=0,1,\cdots, n_\phi$}\, .
  \label{eqb7}
\end{eqnarray}

\subsubsection{Interpolation}

In our numerical calculation, the values of $\tilde {\cal F}_{(n)}$ on the grid points: $\tilde r_m, \phi_{rn}$ and $\tilde v_{zl}$ are calculated from the discretized version of eq. (\ref{eqb3}). In order to calculate the collision kernel, one needs the background quantities calculated from $A_{n,m}$ in (\ref{eqa6}). In terms of the free-streaming coordinates, it is expressed as 
\begin{eqnarray}\label{eq:Anm}
	A_{n,m}(r, \tau)
	&=&\frac{\tau_0}{\tau}\int_{-1}^{1} d \tilde v_z\int_0^{2\pi}\frac{d\phi}{4\pi}\frac{(1-\tilde v_z^2)^{\frac{n}{2}}}{\left[1-\tilde v_z^2+\left(\frac{\tau_0}{\tau} \tilde v_z\right)^2\right]^\frac{n-1}{2}}
	\nonumber \\
	&&\cos(m\phi)\, \tilde F_{(0)}(\tilde x_\perp, \phi,\tilde v_z,\tau)\, .\label{eqb8}
\end{eqnarray}
One also needs the perturbations in (\ref{eqa13}), (\ref{eqa14}), (\ref{eqa15}) as a function of $r$, which depend on all the three phase-space variables and the time $\tau$ according to (\ref{eq11}) and (\ref{eq12}). They are rewritten here in terms of free-streaming coordinates as 
 \begin{eqnarray} %\label{eq:ObvPert}
  \varepsilon_{(n)}(r)&=&\frac{\tau_0}{\tau}\int_0^1d\tilde v_z \int_0^\pi\frac{d\phi_r}{\pi}\left[1-\tilde v_z^2+\left(\frac{\tau_0}{\tau} \tilde v_z\right)^2\right]^\frac{1}{2}
  	\label{eqb9} \\
 && [v\cdot u_{(0)}(r,\phi_r)]^2 \text{Re}~\tilde F_{(n)}(\tilde r,\phi_r-\tilde \theta,\tilde v_z,\tau,\tau_0)\, ,\nonumber  \\
  u_{(n)\theta} &=&-\frac{\tau_0}{\tau}\frac{i}{\varepsilon_{(0)}+P_{(0)\phi}}\int_0^1d\tilde v_z \int_0^\pi\frac{d\phi_r}{\pi}\, , \label{eqb10}\\
  &&
  \left[1-\tilde v_z^2+\left(\frac{\tau_0}{\tau} \tilde v_z\right)^2\right]^\frac{1}{2}  (v\cdot u_{(0)})(v\cdot\hat{\theta}) ~\text{Im}~\tilde F_{(n)}\, , \nonumber\\
  u_{(n)r} &=& -\frac{\tau_0}{\tau}\frac{1}{\varepsilon_{(0)}+P_{(0)r}}\int_0^1d\tilde v_z \int_0^\pi\frac{d\phi_r}{\pi} \label{eqb11} \\
  && \left[1-\tilde v_z^2+\left(\frac{\tau_0}{\tau} \tilde v_z\right)^2\right]^\frac{1}{2}  (v\cdot u_{(0)})(v\cdot\hat{R}) ~ \text{Re}~\tilde F_{(n)}\, . \nonumber 
 \end{eqnarray}
The calculation of these quantities  involves the mapping of $(r,\theta=0)$ to $\tilde r$ with a given $\tilde v_z$. As a result, our numerical integration using the trapezoidal rule over $\tilde v_z$ and $\phi_r$ involves the values of $\tilde F_{(n)}$ which are not on the grid points. In this case, we use the linear interpolation for $\tilde F_{(n)}$ in $\tilde r$ and $\phi_r$.

For the time evolution, we use the 4th order Runge-Kutta algorithm.

 \newpage 
 
\end{appendix}

\end{document}